\documentclass[fleqn,usenatbib,useAMS]{mnras}
\usepackage{newtxtext,newtxmath}

\usepackage[T1]{fontenc}
\usepackage{ae,aecompl}

\usepackage{graphicx}	
\usepackage{amsmath}	
\usepackage{amssymb}	
\usepackage{multirow}   
\usepackage{pdflscape}  

\usepackage{hyperref}

\title[mJIVE--20 gravitational lens search]{A novel search for gravitationally lensed radio sources in wide-field VLBI imaging from the mJIVE--20 survey}

\author[C. Spingola et al.]{C. Spingola,$^{1}$\thanks{E-mail: spingola@astro.rug.nl}
J. P. McKean,$^{1,2}$
Minju Lee,$^{3}$
A. Deller$^{4}$
and J. Moldon$^{5}$\\
$^{1}$Kapteyn Astronomical Institute, University of Groningen, Postbus 800, NL$-$9700 AV Groningen, the Netherlands \\
$^{2}$ASTRON, Netherlands Institute for Radio Astronomy, Oude Hoogeveensedijk 4, 7991 PD Dwingeloo, the Netherlands\\
$^{3}$Nagoya University, National Astronomical Observatory of Japan Department of Physics Graduate School of Science,\\ Furocho, Chikusa ward, Nagoya city, Aichi Pref., 464-8601, Japan\\
$^{4}$Centre for Astrophysics and Supercomputing, Swinburne University of Technology, John St, Hawthorn, VIC 3122, Australia\\
$^{5}$Jodrell Bank Centre for Astrophysics, Alan Turing Building, The University of Manchester, Oxford Road, Manchester, M13 9PL, UK
}

\date{Accepted 2018 November 21. Received 2018 November 09; in original form 2018 September 25}

\pubyear{2018}

\hypersetup{draft} 
\begin{document}
\label{firstpage}
\pagerange{\pageref{firstpage}--\pageref{lastpage}}
\maketitle

\begin{abstract}
We present a novel pilot search for gravitational lenses in the mJIVE--20 survey, which observed $24\,903$ radio sources selected from FIRST with the VLBA at an angular resolution of 5~mas. We have taken the visibility data for an initial $3\,640$ sources that were detected by the mJIVE--20 observations and re-mapped them to make wide-field images, selecting fourteen sources that had multiple components separated by $\geq100$~mas, with a flux-ratio of $\leq15$:1 and a surface brightness consistent with gravitational lensing. Two of these candidates are re-discoveries of gravitational lenses found as part of CLASS. The remaining twelve candidates were then re-observed at 1.4 GHz and then simultaneously at 4.1 and 7.1 GHz with the VLBA to measure the spectral index and surface brightness of the individual components as a function of frequency. Ten were rejected as core-jet or core-hotspot(s) systems, with surface brightness distributions and/or spectral indices inconsistent with gravitational lensing, and one was rejected after lens modelling demonstrated that the candidate lensed images failed the parity test. The final lens candidate has an image configuration that is consistent with a simple lens mass model, although further observations are required to confirm the lensing nature. Given the two confirmed gravitational lenses in the mJIVE--20 sample, we find a robust lensing-rate of 1:($318\pm225$) for a statistical sample of 635 radio sources detected on mas-scales, which is consistent with that found for CLASS.
\end{abstract}




\section{Introduction}
\label{Sec:Introduction}

Gravitational lensing is the deflection of light from a distant background object (the source) by a foreground mass distribution (the lens), which is typically a galaxy or cluster of galaxies (see \citealt{Treu2010} for a review). If the surface mass density of the lens is above some critical value, then multiple images of the background source can be formed. The positions and relative magnifications of these images give valuable information about the structure of the lens that can be used to investigate the lensing mass distribution and test models for galaxy formation; for example, by precisely measuring the inner mass profile of galaxy-scale dark matter haloes (e.g. \citealt{Wucknitz2004,More2008,Suyu2012,Spingola2018}), placing limits on the mass of their central supermassive black hole (e.g. \citealt{Winn2004,Zhang2007,Quinn2016}), constraining the properties of their interstellar medium (e.g. \citealt{Mittal2007,Mao2017}) or through determining the level of dark matter substructure within them (e.g. \citealt{Mao1998,Bradac2002,Dalal2002,Metcalf2002,McKean2007a,More2009,Macleod2013,Hsueh2016,Hsueh2017}). Gravitational lenses are also a powerful astrophysical tool for determining the cosmological parameters, which include precise measurements of the Hubble constant and competitive tests of dark energy (e.g. \citealt{Biggs1999,Koopmans2000,Fassnacht2002,Suyu2010,Biggs2018}). Finally, gravitational lenses magnify the high redshift Universe, allowing detailed studies of galaxies that otherwise would not be detectable with current instruments (e.g. \citealt{Barvainis2002,Impellizzeri2008,Riechers2011,Sharon2016,Stacey2018}). 

The examples cited above are for those cases of gravitational lensing where the background source is also radio-loud, which are at a premium for several reasons. First, the radio emission is not obscured by dust or the bright optical emission from the lens. Moreover, the source emission is also extended (generally several pc in size), and therefore, is not affected by micro-lensing from stars in the lensing galaxy; this allows any intrinsic variability of the background object to be observed or for accurate flux-ratios between the different images to be measured. Furthermore, the high angular resolution imaging that is achievable with very long baseline interferometric (VLBI) observations provides precise positions for the lensed images, with an astrometric precision of tens of $\mu$as. Finally, monitoring with interferometers for time variability between the lensed images can be easily carried out. 

However, even though the first gravitational lens was found at radio wavelengths \citep*{Walsh1979}, and the first systematic all-sky surveys were carried out with the Very Large Array (VLA) over two decades ago (e.g. \citealt{Hewitt1988,King1999,Winn2000,Browne2003,Myers2003}), there are only around 35 gravitational lens systems currently known where the background source is a radio-loud active galactic nucleus (AGN). When compared with the $\sim 200$ gravitationally lensed quasars and star forming galaxies found with the Sloan Digital Sky Survey (e.g. \citealt{Inada2010,Auger2010}), and the $\sim 50$ gravitationally lensed sub-mm galaxies found with the \textsl{Herschel Space Observatory} \citep{Negrello2017} and the South Pole Telescope \citep{Vieira2013}, the paucity of known radio-loud gravitationally lensed objects represents a missed opportunity, particularly given the unique advantages radio datasets have over optical and sub-mm observations. In the future, surveys with the Square Kilometre Array (SKA) have the potential to discover more than $10^5$ galaxy-scale gravitational lenses at radio wavelengths \citep{Koopmans2004SKA,McKean2015SKA}. The success of these surveys will depend on improving new search strategies over the coming years. 

The most successful search to date for gravitationally lensed radio sources is the Cosmic Lens All-Sky Survey (\citealt{Myers2003,Browne2003}), which found 22 gravitational lenses with a maximum image separation between 0.3 and 6 arcsec from a sample of $11\,685$ radio sources that were initially selected based on their flat radio-spectra at cm-wavelengths (13 gravitational lenses were found within a statistically well defined sample of 8958 radio sources; \citealt{Chae2002}). The CLASS parent sample was first observed with the VLA at 8.46 GHz ($\sim 170$ mas resolution) and then followed-up at progressively higher angular resolution at other frequencies with the Multi-Element Radio Linked Interferometer Network (MERLIN) and the Very Long Baseline Array (VLBA) to confirm that the radio spectral energy distribution, surface brightness and polarization of the candidate multiple images were consistent with gravitational lensing. Given that the selection criteria focused on objects with a flat radio-spectrum, which are typically beamed radio sources, almost all of the lensed objects found by CLASS are unresolved or have only slightly extended jet emission when observed on VLBI-scales (e.g. \citealt{Biggs2003,Biggs2004,McKean2007a,More2008}). However, the MIT-Green Bank (MG) survey \citep{Hewitt1988}, which also used the VLA to identify gravitationally lensed radio sources, targeted objects that were extended, and these have been found to have large gravitational arcs/extended images that are 100 to 800 mas in size when observed with VLBI \citep{More2009,Macleod2013,Spingola2018}. Therefore, both the CLASS and MG lens surveys demonstrate the potential of finding new gravitationally lensed radio sources directly from VLBI observations.

In addition, lens surveys with VLBI observations have recently become feasible with the development of wide-band receivers and higher data-recording rates that increase the imaging sensitivity, and new correlation and data processing methods that allow wide-field imaging to be carried out efficiently. Such studies have been used to investigate the relative contributions of AGN and star-formation activity within radio galaxies at cosmological distances, and have typically focused on single fields that are limited by the primary beam of the individual antennas of the VLBI-array \citep{Garrett2001,Wrobel2004,Garrett2005,Morgan2011,Chi2013,Cao2014,Radcliffe2016}, but are now also being carried out over larger areas of the sky \citep{Herrera-Ruiz2017}. In particular, the mJIVE--20 programme (mJy Imaging VLBA Exploration at 20 cm; \citealt{Deller2014}) is the largest survey of the radio sky with a VLBI array, detecting $4\,965$ radio sources in around 200~deg$^2$.

The detectability of radio sources on VLBI-scales and the recent advances in wide-field VLBI imaging techniques can now be combined to open up a new and efficient method to increase the number of known gravitational lenses for studies of galaxy formation and cosmology. In this paper, we present the first pilot survey for such gravitational lens systems by using the large sample of sources found during the mJIVE--20 programme. Our paper is organized as follows. In Section \ref{Sec:Criteria}, we describe the steps for the selection of the lens candidates from the mJIVE--20 parent sample. Section \ref{Sec:Observations} describes the high angular resolution multi-frequency follow-up observations with the VLBA and the data reduction processes. We review the lensing hypothesis for the sample and discuss our results in Section \ref{Sec:Results}. The gravitational lensing statistics of the mJIVE--20 survey and the future prospects for wide-field VLBI lens searches are discussed in Sections \ref{Sec:Statistics} and \ref{Sec:Future}, respectively. We summarize our conclusions in Section \ref{Sec:Conclusions}.

Throughout this paper, we assume $H_0=67.8\; \mathrm{km\,s^{-1}~Mpc^{-1}}$, $\Omega_{\rm M}=0.31$, and $\Omega_{\Lambda}=0.69$ \citep{Planck2016}. The spectral index $\alpha$ is defined as $S_{\nu} \propto \nu^{\alpha}$, where $S_{\nu}$ is the flux density as a function of frequency $\nu$.

\section{Lens candidate selection criteria}
\label{Sec:Criteria}

The parent sample of our lens search has come from the mJIVE--20 survey (Project IDs: BD161 and BD170; PI: Deller), which was a 408-h VLBA filler-time project to image a large  number of radio sources at 1.4 GHz with mas angular resolution \citep{Deller2014}. This survey targeted regions of the sky with a VLBI calibrator radio source, as this allowed in-beam calibration of the antenna complex gains to be used and a more efficient use of the telescope time. A series of four sub-pointings around the calibrator radio source were used to achieve an effective field-of-view of 1~deg$^2$ per field. The total observing time per field was 1 h, with about 15 min per sub-pointing. The data were taken using 64 MHz bandwidth  and with dual polarization (a recording rate of 512 Mbit s$^{-1}$). This observing set-up gave a close to uniform rms noise of about 150~$\mu$Jy~beam$^{-1}$ within a radius of $\sim$20 arcmin around the central calibrator source when using a natural visibility weighting, while retaining a reduced sensitivity to sources out to separations of 35 to 40 arcmin. As imaging the full effective primary beam is not practical, known radio sources were first identified from the Faint Images of the Radio Sky at Twenty cm (FIRST; \citealt{Becker1995}) survey and their positions were used as the phase centres of the multi-field correlations that are provided by the DiFX software correlator \citep{Deller2007,Deller2011}. The multi-field correlations for each pointing used a spectral resolution of 1~MHz~channel$^{-1}$ and an averaging time of 3.2~s for the visibilities. However, for the public data release\footnote{http://safe.nrao.edu/vlba/mjivs/products.html} and imaging, the visibility data were averaged to 16 MHz~channel$^{-1}$ and 20~s, which is sufficient to image a field-of-view of around $0.75\times 0.75$~arcsec. In total, 24\,903 sources from FIRST have been observed as part of the mJIVE--20 survey, with 4\,965 sources detected on mas-scales above a signal-to-noise ratio of 6.75$\sigma$.

Although the field-of-view of the averaged datasets is sufficient to find galaxy-scale gravitational lenses (the average image separation of the CLASS lens sample is around 1.2 arcsec; \citealt{Browne2003}), radio-loud lensed objects have been found with image separations as large as 4.6 arcsec within CLASS \citep{McKean2005}. Note that a search for wide-separation gravitational lens systems ($6 \leq \theta_{\rm sep} \leq15$~arcsec) in the CLASS data returned a null result \citep{Phillips2001,McKean2011}. This is because such wide-separation images require massive lenses and the lensing optical depth of galaxy groups and clusters is over an order of magnitude less than for galaxies. Therefore, we retrieved the un-averaged mJIVE--20 survey {\it uv}-data and re-mapped each pointing to make a wide-field image that was $3.5\times3.5$~arcsec in size. We then ran the mJIVE--20 object detection software on the re-imaged data (note that it is these data that are now available in catalogue form from the mJIVE--20 survey archive).

To identify lens candidates in the wide-field catalogue available on 30 June 2013, sources were selected when there was the detection of at least two radio components at the $>6.75\sigma$-level, with a separation $>100$~mas and with an integrated flux-ratio $<15$:1. Given the resolution of our data (the average synthetized beam is about $11 \times 5$~mas$^2$), we could in principle separate components on 5 to 20 mas-scales. However, this would have severely increased the contamination from core-jet sources and we also do not expect image separations on this scale from galaxy-sized gravitational lenses. The lower size limit that we selected is about 3 times smaller than the smallest separation lens known (i.e., $\theta_{\rm sep} = 334$~mas; \citealt{Wucknitz2004}), but this small size was chosen to ensure that the merging pair from any potential four image gravitational lens system, which are highly magnified with respect to the counter-images, would be identified. We chose a less stringent flux-ratio between the components than CLASS as we were also interested in finding asymmetric lens systems, which are useful for studying the inner parts of lensing galaxies. Using these criteria, we identified 81 radio sources that are then manually classified with grades from A to D, where grade A corresponds to the highest possibility to be a gravitational lensing system. By visually inspecting and/or remapping these sources and by cross-correlating with optical imaging and spectroscopy from the SDSS, we remove those sources with a clear core-jet morphology or that are mis-identified residual side-lobes in our catalogue. 

From this process, we obtain a sample of fourteen good gravitational lens candidates with grade A and B, that have multiple compact components with a similar surface brightness and, therefore, are not likely to be simply core-jet type radio sources. Interestingly, we find that two of the class A candidates (MJV02639 and MJV03238) are already confirmed as gravitational lenses by CLASS (B1127+385 and B2319+051; \citealt{Koopmans1999,Rusin2001}). Neither the calibrator list nor the FIRST sub-sample used for mJIVE--20 was selected to specifically include (or exclude) gravitationally lensed radio sources that were found by CLASS; we note that neither of the two detected CLASS lenses are VLBA calibrators.
This already demonstrates that wide-field VLBI observations can be used to identify radio-loud lensed sources. To determine the lensing nature of the remaining twelve candidates, we follow a similar strategy to CLASS by observing the sources at a higher angular resolution (to test the conservation of surface brightness) and at different frequencies (to test whether the radio-spectra are identical). If a lens candidate does have a radio spectrum and surface brightness consistent with gravitational lensing, we then test whether the high resolution structure of the images is consistent with a plausible lensing mass model. For example, the two known lensed sources, MJV02639 and MJV03238, have been previously observed with VLBI instruments between 1.7 and 15 GHz \citep{Koopmans1999,Rusin2001}. Therefore, we can use already-obtained information on what the follow-up of these systems would reveal. The lensed images in these two cases show nearly identical surface brightness and radio spectra. Moreover, the flux density ratio of the two lensed images does not show any dependency on the frequency. The high frequency imaging resolves the lensed images of both the sources into multiple sub-components, which can be fit by a simple lens mass model. Consequently, they would have passed all of the selection criteria based on the radio imaging.

\begin{figure*}
\centering
\includegraphics[scale=0.72]{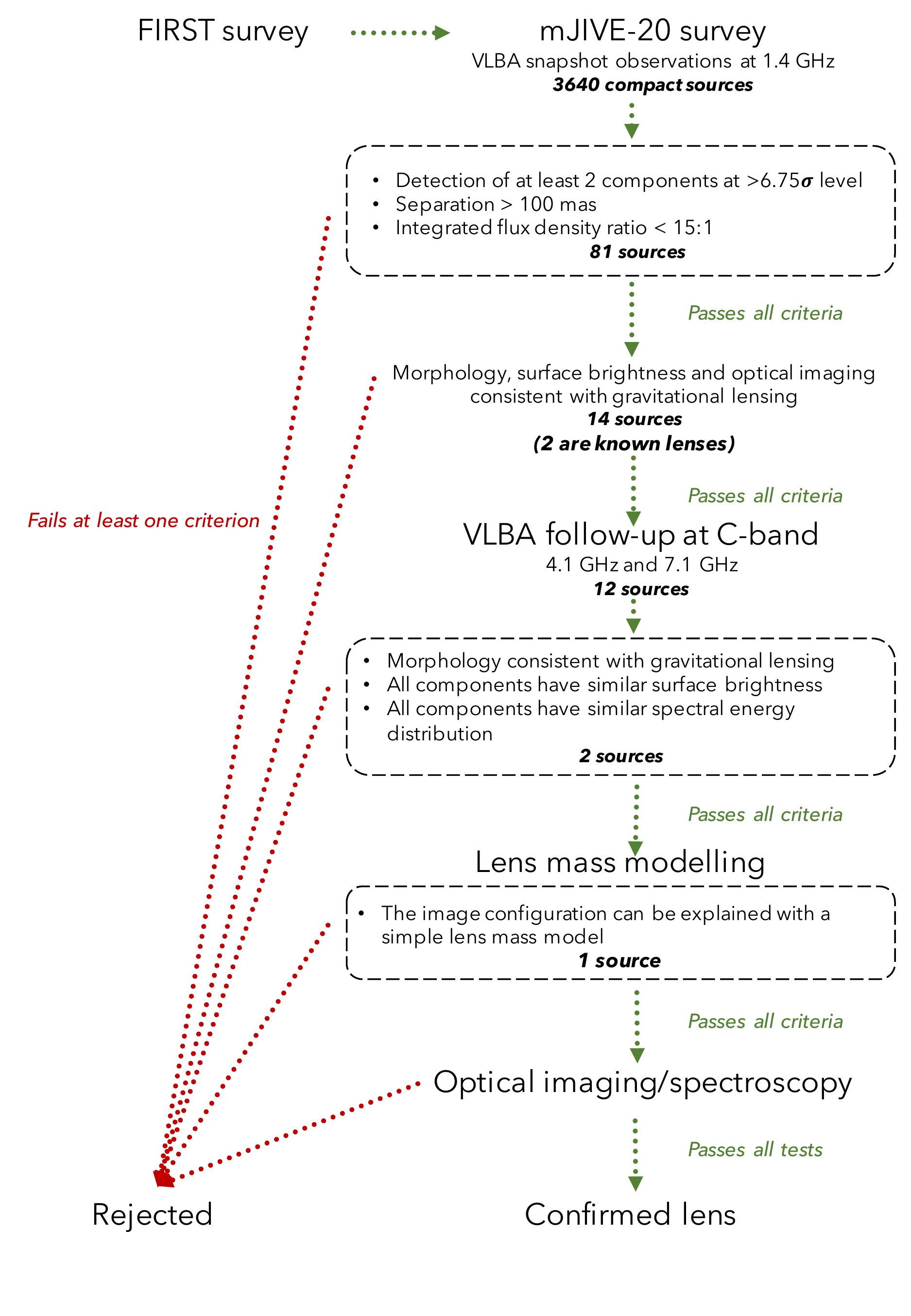}
\caption{Flow chart diagram that summarizes the steps of the gravitational lensing search within the mJIVE--20 survey.}
\label{Fig:flowchart}
\end{figure*}

\begin{figure*}
\centering
	\includegraphics[scale = 0.95]{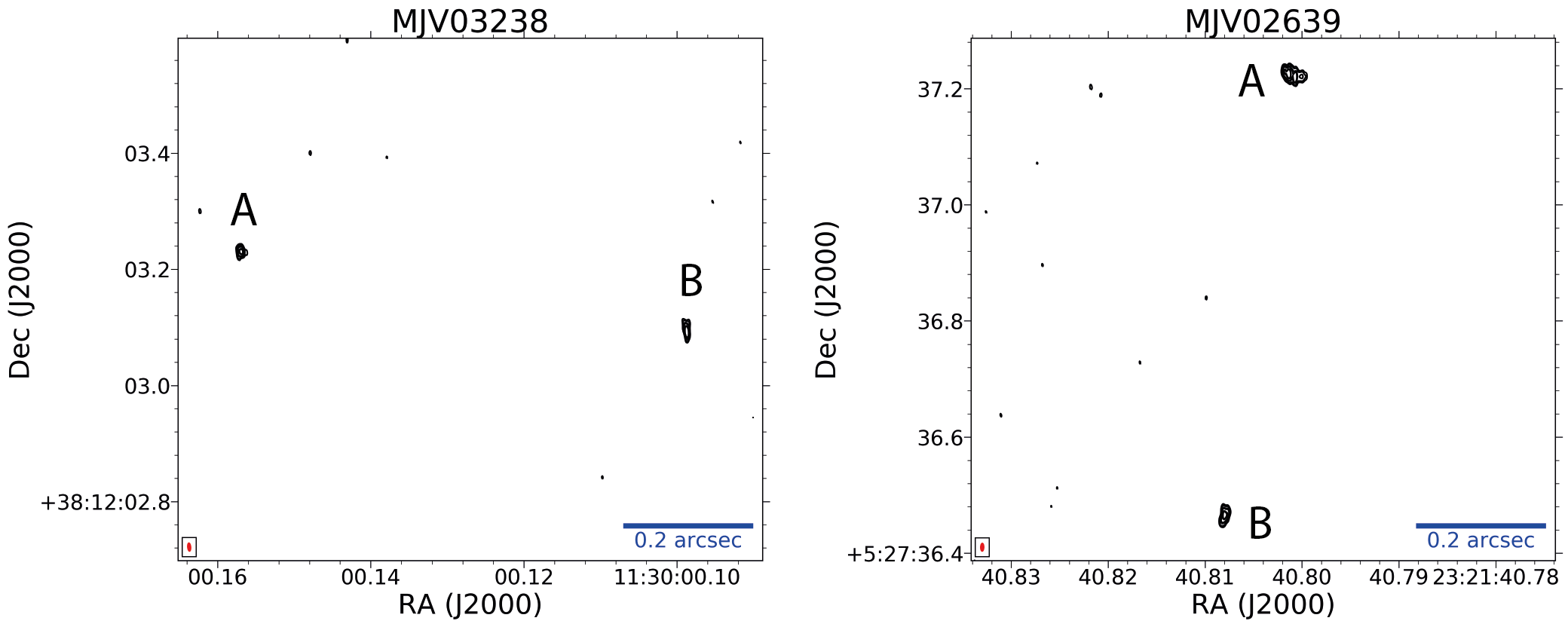} \caption{The self-calibrated images at 1.4 GHz of two mJIVE--20 sources that were previously confirmed as being gravitationally lensed by CLASS (B1127+385 is on the left and B2319+051 is on the right). Contours are at $(-3, 3, 6, 9, 12, 15, 27) \times \sigma_{\rm rms}$, the off-source rms noise. The beam size is shown in the bottom left corner, which is on average $11 \times 5$ mas$^2$; north is up and east is left. The blue scale bar in each image represents 0.2 arcsec.}
	\label{Fig:class-lenses}
\end{figure*}

\begin{figure}
\centering
	\includegraphics[scale = 0.55]{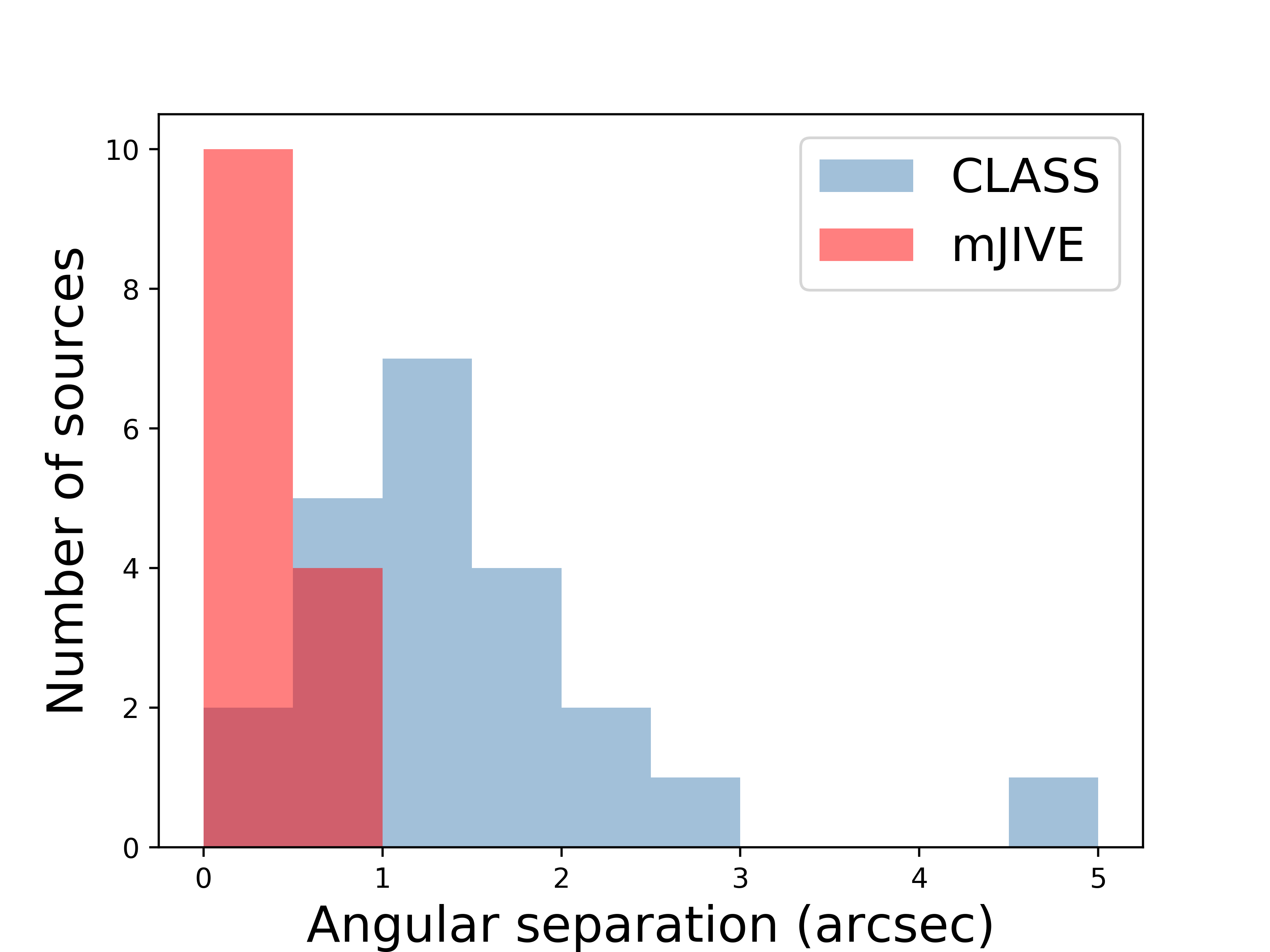}
	\caption{A comparison between the maximum image separation of the mJIVE--20 gravitational lens candidates (red) and the CLASS gravitational lenses (blue). The bins are 0.5 arcsec wide.}
	\label{Fig:angular-sep}
\end{figure}

The follow-up observations are discussed in the next section and all these steps for selecting and confirming the lens candidates are summarized in Fig.~\ref{Fig:flowchart}. The 1.4 GHz imaging from the mJIVE--20 observations of the twelve best lens candidates and the two confirmed gravitationally lensed radio sources are presented in Figs.~\ref{Fig:L-band-1} and \ref{Fig:class-lenses}, respectively. The maximum image separation of the candidate lensed images, as compared to those from the CLASS survey, is shown in Fig.~\ref{Fig:angular-sep}.

\begin{figure*}
\centering
    \includegraphics[width = 0.99\textwidth]{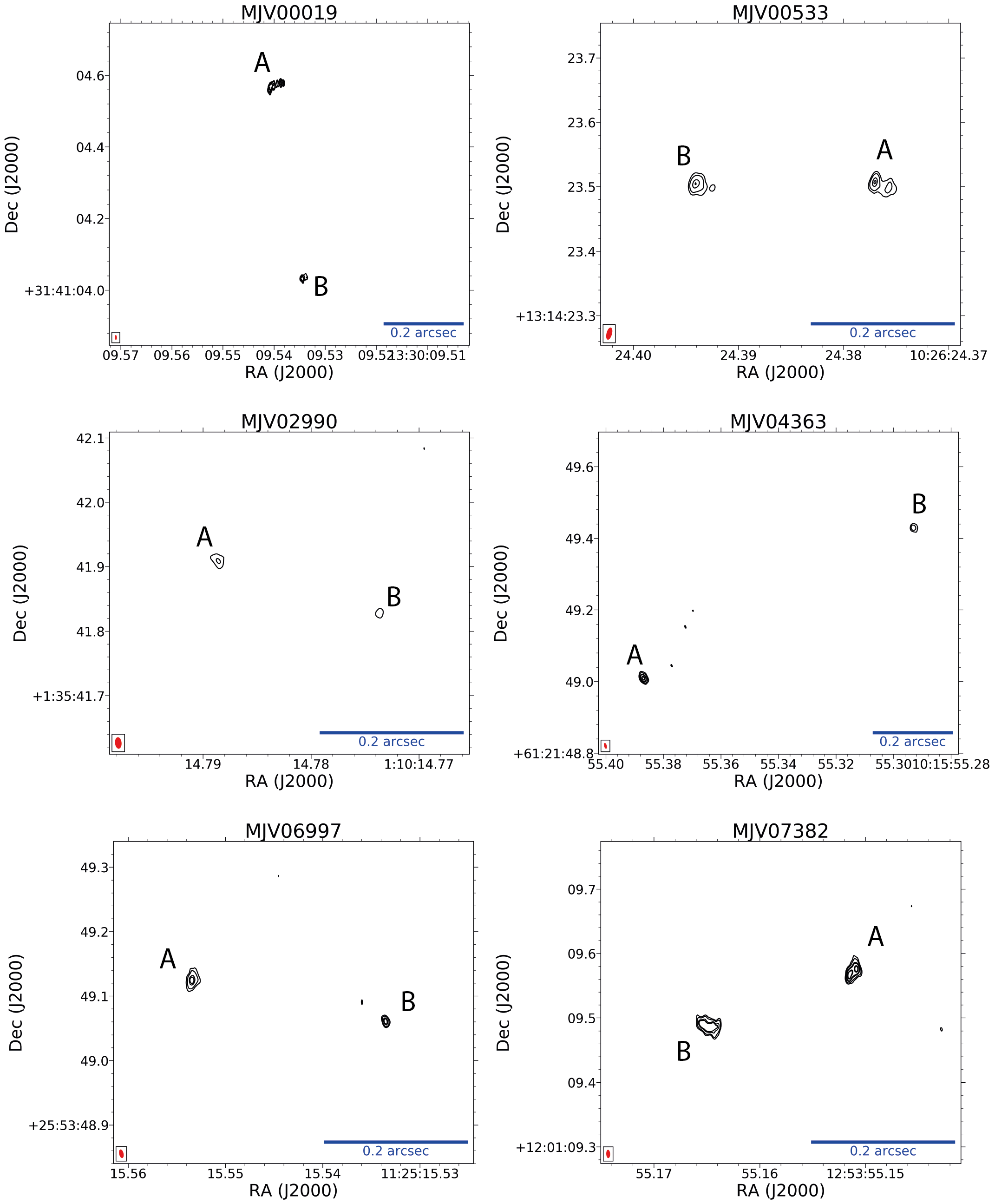}
	\caption{The {\sc clean}ed images at 1.4 GHz of the twelve lens candidates from the mJIVE--20 survey observations. Contours are at $(-3, 3, 6, 9, 12, 15, 27) \times \sigma_{\rm rms}$, the off-source rms noise. The beam size is shown in the bottom left corner, which is on average $11 \times 5$ mas$^2$; north is up and east is left. The blue scale bar in each image represents 0.2 arcsec.}
	\label{Fig:L-band-1}
\end{figure*}

\begin{figure*}
\centering
    \includegraphics[width = 0.99\textwidth]{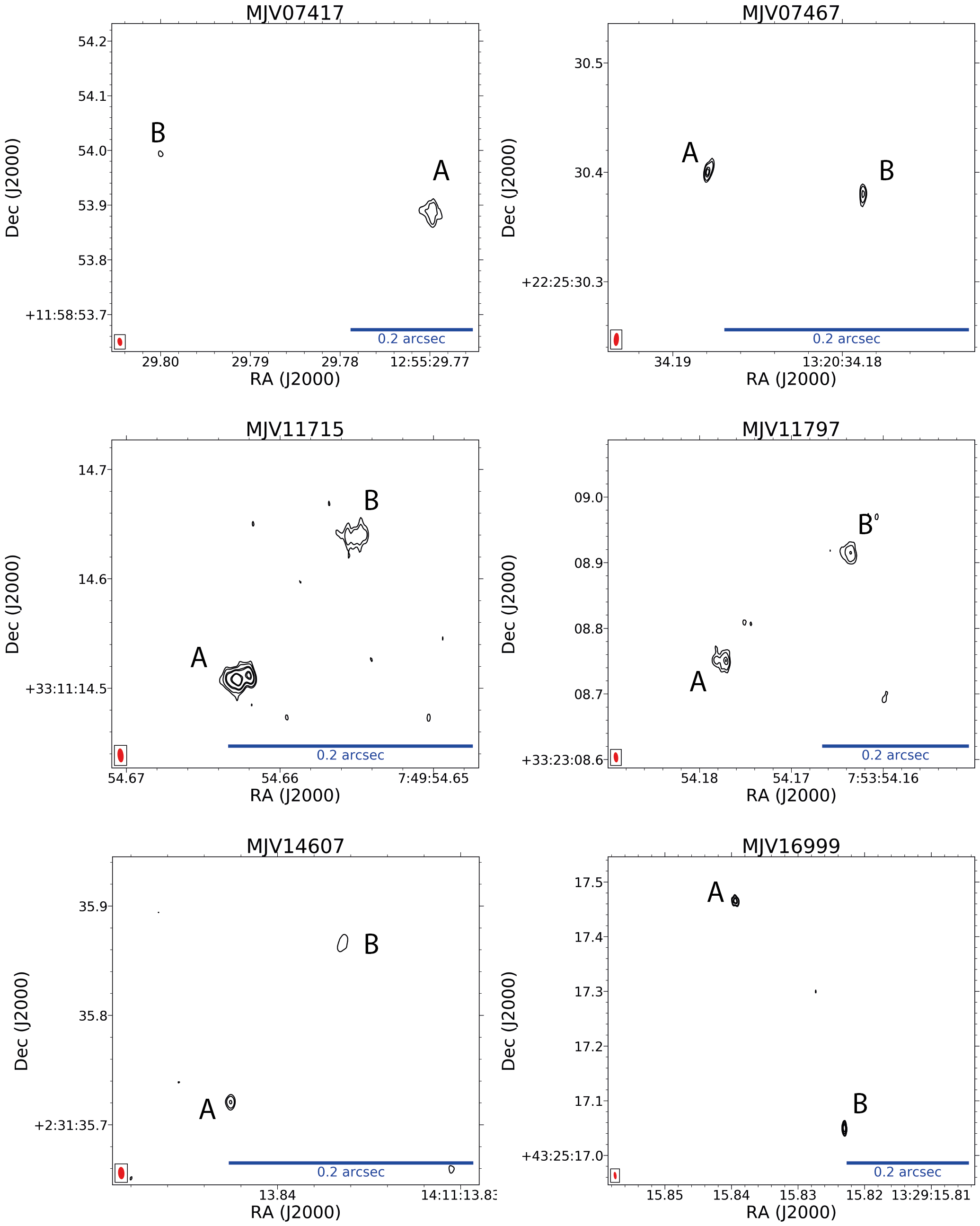}
\contcaption{}\label{Fig:L-band-2}
\end{figure*}

\section{Follow-up observations \& data reduction}
\label{Sec:Observations}

\begin{figure*}
\centering
	\includegraphics[width = \textwidth]{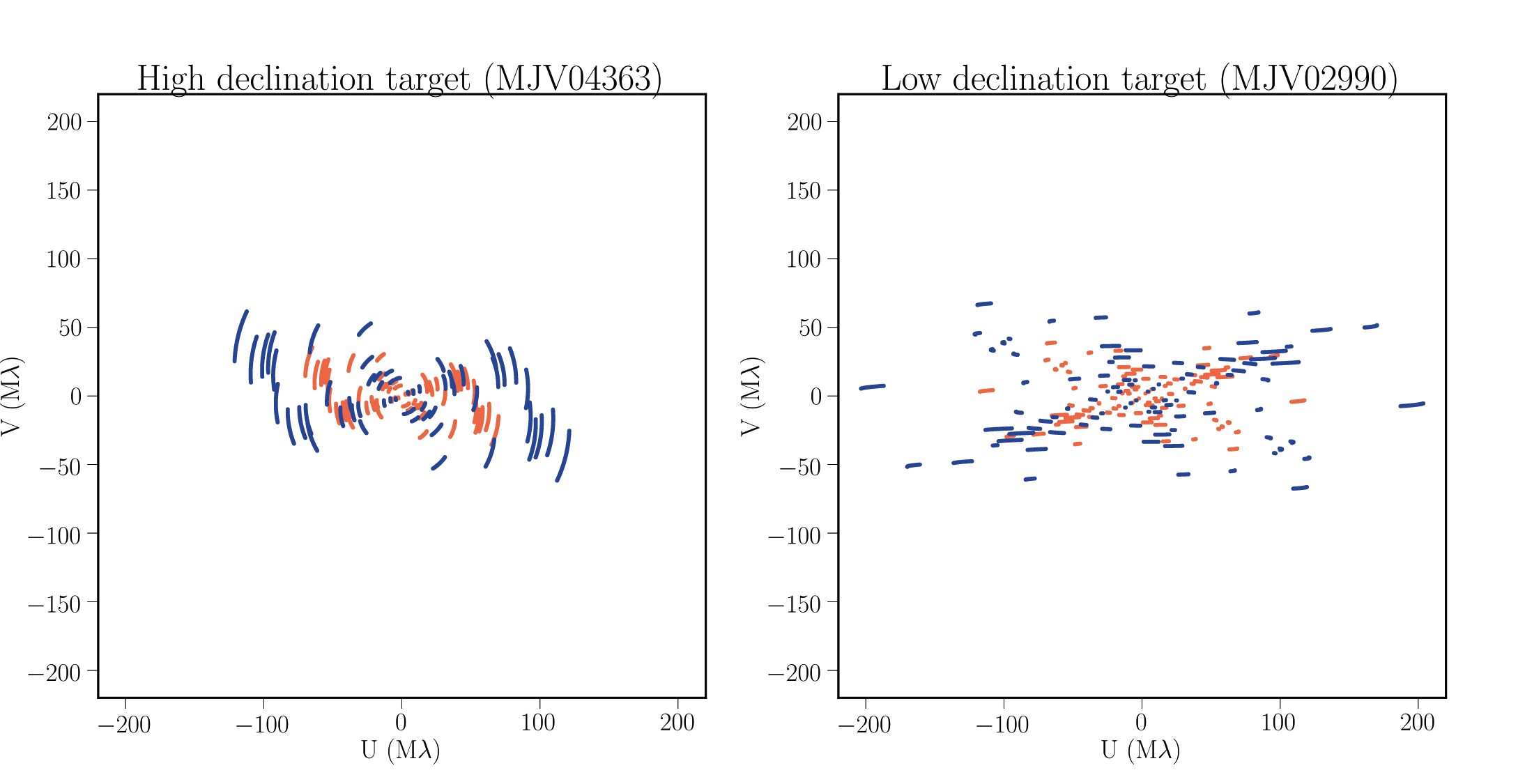}
	\caption{Examples of the \textsl{uv}-coverage for the highest (left) and lowest declination (right) lens candidates at 4.1 (orange points) and 7.1 GHz (blue points). }
	\label{Fig:uvcoverage}
\end{figure*}

We followed-up the twelve remaining lens candidates with the VLBA at C-band between November 2013 and April 2014 (Project IDs: BM397 and BM398; PI: McKean). This upgraded wide-band receiver allows observations to be carried out between 4 and 8 GHz, with two separate spectral windows that can be separated over the band to provide simultaneous multi-frequency imaging. We selected central observing frequencies of 4.1 and 7.1 GHz, based on the advice of NRAO staff. The bandwidth of each spectral window was 128 MHz through dual polarization at each frequency, and the data were recorded at 2048~Mbit\,s$^{-1}$. Given the increase in the observing frequency and the resulting decrease in the primary beam size (6 to 11 arcmin), we could no longer use in-beam calibration, and so, we switched between the target and the phase-reference source every 5 mins, with $\sim3$~min scans on source and $\sim1$ min scans on the calibrator. The observations were typically $\sim 1.5$ h in total for each lens candidate. The rms map noise of the naturally weighted images was typically between 40 and 70~$\mu$Jy~beam$^{-1}$ at both frequencies, where the upper noise values were due to the difficulty in modelling diffuse structure in those cases that were only detected with the shortest baselines of the VLBA. Such cases resulted in a residual fringe pattern in the images, which also helped us establish if a candidate lensed image was extended or not. A summary of the observations is given in Table~\ref{Tab:targets}. Two examples of the \textsl{uv}-coverage are shown in Fig.~\ref{Fig:uvcoverage} for a low- and high-declination candidate at the two observing frequencies.

We apply the standard VLBA calibration procedure for phase-referenced observations using the Astronomical Image Processing System ({\sc aips}). We start by applying the Earth orientation and ionospheric corrections. After an initial flagging of bad data, we apply the a priori amplitude calibration by using the gain curves and system temperatures of the individual antennas, and correcting for the sampler offsets. Next, we correct for the instrumental delay and parallactic angle variation. We then perform global fringe fitting to correct for the residual fringe delays and rates, which are measured from observations of the phase-reference calibrator using solution intervals of between 3 and 5 min. We use the fringe-finder calibrator to perform the bandpass calibration and we split out the data at the two frequencies separately in order to perform the imaging. Most of the targets are detected at a relatively low significance (3 to 10$\sigma$), therefore when attempting phase-only self-calibration more than 60~per cent of the solutions failed, even when a solution interval much longer than the coherence time was used. For this reason, we do not perform self-calibration. We {\sc clean} the images by using a threshold that is three times the theoretical rms noise level and a natural weighting scheme. We assume a conservative uncertainty of $\sim 10$ per cent on the estimate of the absolute flux density, which takes into account calibration errors and possible variability in the candidate lensed images when we consider the relative flux-ratios.

\begin{table*}
	\centering
	\caption{A summary of the 4.1 and 7.1 GHz follow-up observations of the mJIVE--20 lens candidates with the VLBA and the rejection criteria. The columns (from left to right) give the mJIVE--20 identification number, the VLBA observation date, the lens confidence category (A or B), the rejection criteria, and whether there is an optical ID in Pan-STARRS. The rejection criteria are based on the spectral index and the surface brightness of the individual candidate lensed images, and whether a simple lens mass model can explain the structure (see Fig.~\ref{Fig:flowchart}). The symbol $\times$ is used when a criterion for being gravitationally lensed has not been fullfilled. }
	\label{Tab:targets}
	\begin{tabular}{lllcccc} 
		\hline
		mJIVE--20 ID & Obseravtion Date   & Lens Cat.	&\multicolumn{3}{c}{Rejection criteria} & Optical ID\\
        	& &  & Spectral index & Surface brightness & Lens model &  \\
		\hline
        MJV00019 & 2013 Nov 01		& A	&				  & 				& $\times$	& no\\
 	    MJV00533 & 2014 Apr 12		& B	&				  & $\times$&		& no\\
        MJV02990 & 2014 Feb 22		& B	& $\times$& $\times$& 		& no \\
        MJV04363 & 2014 Mar 23		& B	& $\times$& 				& 		& yes \\
        MJV06997 & 2013 Nov 15		& A	& $\times$& $\times$&  	& no\\
        MJV07382 & 2014 Mar 11		& B	& $\times$& $\times$& 		& no \\
        MJV07417 & 2014 Mar 29		& B	& 	 	         &   				& 		& no\\
        MJV07467 & 2014 Mar 05		& B	& $\times$& $\times$& 		& no \\
        MJV11715 & 2014 Apr 24		& B	& $\times$& $\times$& 		& no\\
        MJV11797 & 2014 Apr 19		& B	& $\times$& $\times$& 		& no \\
        MJV14607 & 2014 Mar 02		& B	& $\times$& $\times$& 		& no  \\
        MJV16999 & 2013 Dec 07		& A	& $\times$& $\times$& 		& no \\ 
		\hline
	\end{tabular}
\end{table*}

\section{Results}
\label{Sec:Results}

We now summarize the results of our lens search. In Section \ref{Sec:individual_systems}, we describe the properties of the remaining twelve lens candidates at all frequencies. We also give a discussion on whether they meet the gravitational lensing criteria or not, as illustrated in Fig. \ref{Fig:flowchart}, which is summarized in Table~\ref{Tab:targets}. In Section \ref{Sec:lens_modelling}, we present lens mass models for those candidates that pass the VLBA imaging tests. In the Appendix, we show the maps from the 4.1 and 7.1 GHz VLBA observations in Figs.~\ref{Fig:mjv00019-composite} to \ref{Fig:mjv16999-composite}, the radio spectral energy distributions and flux-ratios as a function of frequency are shown in Figs.~\ref{Fig:radio-spectra-1} and \ref{Fig:flux-ratio-1}, respectively, and we show the optical imaging of the candidates from Pan-STARRS in Fig~\ref{Fig:Panstarss}.  Also, a summary of the properties of the fourteen targets from our lens search is presented in Table \ref{Tab:fluxdensities}. We would like to highlight that the large negative spectral indices derived by our measurements must be taken with caution. This is because the spectral index can be biased towards more negative values by angular resolution and surface brightness sensitivity effects, especially when the images are very resolved at high frequencies. This does not affect our ability to discriminate between lensed and non-lensed sources, as we use the spectral indices as just an indication for comparing the spectral energy distribution of the two putative lensed images. 

\subsection{Description of the individual lens candidates}
\label{Sec:individual_systems}

We now give a brief review of the radio and optical data for our lens candidates. In each case, we denote the candidate lensed image with the highest flux-density at 1.4~GHz as component A, and the second and third highest as components B and C, respectively.

\subsubsection{MJV00019} 

This lens candidate shows two components separated by 537 mas at 1.4 GHz (see Fig. \ref{Fig:L-band-1}). Component A has extended structure in the east--west direction with a curved morphology, while component B is also resolved, but has a smaller angular-extent than component A. Therefore, their surface brightness agrees with gravitational lensing. At 4.1 and 7.1 GHz, the two components are detected and resolved into multiple sub-components and their surface brightnesses are consistent (see Fig. \ref{Fig:mjv00019-composite}). The spectral index of components A and B between 4.1 and 7.1 GHz are comparable within the uncertainties and it is steep ($\alpha_{4.1}^{7.1} \sim -1.9$; see Table~\ref{Tab:fluxdensities}, and Figs.~\ref{Fig:radio-spectra-1} and \ref{Fig:flux-ratio-1}). For these reasons, MJV00019 passes the observational tests and remains a lens candidate. However, we note that 55 per cent of the FIRST emission was not recovered on VLBI-scales, which suggests there is a significant undetected component of extended emission associated with this source. Due to its steep spectrum between 1.4 and 4.1 GHz, and low flux-density at 5 GHz, this object was not selected to be observed during CLASS. No optical emission from a putative lensing galaxy or the candidate lensed images is detected at the location of the system (see Fig. \ref{Fig:Panstarss}).

\subsubsection{MJV00533} MJV00533 has two resolved radio components separated by 248 mas at 1.4 GHz, where each component shows evidence for an extension in the east--west direction (see Fig. \ref{Fig:L-band-1}). At 4.1 and 7.1 GHz, component A is unresolved and there is no detection of the sub-component to the west, while component B shows faint resolved extended structure at 4.1 GHz and is completely resolved out at 7.1 GHz (see Fig. \ref{Fig:mjv00533-composite}). Therefore, the surface brightness of the candidate lensed images is not conserved between the two frequencies (see Table~\ref{Tab:fluxdensities}). We note that their spectral indices are similar between 1.4 and 4.1 GHz (see Figs.~\ref{Fig:radio-spectra-1} and \ref{Fig:flux-ratio-1}), which highlights the importance of the higher-frequency data in determining the status of such lens candidates. The surface brightness and morphology of this system suggest that it is a core-jet (or core-hotspot) source. There is no evidence for an optical counterpart for this object (see Fig.~\ref{Fig:Panstarss}).

\subsubsection{MJV02990} At 1.4 GHz, this lens candidate consists of two resolved components that are separated by 240~mas (see Fig. \ref{Fig:L-band-1}), where component A also has the larger angular-size, respecting the surface brightness criterion for being a lens system. At 4.1 GHz, component A is detected and found to be unresolved, while component B is not detected (see Fig. \ref{Fig:mjv02990-composite}). Given the flux ratio between components A and B at 1.4 GHz, the fainter component should have been detected at the $7.5\sigma$ level at 4.1 GHz, if it was also unresolved (see Table~\ref{Tab:fluxdensities}, and Figs.~\ref{Fig:radio-spectra-1} and \ref{Fig:flux-ratio-1}). Neither component is detected at 7.1 GHz. Overall, the multi-frequency radio data suggests that MJV02990 is not a gravitational lens system, but is more likely a core-jet, mainly due to component B being undetected at 4.1 and 7.1 GHz, which implies that the surface brightness is not conserved. There is no detection of optical emission at the location of the two radio components (see Fig.~\ref{Fig:Panstarss}).

\subsubsection{MJV04363} MJV04363 has two radio components separated by 791 mas at 1.4 GHz, with component A being slightly resolved in the east--west direction, whereas component B is found to be unresolved (see Fig. \ref{Fig:L-band-1}). At 4.1 and 7.1 GHz, a third unresolved component (C) is detected between the two candidate lensed images (see Fig. \ref{Fig:mjv04363-composite}). This third radio image is closer in projection to component A and it could be the emission from a possible lensing galaxy, as for example in the lensing systems CLASS B2045+265, CLASS B2108+213 and CLASS B2114+022 \citep{Augusto2001b, McKean2007a, McKean2005}. If this were the case, then the position of component C suggests that the potential lensing galaxy should be highly elliptical to reproduce the observed image configuration. However, the spectral energy distribution of the two putative lensed images is significantly different, indicating that they are not gravitationally lensed images of the same background object (see Table~\ref{Tab:fluxdensities}, and Figs.~\ref{Fig:radio-spectra-1} and \ref{Fig:flux-ratio-1}). Therefore, the surface brightness and spectral energy distribution of the components suggest that this system more likely has an optically thick radio core (component C) with two jets or hotspots (components A and B). Also, only 7 per cent of the FIRST emission is detected by the mJIVE--20 observations. Finally, there is faint optical emission at the location of this source in both Pan-STARSS and SDSS (see Fig. \ref{Fig:Panstarss}; \citealt{Flewelling2016Panstarss,Abazajian2009SDSS}). The optical emission is classified as a galaxy, with a magnitude in the \textsl{r}-band of 21.95; there is no spectroscopic information available, so the redshift is unknown.

\subsubsection{MJV06997} This candidate shows two resolved components separated by 276 mas at 1.4 GHz (see Fig. \ref{Fig:L-band-1}). However, we detect only the fainter component B at both 4.1 and 7.1 GHz, and at these frequencies it is unresolved (see Fig. \ref{Fig:mjv06997-composite}). Moreover, given the flux ratio between the two components at 1.4 GHz, component A should have been detected at the $22$-$\sigma$ level at 4.1 GHz (see Table~\ref{Tab:fluxdensities}, and Figs.~\ref{Fig:radio-spectra-1} and \ref{Fig:flux-ratio-1}). The images at 4.1 and 7.1~GHz also show a strong fringe pattern passing through the non-detected component A. This indicates that emission is detected at this position on the shortest baselines, but not on the majority of the VLBA baselines, where it is resolved out.  Therefore, based on the spectral energy distribution and the surface brightness of the components, this candidate is ruled out as a gravitationally lensed radio source. We conclude that this system is likely a core-jet source, where the jet (component A) has a higher flux-density at lower frequencies than the flat-spectrum radio core (component B). There is no optical emission detected at the position of the two radio components (see Fig.~\ref{Fig:Panstarss}).
 
\subsubsection{MJV07382} MJV07382 is one of the most interesting lens candidates that was selected from the mJIVE--20 parent catalogue. The object is comprised of two extended objects around 30~mas in size that are separated by 221~mas at 1.4 GHz (see Fig. \ref{Fig:L-band-1}). The morphology and small separation between the components suggested that it may be a quadruply imaged radio source, where the mJIVE--20 observations detected only the highly magnified merging pair. The additional 1.4 GHz imaging taken as part of the mJIVE--20 programme failed to detect any counter-image emission down to a flux-ratio limit of $>27$ (for a $5\sigma$ detection), which should have been sufficient if this is a quadruply imaged radio source. At 4.1 GHz, we detect only component A, but there is a strong side-lobe pattern corresponding to component B, indicating that there is resolved extended structure associated with this component (see Fig. \ref{Fig:mjv07382-composite}). We do not detect either component at 7.1 GHz, but again, there is a fringe structure across the components at this frequency. In principle, given the flux density of component B at 1.4 GHz, it should have been detected at the 3.5$\sigma$ level at 4.1 GHz, 
although the sparse $uv$-coverage may have affected the detectability (see Table~\ref{Tab:fluxdensities}, and Figs.~\ref{Fig:radio-spectra-1} and \ref{Fig:flux-ratio-1}). Given the differences in the surface brightness of the two components, MJV07382 is likely a core-jet object, where component A, with the highest surface brightness, being the core. There is no evidence for optical emission corresponding to this source in Pan-STARRS or SDSS (see Fig. \ref{Fig:Panstarss}).

\subsubsection{MJV07417} At 1.4 GHz, this candidate shows a bright extended component A and an unresolved, marginally detected fainter component B that are separated by 458 mas (see Fig. \ref{Fig:L-band-2}). The map at 1.4 GHz also shows a strong fringe pattern associated with component A, which signifies that it is partly resolved out. There is no detection of either component in the 4.1 or 7.1~GHz observations (see Fig. \ref{Fig:mjv07417-composite}). Therefore, it is not possible to compare the spectral energy distributions of components A and B, or their surface brightness as a function of frequency.  At 1.4 GHz, the surface brightness of components A and B are 44 and 21~Jy~arcsec$^{-2}$, respectively. However, as component B is not resolved we do not have an accurate estimate of its surface brightness. There is extensive extended emission from this target that is not recovered by the 1.4 GHz VLBA imaging; the flux-density measured by FIRST is almost seven times higher than the mJIVE--20 flux density. There is also no optical emission detected from this source (see Fig. \ref{Fig:Panstarss}). Given the observational data in hand, we currently cannot exclude MJV07417 from our candidate list, even though it is unlikely to be a gravitational lensing system given the extended nature of component A and the marginal detection of component B. 

\subsubsection{MJV07467} This candidate has two components separated by 221 mas that are resolved at 1.4~GHz into two sub-components in a north-south direction, and where almost all of the low-resolution FIRST emission is recovered on VLBI-scales by the mJIVE--20 observations (see Table~\ref{Tab:fluxdensities} and Fig. \ref{Fig:L-band-2}). At 4.1 GHz, component A is resolved into two sub-components with a more east-west morphology, while component B is marginally detected, showing a very diffuse and extended structure around a more compact central emission (see Fig. \ref{Fig:mjv07467-composite}). Therefore, the surface brightness is not conserved at 4.1 GHz, because the fainter image is also the most extended. At 7.1 GHz, only component A is detected, while component B is completely resolved out.  Given the flux density ratio between the two components at 4.1 GHz (see Fig. \ref{Fig:flux-ratio-1}), component B should have been detected at the $4\sigma$ level at 7.1~GHz if MJV07467 were a gravitationally lensed source. From the surface brightness and spectral energy distribution of the components, we reject this candidate as a gravitational lensing system. There is also no optical emission detected from this source (see Fig. \ref{Fig:Panstarss})

\subsubsection{MJV11715} This lens candidate shows two components that are resolved in an east-west direction at 1.4 GHz, and although they have a similar size, their flux-densities differ by a factor of about 4 (see Fig. \ref{Fig:L-band-2}). Since their morphology is consistent with gravitational lensing at 1.4 GHz, but their surface brightness is not, this candidate was followed-up at 1.4 GHz with deeper observations as part of mJIVE--20. However, the size of the two components remained very similar, and so their surface brightness is not consistent with gravitational lensing (see Table \ref{Tab:fluxdensities}). These properties rule out this candidate as a lensed source. Moreover, at 4.1 and 7.1~GHz, only component A is detected, and this is resolved into three compact sub-components with a clear core-jet morphology (see Fig.  \ref{Fig:mjv11715-composite}). There is no evidence of a residual fringe pattern in the images, which is a strong indication that component B is completely resolved out at these frequencies, confirming that the surface brightness between the two components is different. There is no optical counterpart of this object (see Fig. \ref{Fig:Panstarss}).

\subsubsection{MJV11797} At 1.4 GHz, this lens candidate consists of two resolved components separated by 238 mas that are elongated in an east-west direction (see Fig. \ref{Fig:L-band-2}). When following them up at 4.1 GHz, we find that the two components are slightly resolved, also in the same direction as seen at 1.4 GHz. This morphology at both 1.4 and 4.1 GHz would be consistent with lensing, as seen on VLBI-scales in the gravitational lensing system JVAS~B0218+357 for example \citep{Biggs2003}. At 7.1 GHz, we clearly detect component A, while component B is detected only at the $3\sigma$-level (see Fig. \ref{Fig:mjv11797-composite}). However, there is a clear fringe pattern crossing through the location of component B, indicating that there is extended resolved structure associated with this component. For this reason, we smooth the 4.1 and 7.1~GHz images to the same resolution as the 1.4 GHz data by using a {\it uv}-taper, in order to verify the marginal detection of component B (see Fig. \ref{Fig:mjv11797-smooth}). The surface brightness and morphology seem consistent with lensing, although component B is more resolved at 7.1~GHz. Also, the overall spectral energy distributions are similar, particularly at the highest frequencies, but the flux-ratio changes from 1.4 GHz to 4.1 and 7.1~GHz, in that the flux-density of component B is too high at 1.4~GHz. It may be that due to the {\it uv}-coverage, parts of component A are resolved out at 1.4 GHz, or alternatively, the source is highly variable. However, we find the total flux-density that we measure on VLBI-scales is within about 10 percent of the flux-density found by FIRST at a resolution of around 5 arcsec. Therefore, given the change in the flux-ratios with frequency and the resolved nature of component B, MJV11797 is likely not a gravitationally lensed source; further long-track observations of the target to improve {\it uv}-coverage are needed. As with almost all of the mJIVE--20 targets, no optical emission is associated with this object (see Fig. \ref{Fig:Panstarss}).

\subsubsection{MJV14607} MJV14607 has two components A and B that are separated by 215 mas when observed at 1.4~GHz (see Fig. \ref{Fig:L-band-2}). However, the higher flux-density component A is more compact than component B, violating the required conservation of surface brightness. This is confirmed by deeper follow-up observations at 1.4 GHz as part of mJIVE--20. At 4.1 and 7.1~GHz, only component A has been detected, while component B is completely resolved out, as indicated by the uniform structure of the noise (see Fig. \ref{Fig:mjv14607-composite}). Moreover, component A has a flat spectral index across the three observing frequencies (see Table~\ref{Tab:fluxdensities}, and Figs.~\ref{Fig:radio-spectra-1} and \ref{Fig:flux-ratio-1}). Therefore, this system is not a gravitational lens, but is likely to be a core-jet, where the jet component is resolved out during the high frequency observations. There is no optical emission detected for this system (see Fig. \ref{Fig:Panstarss}).

\subsubsection{MJV16999} This lens candidate shows two components separated by 452 mas at 1.4 GHz (see Fig. \ref{Fig:L-band-2}).  However, the surface brightness is not conserved as component B is slightly resolved, whereas component A is compact. Consistent with this, only component A is detected at 4.1 and 7.1 GHz (see Fig. \ref{Fig:mjv16999-composite}). Several sub-components are detected at the $3\sigma$-level around component A at both 4.1 and 7.1 GHz. These unresolved components indicate a core-jet structure. Moreover, given the flux density ratio between components A and B at 1.4 GHz, and the spectral index of component A between 1.4 and 4.1 GHz, component B should have been detected at the $21\sigma$-level at 4.1 GHz (see Table~\ref{Tab:fluxdensities}, and Figs.~\ref{Fig:radio-spectra-1} and \ref{Fig:flux-ratio-1}). Therefore, MJV16999 is rejected as a gravitational lensing candidate.

\begin{landscape}
\begin{table}
	\caption{Observed properties of the final sample of fourteen lens candidates. Column 1: mJIVE--20 ID of the lens candidate. Column 2: name of the components. Column 3 and 4: Right Ascension and Declination of the component. Column 5: flux density from the FIRST survey at 1.4 GHz. The uncertainty on this flux density is around 5 per cent. Column 6: flux density from the GB6 survey at 4.85 GHz. Column 7: flux density from VLBA 1.4 GHz observations. Column 8: flux density from VLBA 4.1 GHz observations.  The $3\sigma$ detection limit is estimated as three times the rms noise within the same area of the 1.4 GHz detection of that image. Column 9: flux density from VLBA 7.1 GHz observations. The $3\sigma$ detection limit is estimated as three times the rms noise within the same area of the 1.4 GHz detection of that image. Column 10: power-law spectral index $\alpha$ between 1.4 GHz and 4.1 GHz. Column 11: power-law spectral index $\alpha$ between 4.1 GHz and 7.1 GHz. Column 12: surface brightness at 4.1 GHz. Column 13: surface brightness at 7.1 GHz.}
	\label{Tab:fluxdensities}
	\begin{tabular}{lllllllllllll} 
		\hline
		\multirow{2}{*}{mJIVE--20 ID} & \multirow{2}{*}{Component} & RA & Dec & $S_{\rm{FIRST}}$ & $S_{\rm{GB6}}$ & $S_{1.4~\rm{GHz}}$ & $S_{4.1~\rm{GHz}}$ & $S_{7.1~\rm{GHz}}$ & \multirow{2}{*}{ $\alpha_{\rm{1.4}}^{\rm{4.1}}$} & \multirow{2}{*}{ $\alpha_{\rm{4.1}}^{\rm{7.1}}$} & SB$_{4.1~\rm{GHz}}$ & SB$_{7.1~\rm{GHz}}$  \\ 
         & & (J2000) & (J2000) & (mJy) & (mJy) & (mJy) & (mJy) &  (mJy) & & & (Jy arcsec$^{-2}$) &  (Jy arcsec$^{-2}$) \\
        \hline
		\multirow{2}{*}{MJV00019} & A & 13:30:09.541 & +31:41:04.563 & \multirow{2}{*}{96.0} &\multirow{2}{*}{n.d.} & $33\pm3$ &  $3\pm1$ & $1.0\pm0.5$ & $-2.2\pm0.1$ & $-1.9\pm 0.2$ & $1.98$ & $2.75$ \\
		& B & 13:30:09.534 & +31:41:04.032 &  & &$12\pm1$ & $1.3\pm0.5$ & $0.45\pm 0.20$ & $-2.2\pm 0.1$ & $-1.7\pm 0.2$ & $2.07$ & $3.24$ \\
		\hline
		\multirow{2}{*}{MJV00533} & A  & 10:26:24.377 & +13:14:23.507 & \multirow{2}{*}{60.5} & \multirow{2}{*}{n.d.} & $19\pm2$ &  $1.3\pm0.2$ & $0.7\pm0.1$ & $-2.5\pm0.1$ &  $-1.2\pm 0.2$ & $1.32$ & $12.61$\\
		& B &10:26:24.394 & +13:14:23.505 & & & $12\pm1$ & $1.1\pm0.2$ & $<0.07$ & $-2.2\pm0.1$ &  & $0.64$ &\\
		\hline
        \multirow{2}{*}{MJV02990} & A & 01:10:14.788 & +01:35:41.909 & \multirow{2}{*}{16.6} &\multirow{2}{*}{n.d.} & $7.7\pm0.8$ & $0.49\pm 0.07$  & $<0.09$ & $-2.6\pm0.1$ & & $4.57$ &  \\
		& B & 01:10:14.774 & +01:35:41.828 & & & $3.7\pm0.4$  & $<0.3$ & $<0.09$ & & & &  \\
		\hline
		\multirow{3}{*}{MJV04363} & A & 10:15:55.387 & +61:21:49.010 & \multirow{3}{*}{286.9} & \multirow{3}{*}{$83\pm8$} & $15\pm2$ &  $7.5\pm0.8$ & $6.6\pm0.2$ & $-0.6\pm0.1$ & $-0.25\pm0.19$ & $1.09$ & $2.09$ \\
		& B & 10:15:55.293 & +61:21:49.429 & & & $4.3\pm0.4$ & $0.8\pm0.2$ & $0.18\pm0.04$ & $-1.5\pm0.1$ & $-2.7\pm 0.2$ & $0.92$ & $1.31$\\
		& C & 10:15:55.350 & +61:21:49.190 & & & $<0.3$ & $0.7\pm0.2$ & $0.4\pm0.08$ &  & $-0.9\pm0.2$ & $1.06$ & $1.96$ \\
        \hline
          \multirow{2}{*}{MJV06997} & A & 11:25:15.553 & +25:53:49.125 &  \multirow{2}{*}{32.7} & \multirow{2}{*}{n.d.} & $5.7\pm0.6$ &  $<0.7$ & $<0.3$ &  & & &  \\
		& B & 11:25:15.534 & +25:53:49.061 & & & $2.2\pm0.2$ &  $1.7\pm0.2$ &  $1.7\pm 0.2$ & $-0.2\pm0.1$  &  $-0.01\pm0.19$ & $30.15$ & $5.12$ \\
		\hline
        \multirow{2}{*}{MJV07382} & A & 12:53:55.151 & +12:01:09.568 & \multirow{2}{*}{31.2} & \multirow{2}{*}{n.d.} & $11\pm1$ &  $2.7\pm0.5$ &  $<2.4$ & $-1.3\pm0.1$ & & $15.41$ &  \\
		& B & 12:53:55.165 & +12:01:09.491 & & &  $9 \pm 1$ & $<1.0$ & $<0.7$ & & & & \\
		\hline
        \multirow{2}{*}{MJV07417} & A & 12:55:29.770 & +11:58:53.890 & \multirow{2}{*}{93.4} & \multirow{2}{*}{n.d.} & $13\pm1$ & $<0.5$ & $<0.43$  & & & &  \\
		& B & 12:55:29.800 & +11:58:53.993 & & &  $1.2\pm0.1$  & $<0.07$ & $<0.41$  & & & & \\
		\hline
        \multirow{2}{*}{MJV07467} & A & 13:20:34.188 & +22:25:30.401 & \multirow{2}{*}{6.5} & \multirow{2}{*}{n.d.} & $4.1\pm0.4$ & $1.0\pm 0.1$ &$0.8\pm0.1$& $-1.3\pm0.1$ & $-0.6\pm0.2$ & $0.81$ & $1.85$ \\
		& B & 13:20:34.179 & +22:25:30.381 & &  & $2.3\pm0.2$ & $1.0\pm0.1$ & $<0.39$ & $-0.9\pm0.1$ & & $0.96$ &  \\
		\hline
        \multirow{2}{*}{MJV11715} & A & 07:49:54.662 & +33:11:14.510 & \multirow{2}{*}{35.0} & \multirow{2}{*}{n.d.} &  $23\pm 2$ &  $5\pm1$ &$4\pm1$ &  $-1.4\pm0.1$ & $ -0.5\pm0.2$ & $3.84$ & $4.53$ \\
		& B & 07:49:54.655 & +33:11:14.638 & & &  $6.1\pm 1.2$& $<3.7$ & $<3.9$ & & & & \\
		\hline
        \multirow{2}{*}{MJV11797} & A & 07:53:54.178 & +33:23:08.749 & \multirow{2}{*}{27.4} & \multirow{2}{*}{n.d.} &  $14\pm1$  & $1.1\pm0.1$  & $0.9\pm 0.2$ & $-2.4\pm0.1$ & $ -0.4\pm0.2$ & $1.84$ & $2.89$ \\
		& B & 07:53:54.164 & +33:23:08.914 & & & $10\pm2$ & $0.36\pm0.07$ & $0.33\pm0.07$ & $-3.1\pm0.1$ & $-0.2\pm0.2$& $1.73$ &\\
        \hline
        \multirow{2}{*}{MJV14607} & A & 14:11:13.843 & +02:31:35.721 & \multirow{2}{*}{26.8} &  \multirow{2}{*}{n.d.} & $10\pm1$ & $8.9\pm 1.8$ & $8.5\pm1.7$& $-0.1\pm 0.1$ & $-0.08\pm0.19$ & $34.94$ & $57.81$\\
		& B & 14:11:13.836 & +02:31:35.867 & & &  $8.5\pm 1.7$ & $<0.23$ & $<0.20$ & & & & \\
        	\hline
        \multirow{2}{*}{MJV16999} & A & 13:29:15.823 & +43:25:17.050 & \multirow{2}{*}{23.6} &  \multirow{2}{*}{n.d.} & $10.1\pm1.0$ & $1.6\pm0.3$ &$1.5\pm0.3$ & $-1.7\pm0.1$ &$-0.07\pm0.19$ & $3.04$ & $10.54$  \\
		& B & 13:29:15.839 & +43:25:17.466 & &&  $6.0\pm1.2$ & $<0.23$ & $<0.20$ & & &\\
        \hline
        \hline
        \multirow{2}{*}{MJV03238} & A & 11:30:00.099  & +38:12:03.091 & \multirow{2}{*}{28.9} &  \multirow{2}{*}{$29\pm4$} & $16\pm2$ & &  &  &  &  & \\
		& B & 11:30:00.157 & +38:12:03.230 & & & $10\pm1$ & & & & &\\
        \hline
        \multirow{2}{*}{MJV02639} & A & 23:21:40.801 & +05:27:37.225 & \multirow{2}{*}{82.3} &  \multirow{2}{*}{$76\pm8$} & $63\pm6$ & &  &  &  &  & \\
		& B &23:21:40.808 &  +05:27:36.466 & & & $19\pm2$  & & & & &\\
        \hline

	\end{tabular}
\end{table}	
\end{landscape}

\subsection{Testing lens models for the remaining candidates}
\label{Sec:lens_modelling}

From our analysis of the multi-frequency VLBI data obtained for the mJIVE--20 lens candidates, we have two targets remaining that satisfy the surface brightness requirements, or there is insufficient information in the data to completely rule out a particular target; these are MJV00019 and MJV07417. Here, we now test whether the structure of the sub-components in the candidate lensed images can be explained by a simple lens mass model. 

To model the mass distribution of the remaining gravitational lens candidates, we use the publicly available software {\sc gravlens}, which applies a parametric lens modelling approach \citep{Keeton2001a,Keeton2001b}. We approximate the mass density distribution of the lensing galaxy to be a singular isothermal sphere (SIS), which is a simple, but not unrealistic lens mass model that can straightforwardly produce two images of the background source. The SIS is described by only three parameters, the lensing galaxy position ($x_l$, $y_l$) and its mass strength ($b$). The background object is assumed to be point-like or a collection of point-like sources where there are separate sub-components observed. Here, we only aim to test whether the relative positions and flux-densities of the data are consistent with this mass model. 

We assume the lens and source redshifts to be $z_l=0.5$ and $z_s =2$, respectively, which are the mean redshift of the lensing galaxy population for galaxy-galaxy lenses (e.g. \citealt{Collett2015}) and the mean redshift of the CLASS lensed sources (e.g. \citealt{Browne2003}).  This choice of lens and source redshift has no impact on testing whether the mJIVE--20 objects are gravitationally lensed or not, but is needed to estimate the physical enclosed mass of the lensing galaxy. Also, as none of the potential lensing galaxies are detected at optical wavelengths, neither in SDSS nor in Pan-STARSS, for both remaining candidates, we do not have any information about their position (see Fig. \ref{Fig:Panstarss}). As constraints to the lens mass model, we use the relative positions of the two candidate lensed images and their sub-components, if detected, from each target. These positions are measured from a Gaussian fit to the observed emission in the image-plane by using the task {\sc imfit} in {\sc casa}.

\subsubsection{MJV00019}
This candidate did not fail the selection criteria from the follow-up VLBA imaging; components A and B show a similar surface brightness and spectral energy distribution (see Table \ref{Tab:fluxdensities} and Fig. \ref{Fig:radio-spectra-1}). Moreover, there was no evidence of a clear core-jet or core-hotspots morphology (see Fig.~\ref{Fig:mjv00019-composite}). Since the two components are resolved, we parameterize the extended emission to provide additional constraints to the mass model, which otherwise would have been under-constrained. In total, we have two sub-components in the candidate lensed images to represent the likely core and extended jet emission, respectively. However, the image configuration of this system fails the parity test while performing the lens modelling. For a doubly imaged lensed object, such as is proposed here, the lensed images should have opposite parity \citep[e.g.][]{Wambsganss1998}. Moreover, the relative orientation of the extended structure is expected to be different.

However, at both 1.4 and 4.1 GHz, the most diffuse emission of components A and B is extended along the same direction (see Fig. \ref{Fig:mjv00019-composite}). Although such strange morphologies have been observed before on VLBI-scales, for example, in the case of CLASS B0128+437 that has four lensed images where one shows a shift in the jet position angle of around 90 degrees \citep{Biggs2004}, this would require an unusually complex model that we cannot test with the data in hand. Therefore, we cannot confirm that MJV00019 is a doubly imaged lensed source, given the extended emission fails the parity test.

\subsubsection{MJV07417}

\begin{figure}
\centering
\includegraphics[width = 0.47\textwidth]{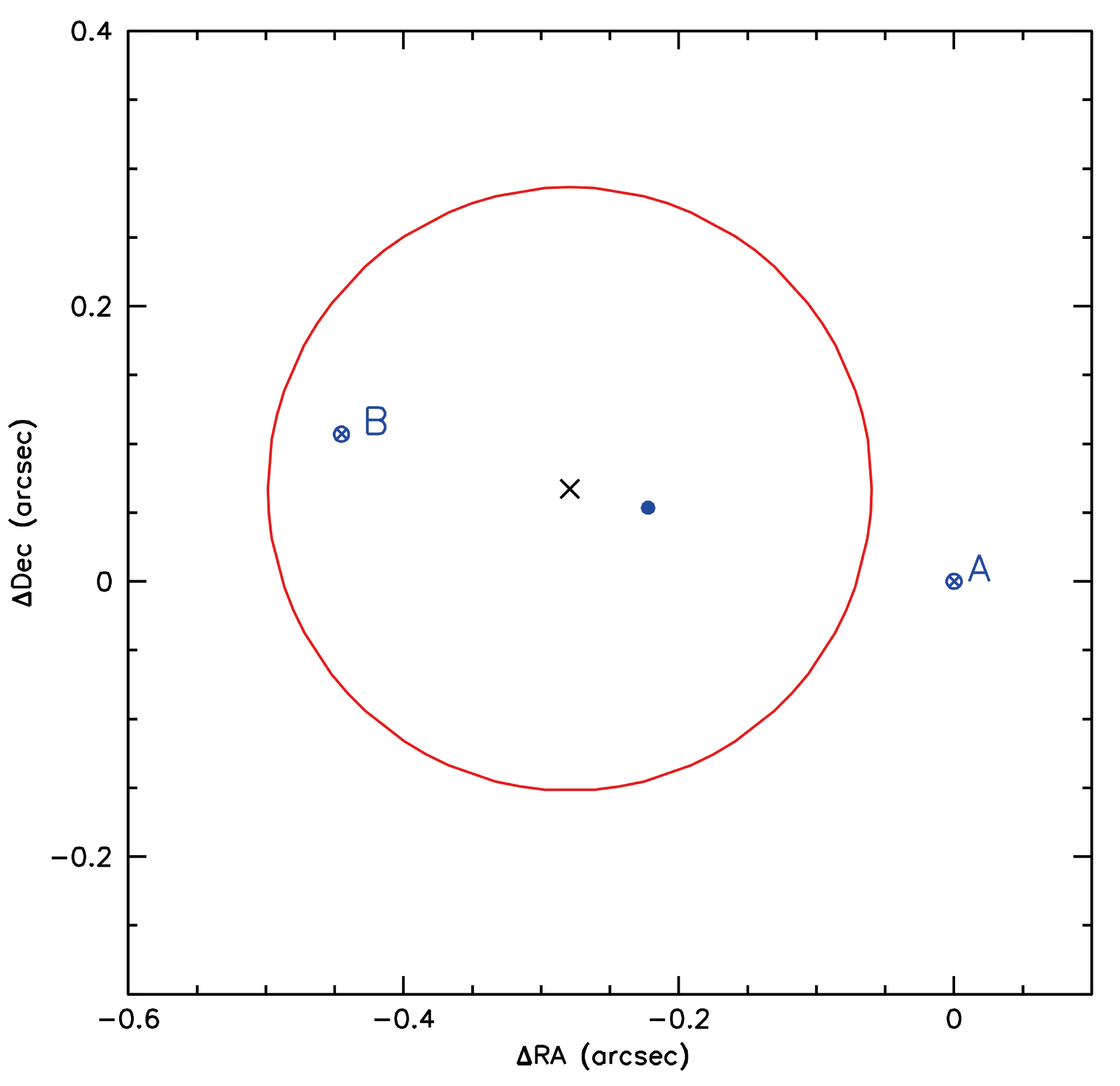}
\caption{An SIS lens mass model that reproduces the image configuration of the candidate MJV07417. The observed positions are the open circles, while the model-predicted positions are indicated by the crosses. The critical curve is represented by the red solid line and the black cross indicates the position of the lensing galaxy. The filled blue circle represents the source position.}\label{Fig:model-mjv07417}
\end{figure}

This lens candidate is detected only at 1.4 GHz and, therefore, could not be ruled out through the multi-frequency imaging. Since component B is unresolved, we do not have any constraints to test the parity of the candidate lensed images. Moreover, the positions of the two components do not give enough constraints to test a realistic lensing mass model. For this reason, we also use the flux density of components A and B to constrain the SIS mass model; this gives 5 observational constraints to a model that has 5 free parameters (the lensing galaxy position, its mass strength and the position of the source). In general, the image flux density may not be reliable, either because of the possible intrinsic variability of compact radio sources (e.g. \citealt{Biggs1999}), because of the presence of substructures (in the lens plane or along the line-of-sight; e.g. \citealt{Mao1998}), or because of possible propagation effects due to the interstellar medium (e.g. \citealt{Koopmans2003, Mittal2007}). Therefore, we assume an uncertainty of 20 percent on the flux density of the candidate lensed-images to take into account these possible effects. Our SIS mass model that reproduces the image positions is shown in Fig. \ref{Fig:model-mjv07417}. The SIS has a mass strength $b$ that corresponds to half the angular separation between the two images ($b = 0.23 \pm 0.03$) and the lensing galaxy is at ($x_l,y_l$) = ($-0.28\pm0.02$,  $0.06\pm0.04$), with respect to image A. Based on lens modelling, it is not possible to rule out this candidate as a gravitational lens. However, given the simplicity of the model, in order to confirm the lensing nature, further multi-frequency observations that detect both components are need. Also, deeper optical imaging may uncover the lensing galaxy or detect the AGN host galaxy if MJV07417 is not gravitationally lensed.

\section{The lensing statistics of the mJIVE--20 gravitational lens survey}
\label{Sec:Statistics}

From our follow-up of an initial 3\,640 radio sources that were detected on mas-scales during the mJIVE--20 survey, we have re-discovered two gravitational lenses that were previously found by CLASS. This gives a lensing rate of at least 1:($1820\pm 1287$) for our pilot lens search, where the uncertainty is based on the Poisson statistics for the number of lensed sources discovered and the total number of sources in the parent sample. However, we have one remaining lens candidate that we cannot completely rule out given the data in hand. If this candidate were to be confirmed as a genuine gravitational lens, then this will increase our lensing rate to at least 1:($1213\pm 700$), which is fairly typical for galaxy-scale gravitational lensing. Note that we also checked in the mJIVE--20 catalogue for any other previously known radio-loud gravitational lenses that we may have missed during our search, but there are none. An in-depth analysis of these lensing statistics is beyond the scope of this paper, particularly given the small number statistics that we currently have. Nevertheless, we qualitatively discuss here the statistical consequences of these two lensing rates.

If we consider a source at redshift $z_s$, the lensing probability, $\tau(z_s)$, can be derived by integrating along the light path,
	\begin{equation}
		\tau(z_s) = \int_0^{z_s} n(z)\, \sigma(z) \,\frac{c\,dt}{dz}\, dz,
		\label{eq.optical-depth}
	\end{equation}
where $n(z)$ is the number density of the potential lens population; $\sigma(z)$ is the area in the image-plane where the source must be in order to be strongly gravitationally lensed (the lensing cross-section), which is dependent on the mass of the lens, and the redshifts of the lens and source; and $c\,dt$ is the path-length to the background source, which is dependent on the cosmology and the source redshift (e.g. \citealt*{Turner1984}). The total number of gravitational lenses that are found from a survey, $N_t$, is calculated by integrating the lensing probabilities of all sources in the parent source population,
	\begin{equation}
		N_t = \int_0^{z_{\rm max}} \tau(z) \,\frac{dN(z_s)}{dz_s}\,dz_s,
		\label{eq.total_lens}
	\end{equation}
where $dN(z_s)/dz_s$ is the differential number counts of the source parent population as a function of redshift. From equations (\ref{eq.optical-depth}) and (\ref{eq.total_lens}), it is clear that the number of possible gravitational lenses that are found is a function of the lens (mass, redshift) and source (redshift) populations. However, the actual number of gravitational lenses that are found from a survey will also have to take into account the observational selection effects of that survey, such that the actual number of gravitational lenses, $N_a$, that we would expect to find is given by,
        \begin{equation}
		N_a = N_t \times C(\Delta\theta, S_{A/B}) \times B(S_{\nu}, z_s).
		\label{eq.actual_lens}
	\end{equation}
The survey selection function, $C (\Delta \theta, S_{A/B})$, is dependent on the selection criteria of the lensed sources that are found from the parent sample. In equation (\ref{eq.actual_lens}), the maximum image separation, $\Delta \theta$, and the flux-density ratios, $S_{A/B}$, of the lensed images are used, which are dependent on the angular resolution and sensitivity of the survey. A magnification bias is also introduced by the boost on the flux-density that is caused by lensing. This results in sources with an intrinsic flux-density below the survey limit being included in the parent population sample. This bias factor is given by $B (S_{\nu}, z_{s})$, and requires knowledge of the background source number density as a function of their flux density and redshift, which should also be extended below the survey flux-density limit.

For example, in the case of CLASS, a complete sample of 11\,685 flat-spectrum radio sources (defined as $\alpha_{4.85}^{1.4} \geq -0.5$ between 1.4 and 4.85~GHz from low resolution imaging with NVSS and GB6, respectively) with $S_{\rm 4.85~GHz} \geq 30$~mJy were observed using the VLA at 8.46 GHz with an angular-resolution of 170~mas, and to an rms of 170~$\mu$Jy~beam$^{-1}$. From these observations, gravitational lenses could only be confidently identified when the image separation was $\Delta\theta \geq 300$~mas, and when the total flux-density was $S_{\rm 8.46~GHz} \geq 18.7$~mJy (i.e. for at least a $10\sigma$ detection of the weakest image and for a flux-ratio of $S_{A/B} \leq 10$). These criteria gave a statistical parent sample of 8\,958 radio sources, from which 13 gravitational lenses were found. This results in a CLASS lensing rate of 1:($689 \pm 190$) \citep{Browne2003}. 

Given that the sample of 3\,640 radio sources detected on mas-scales by mJIVE--20 will likely be core-dominated, we can use the CLASS lensing rate to estimate that, if our parent sample had the same source properties, then we would expect $5\pm2$ gravitational lenses within mJIVE--20. 

However, when we take into account the sensitivity of the mJIVE--20 survey by imposing our condition of detecting the faintest lensed image at least at the $6.5\sigma$-level, and a minimum flux-ratio of 10:1, we obtain a statistical parent population of 635 radio sources with $S_{\rm 1.4~GHz} \geq 10.7$~mJy. Note that in our lens candidate selection we apply a larger upper limit on the flux ratio between the putative lensed images, and our two confirmed lenses fully satisfy the condition of a flux-ratio larger than 10:1.
This statistical parent sample gives an mJIVE--20 lensing rate of 1:($318\pm225$), given the detection of two confirmed gravitational lenses, which although higher, is still consistent with the CLASS lensing-rate at the $1\sigma$-level, given the small number statistics (i.e. an expectation value of $1\pm1$ gravitational lenses within the mJIVE--20 sample when applying the CLASS lensing rate). This implies that the lensing optical depth between the mJIVE--20 and CLASS parent populations are in agreement within the uncertainties, that is, the mJIVE--20 sources are consistent with the power-law number counts \citep{McKean2007b} and the redshift distribution (mean $z \sim 1.2$; \citealt{Henstock1997,Marlow2000}; McKean et al., in prep.) of flat-spectrum radio sources down to $S_{\rm 4.85~GHz} \geq 5$~mJy.

Finally, we note that even though the mJIVE--20 lens search probes image-separations of the order of tens mas due to the excellent angular resolution of the VLBI observations, this does not necessarily mean that substantially more gravitational lenses will be discovered. For example, the image-separation distribution of the CLASS sample of gravitational lenses (see Fig.~\ref{Fig:angular-sep}) shows a turnover at $\Delta\theta < 1$~arcsec, demonstrating a paucity of gravitational lenses found with small image-separations; for the nine CLASS gravitational lenses in the statistically well-defined sample, where there is only one dominant lensing galaxy, a similar distribution is found (see \citealt{Chae2003}). This is because the lensing cross-section of lower mass galaxies is also lower, and in fact \citet{Chae2003} find that the image-separation distribution from CLASS is consistent with the galaxy luminosity functions derived from the SDSS or the 2dF Galaxy Redshift Survey (see also \citealt{Oguri2006} for a discussion on the image-separation distribution with respect to the halo-mass function). Moreover, the possible mas-scale lens population has been previously investigated by \citet{Wilkinson2001} using targetted VLBI observations of a sample of 300 radio sources, which led to a null detection of lensed sources with image separations between 1.5 and 50 mas. Therefore, it is not too unexpected that the mJIVE--20 lens search has not detected any small image-separation systems with $\Delta\theta < 0.3$~arcsec, even though the data quality is sensitive to such rare gravitational lenses. 

Observations with a much larger sample of sources detected on mas-scales with wide-field VLBI would be needed to both test whether the lensing statistics, and hence the parent population, from the mJIVE--20 and CLASS sources are similar, and to detect rare systems where the image separation statistics could potentially constrain models for the halo mass-function. 

\section{Prospects for future lens searches with wide-field VLBI}
\label{Sec:Future}

Our re-discovery of two gravitational lenses found during CLASS, directly from the mJIVE--20 imaging at 1.4~GHz, has provided an immediate proof-of-concept that wide-field surveys with VLBI arrays can efficiently find gravitationally lensed radio sources. Although these two gravitational lens systems did not go through the same follow-up procedure as the rest of our candidates (to save observing time), such observations (at 5 GHz) were carried out as part of their discovery datasets \citep{Koopmans1999,Rusin2001}. In this section, we now discuss the prospects for future lens surveys with wide-field VLBI. Even though observations at such a high angular resolution of just a few mas are sensitive to high brightness temperature radio sources, we divide our discussion into essentially two classes of radio source; those  where the background source is compact, such as from CLASS, and those that are extended, such as from the MG lens survey.

\subsection{Compact lensed radio sources}

These sources are core-dominated in morphology and have a flat or rising spectrum at cm-wavelengths due to synchrotron self-absorption.  Being intrinsically compact, they are easily identifiable at mas-scales with VLBI, even when the {\it uv}-coverage is sparse, such as in the case of a snapshot survey. For example, all of the gravitational lenses within CLASS would be immediately recovered from a wide-field VLBI survey that has a similar sensitivity (central $\sigma_{\rm rms} = 150~\mu$Jy~beam$^{-1}$) and {\it uv}-coverage (see Fig.~\ref{Fig:uvcoverage}) to mJIVE--20. However, multi-frequency follow-up with VLBI would still be needed to ensure that the surface brightness and the radio spectra of the candidate lensed images are consistent with gravitational lensing. Therefore, some filtering or pre-selection of objects that have a higher likelihood of being lensed would reduce the number of false-positives and limit the amount of follow-up imaging required.

Efficiently selecting such sources, without spectral information, could be done by comparing their low-resolution flux-density at arcsec-scales with the flux-density recovered at mas-scales with a long baseline component of the observing array.  For example, in the cases of MJV03238 (CLASS~B1127+385) and MJV02639 (CLASS~B2319+051), the FIRST and mJIVE--20 flux densities agree to within 10 percent. However, compact radio sources are also sometimes highly variable. In the case of CLASS~B1127+385, radio monitoring has shown that it is not variable at 8.46~GHz \citep{Rumbaugh2015}. Also, although CLASS~B2319+051 has not been extensively monitored with the VLA for variability, monitoring with the Westerbork Synthesis Radio Telescope (WSRT) at 5~GHz, where the lensed images were not resolved, did not detect any rapid-variability in the total flux-density of the system within the measurement uncertainties \citep{Gurkan2014}. We note that as well as the mJIVE--20 flux density being consistent with FIRST (and previous VLA imaging by \citealt{Koopmans1999} and \citealt{Rusin2001}), the flux-ratio between the two lensed images has also not changed in both systems, which also adds to the case that these objects are not variable. Therefore, if only those candidates where the mas- and arcsec-scale flux-densities agree to within 90 percent were selected as part of our search, then our initial sample of fourteen objects would have been reduced to just five candidates for follow-up imaging (including at least two confirmed gravitational lenses). As such, comparing the flux-densities on different angular-scales should help identify core-dominated radio sources and improve the efficiency of any lens search, even in the absence of spectral information.

Another method of pre-selecting VLBI-detected radio sources with a high lensing likelihood is by comparing their very precise radio positions (sub-mas) with optical information. In the vast majority of non-lensed objects, the core-dominated radio emission will be associated with AGN activity that is coincident with the centre of a massive elliptical galaxy. In the case of a gravitationally lensed radio source, with a maximum image separation of 0.5 to 2 arcsec, the radio components will be offset from the position of the lensing galaxy, by around 0.3 to 1 arcsec (depending on the particular image-configuration). Such offsets would be detectable in comparison with optical imaging, which has a typical astrometric uncertainty of $<0.3$~arcsec. This method was first suggested by \citet{Jackson2007}, who compared the astrometric positions of objects within FIRST and SDSS. However, they found no new gravitational lenses as part of a small pilot search with the VLA and MERLIN, mainly due to the positional uncertainties of the respective surveys being too large. Future optical surveys, such as with the Large Synoptic Survey Telescope (LSST) will have a better astrometric precision, particularly when referenced to {\it Gaia}, which will allow for a better pre-selection of likely lensed objects for high resolution follow-up. 

The expected number of compact radio sources that are gravitationally lensed can be estimated using the CLASS lensing rate and assuming that the redshift distribution and number counts of the background source population do not change significantly at low flux densities. There is evidence that the composition of the compact radio source population is changing from being dominated by quasars to radio galaxies towards lower flux-densities, but the overall mean redshift down to $S_{\rm 4.85~GHz} \geq 5$~mJy seems to be fairly consistent at $z \sim 1.2$ (\citealt{Henstock1997,Marlow2000}; McKean et al, in prep.). In addition, our lens search using the mJIVE--20 survey data is consistent with their being no change in the properties of the parent population. The differential number counts of flat-spectrum radio sources with flux-densities $S_{\rm 4.85~GHz} \geq 5$~mJy are well described by the power-law,
\begin{equation}
n(S) = (6.91\pm0.42) \left( \frac{S_{4.85}}{100~{\rm mJy}} \right)^{-2.06\pm0.01}~{\rm mJy^{-1}~sr^{-1}},
\label{number-counts}
\end{equation}
which can be used to estimate the number of compact radio sources that would be detectable at 1.4 GHz with the VLBA (assuming that all of the flux is recovered on mas-scales, and by taking into account a median spectral index of faint flat-spectrum radio sources of $\alpha_{1.4}^{4.85} = -0.15$; \citealt{McKean2007b}). An all-sky survey at declinations $\delta> -30$~deg (9.42~sr area) with a $6.5\sigma$ point-source sensitivity of $S_{\rm 1.4~GHz} \geq 0.91$~mJy (corresponding to a rms at the centre of the primary beam of $\sigma_{\rm rms} = 140$~$\mu$Jy~beam$^{-1}$) is expected to detect $1.15 \times 10^6$ radio sources on VLBI-scales. From the CLASS lensing-rate, it is predicted that 1675 of these radio sources will be gravitationally lensed.

As discussed above, to identify such a sample of gravitationally lensed radio sources would require detecting at least two of the lensed images directly in the survey data. Therefore, assuming a maximum flux ratio of 10:1 between the two lensed images gives a minimum total flux-density of $S_{\rm 1.4~GHz} \geq 10.0$~mJy, for a $6.5\sigma$ detection threshold and a central rms of $\sigma_{\rm rms} = 140$~$\mu$Jy~beam$^{-1}$. This would result in a potential parent sample of $9.1 \times 10^4$ radio sources, which given the CLASS lensing-rate, should provide a sample of 130 gravitational lenses (including the 22 found as part of CLASS). However, this would be for a statistically complete sample. If the flux-ratio limit were relaxed (to around $<2$:1; typical for doubly imaged sources and the merging images of quadruply imaged sources) and the survey limit was correspondingly lowered to $S_{\rm 1.4~GHz} \geq 2.7$~mJy, then around 530 gravitational lenses could be found, potentially increasing the number of known radio-loud lensed sources by over an order of magnitude. Such a survey, in terms of depth and area, would be feasible with the VLBA at 1.4 GHz in around 3000 h (assuming 90-s on-source per pointing and a recording rate of 4096~Mbit\,s$^{-1}$). We note that taking the primary beam attenuation into account would increase the detection threshold in those regions away from the pointing centre by at most a factor of two, and therefore, the number of radio sources and gravitational lenses found from such a survey could decrease by at most a factor of two from those estimated here.

Such a sample of 530 gravitationally lensed radio-loud AGN would be useful for investigating the halo mass function at the low mass end, given the small image separations that could be probed (e.g. \citealt{Chae2003,Oguri2006}), but also for testing different models for dark matter by searching for low-mass substructure or sub-haloes along the line-of-sight (e.g. \citealt{Dalal2002}). Again, assuming a similar lensing-rate and ratio of doubly- to quadruply-imaged sources (about 2:1) to CLASS, around 175 new four-image lens systems are expected from the wide-field snapshot VLBI survey described above. This is close to the around $10^2$ systems needed in order to discriminate between cold and warm dark matter models \citep{Gilman2018}. Finally, the sensitivity to lensed images with extremely small separations ($>5$~mas), would also uniquely constrain the abundance of compact objects (e.g. free-floating black holes with masses $>10^6$~M$_{\odot}$) in the Universe \citep{Wilkinson2001}. Of course, if the number of lensed radio sources within the statistically complete sample differs significantly from what is predicted, then that would suggest the compact radio source population with intrinsic de-magnified flux-densities of 0.2 to 0.5 mJy has likely also changed significantly in comparison to those sources  that were probed by CLASS. 

\subsection{Extended lensed radio sources}

Sources with extended emission, such as those that produce gravitational arcs or rings at radio wavelengths \citep{Biggs2001,More2009,Macleod2013,Hsueh2016,Spingola2018} can provide a wealth of information to determine precise mass models for the lens \citep{Wucknitz2004}. Also, due to the excellent angular resolution provided by VLBI, such lensed sources are sensitive to the direct detection of low mass substructure in the lens \citep{Vegetti2012} or along the line-of-sight \citep{Despali2018}, through the local change they produce in the surface brightness distribution of the extended jet-emission \citep{Metcalf2002}. However, the lensing-rate of such objects is not well known, as the MG lens survey only discovered six gravitational lenses, and the survey was not designed to be statistically complete. Also, the radio sources that have produced the largest gravitational arcs are also the most luminous known, and so, new surveys that extend to fainter flux-densities will likely only detect lensed radio sources with weak jets.

Identifying those lensed radio sources with extended emission would be challenging from a snapshot survey as the {\it uv}-coverage would not be sufficient to detect the full extent of the gravitational arc. However, once a gravitationally lensed radio source has been identified from the survey data, an indication for extended emission that wasn't fully detected from the snapshot observations could come from comparing the total flux density measured on arcsec-scales with that found on VLBI-scales; this method can be confidently used only in the case of quadruply-imaged radio sources, because the VLBI observations alone can already show that the source is gravitationally lensed. Such objects could then be prioritised for deeper, long-track imaging in order to recover the extended gravitational arcs. However, for those cases of doubly-imaged sources with extended emission, this method would likely not be sufficient to distinguish between a lensed radio source with extended emission and a resolved non-lensed core-jet radio source, for example. Therefore, multi-wavelength and spectroscopic observations would still be required to understand if such an object is gravitationally lensed or not.

\section{Conclusions}
\label{Sec:Conclusions}

We have carried out the first wide-field VLBI search for gravitationally lensed radio sources using data from the mJIVE--20 survey. Among the mJIVE--20 sample of $3\,640$ compact radio sources, 81 sources were identified as having multiple radio components that are separated by more than 100 mas and with a flux-density ratio of less than 15:1. Among them, we selected fourteen sources as gravitational lens candidates, based on the morphology and surface brightness of the radio emission at 1.4 GHz. Two of these selected candidates are a re-discovery of the known gravitational lenses CLASS~B1127+385 and CLASS ~B2319+051, which provides an immediate proof-of-concept that this selection method is able to find lensed radio sources. 

We have followed-up the remaining twelve lens candidates at 4.1 and 7.1 GHz with the VLBA, following a strategy that is similar to the CLASS survey. Two of the targets (MJV00019 and MJV07417) have a surface brightness and spectral indices that are either consistent with lensing, or the data in hand are insufficient to rule out the lensing hypothesis; the remaining ten objects are likely core-jet radio sources. We find that in the case of MJV00019, it is not possible to reproduce the morphology of the candidate lensed images with a simple lensing mass model. For the other target, the configuration of the candidate lensed images is found to be compatible with the lensing scenario. However, since MJV07417 is detected only at 1.4 GHz, further deeper observations at 5 GHz are needed in order to confirm the true nature of this system. 

Based on our search, we find a lensing rate for VLBI-detected radio sources at 1.4 GHz of 1:($318\pm225$), which although is almost a factor of two higher than that of CLASS, is still consistent given the large uncertainties due to the small number statistics. The implication of this result is that the lensing optical depth of compact radio sources has not changed significantly toward lower flux densities. We estimate that a wide-field survey carried out with the VLBA, to a sensitivity of 140~$\mu$Jy~beam$^{-1}$ over 9.42~sr, should find around 530 new gravitationally lensed radio sources, given the lensing-rate and source properties of the CLASS survey. Such a survey would be a precursor to what could be done with the SKA, where the $\mu$Jy~beam$^{-1}$ sensitivity should detect thousands of radio-loud lensed AGN \citep{McKean2015SKA}, which could be done most efficiently if the SKA has a VLBI component \citep{Paragi2015}.

Finally, we note that we have looked through only part of the mJIVE--20 survey data, and now that the survey is complete, we may find new gravitationally lensed radio sources in the final 30 percent of the data. Our search will also be refined using the results found here. In addition, we have only searched for gravitationally lensed objects with image separations $\geq 100$~mas. Extending the parameter space to smaller image-separations may uncover close merging pairs of quadruply imaged sources, and also potentially detect the $>5$~mas image splitting by low mass, compact objects along the line-of-sight to the distant radio sources.

\section*{Acknowledgements}

The authors thank the referee Olaf Wucknitz for the careful reading of the manuscript and useful comments, which improved the presentation of this work. CS would like to thank L.~V.~E. Koopmans for useful discussions and comments on this paper. This work is supported in part by NWO grant 629.001.023. ML acknowledges support from the ASTRON/JIVE summer student programme. The National Radio Astronomy Observatory is a facility of the National Science Foundation operated under cooperative agreement by Associated Universities, Inc. This work made use of the Swinburne University of Technology software correlator, developed as part of the Australian Major National Research Facilities Programme and operated under licence.




\bibliographystyle{mnras}
\bibliography{mjive} 

\newcommand{\noop}[1]{}
\begin{thebibliography}{}
\makeatletter
\relax
\def\mn@urlcharsother{\let\do\@makeother \do\$\do\&\do\#\do\^\do\_\do\%\do\~}
\def\mn@doi{\begingroup\mn@urlcharsother \@ifnextchar [ {\mn@doi@}
  {\mn@doi@[]}}
\def\mn@doi@[#1]#2{\def\@tempa{#1}\ifx\@tempa\@empty \href
  {http://dx.doi.org/#2} {doi:#2}\else \href {http://dx.doi.org/#2} {#1}\fi
  \endgroup}
\def\mn@eprint#1#2{\mn@eprint@#1:#2::\@nil}
\def\mn@eprint@arXiv#1{\href {http://arxiv.org/abs/#1} {{\tt arXiv:#1}}}
\def\mn@eprint@dblp#1{\href {http://dblp.uni-trier.de/rec/bibtex/#1.xml}
  {dblp:#1}}
\def\mn@eprint@#1:#2:#3:#4\@nil{\def\@tempa {#1}\def\@tempb {#2}\def\@tempc
  {#3}\ifx \@tempc \@empty \let \@tempc \@tempb \let \@tempb \@tempa \fi \ifx
  \@tempb \@empty \def\@tempb {arXiv}\fi \@ifundefined
  {mn@eprint@\@tempb}{\@tempb:\@tempc}{\expandafter \expandafter \csname
  mn@eprint@\@tempb\endcsname \expandafter{\@tempc}}}

\bibitem[\protect\citeauthoryear{{Abazajian} et~al.,}{{Abazajian}
  et~al.}{2009}]{Abazajian2009SDSS}
{Abazajian} K.~N.,  et~al., 2009, \mn@doi [\apjs]
  {10.1088/0067-0049/182/2/543}, \href
  {http://adsabs.harvard.edu/abs/2009ApJS..182..543A} {182, 543}

\bibitem[\protect\citeauthoryear{{Auger}, {Treu}, {Bolton}, {Gavazzi},
  {Koopmans}, {Marshall}, {Moustakas}  \& {Burles}}{{Auger}
  et~al.}{2010}]{Auger2010}
{Auger} M.~W.,  {Treu} T.,  {Bolton} A.~S.,  {Gavazzi} R.,  {Koopmans}
  L.~V.~E.,  {Marshall} P.~J.,  {Moustakas} L.~A.,   {Burles} S.,  2010,
  \mn@doi [\apj] {10.1088/0004-637X/724/1/511}, \href
  {http://adsabs.harvard.edu/abs/2010ApJ...724..511A} {724, 511}

\bibitem[\protect\citeauthoryear{{Augusto} et~al.,}{{Augusto}
  et~al.}{2001}]{Augusto2001b}
{Augusto} P.,  et~al., 2001, \mn@doi [\mnras]
  {10.1046/j.1365-8711.2001.04764.x}, \href
  {http://adsabs.harvard.edu/abs/2001MNRAS.326.1007A} {326, 1007}

\bibitem[\protect\citeauthoryear{{Barvainis} \& {Ivison}}{{Barvainis} \&
  {Ivison}}{2002}]{Barvainis2002}
{Barvainis} R.,  {Ivison} R.,  2002, \mn@doi [\apj] {10.1086/340096}, \href
  {http://adsabs.harvard.edu/abs/2002ApJ...571..712B} {571, 712}

\bibitem[\protect\citeauthoryear{{Becker}, {White}  \& {Helfand}}{{Becker}
  et~al.}{1995}]{Becker1995}
{Becker} R.~H.,  {White} R.~L.,   {Helfand} D.~J.,  1995, \mn@doi [\apj]
  {10.1086/176166}, \href {http://adsabs.harvard.edu/abs/1995ApJ...450..559B}
  {450, 559}

\bibitem[\protect\citeauthoryear{{Biggs} \& {Browne}}{{Biggs} \&
  {Browne}}{2018}]{Biggs2018}
{Biggs} A.~D.,  {Browne} I.~W.~A.,  2018, \mn@doi [\mnras]
  {10.1093/mnras/sty565}, \href
  {http://adsabs.harvard.edu/abs/2018MNRAS.476.5393B} {476, 5393}

\bibitem[\protect\citeauthoryear{{Biggs}, {Browne}, {Helbig}, {Koopmans},
  {Wilkinson}  \& {Perley}}{{Biggs} et~al.}{1999}]{Biggs1999}
{Biggs} A.~D.,  {Browne} I.~W.~A.,  {Helbig} P.,  {Koopmans} L.~V.~E.,
  {Wilkinson} P.~N.,   {Perley} R.~A.,  1999, \mn@doi [\mnras]
  {10.1046/j.1365-8711.1999.02309.x}, \href
  {http://adsabs.harvard.edu/abs/1999MNRAS.304..349B} {304, 349}

\bibitem[\protect\citeauthoryear{{Biggs}, {Browne}, {Muxlow}  \&
  {Wilkinson}}{{Biggs} et~al.}{2001}]{Biggs2001}
{Biggs} A.~D.,  {Browne} I.~W.~A.,  {Muxlow} T.~W.~B.,   {Wilkinson} P.~N.,
  2001, \mn@doi [\mnras] {10.1046/j.1365-8711.2001.04176.x}, \href
  {http://adsabs.harvard.edu/abs/2001MNRAS.322..821B} {322, 821}

\bibitem[\protect\citeauthoryear{{Biggs}, {Wucknitz}, {Porcas}, {Browne},
  {Jackson}, {Mao}  \& {Wilkinson}}{{Biggs} et~al.}{2003}]{Biggs2003}
{Biggs} A.~D.,  {Wucknitz} O.,  {Porcas} R.~W.,  {Browne} I.~W.~A.,  {Jackson}
  N.~J.,  {Mao} S.,   {Wilkinson} P.~N.,  2003, \mn@doi [\mnras]
  {10.1046/j.1365-8711.2003.06050.x}, \href
  {http://adsabs.harvard.edu/abs/2003MNRAS.338..599B} {338, 599}

\bibitem[\protect\citeauthoryear{{Biggs}, {Browne}, {Jackson}, {York},
  {Norbury}, {McKean}  \& {Phillips}}{{Biggs} et~al.}{2004}]{Biggs2004}
{Biggs} A.~D.,  {Browne} I.~W.~A.,  {Jackson} N.~J.,  {York} T.,  {Norbury}
  M.~A.,  {McKean} J.~P.,   {Phillips} P.~M.,  2004, \mn@doi [\mnras]
  {10.1111/j.1365-2966.2004.07701.x}, \href
  {http://adsabs.harvard.edu/abs/2004MNRAS.350..949B} {350, 949}

\bibitem[\protect\citeauthoryear{{Brada{\v c}}, {Schneider}, {Steinmetz},
  {Lombardi}, {King}  \& {Porcas}}{{Brada{\v c}} et~al.}{2002}]{Bradac2002}
{Brada{\v c}} M.,  {Schneider} P.,  {Steinmetz} M.,  {Lombardi} M.,  {King}
  L.~J.,   {Porcas} R.,  2002, \mn@doi [\aap] {10.1051/0004-6361:20020559},
  \href {http://adsabs.harvard.edu/abs/2002A%26A...388..373B} {388, 373}

\bibitem[\protect\citeauthoryear{{Browne} et~al.,}{{Browne}
  et~al.}{2003}]{Browne2003}
{Browne} I.~W.~A.,  et~al., 2003, \mn@doi [\mnras]
  {10.1046/j.1365-8711.2003.06257.x}, \href
  {http://adsabs.harvard.edu/abs/2003MNRAS.341...13B} {341, 13}

\bibitem[\protect\citeauthoryear{{Cao}, {Frey}, {Gurvits}, {Yang}, {Hong},
  {Paragi}, {Deller}  \& {Ivezi{\'c}}}{{Cao} et~al.}{2014}]{Cao2014}
{Cao} H.-M.,  {Frey} S.,  {Gurvits} L.~I.,  {Yang} J.,  {Hong} X.-Y.,  {Paragi}
  Z.,  {Deller} A.~T.,   {Ivezi{\'c}} {\v Z}.,  2014, \mn@doi [\aap]
  {10.1051/0004-6361/201323328}, \href
  {http://adsabs.harvard.edu/abs/2014A%26A...563A.111C} {563, A111}

\bibitem[\protect\citeauthoryear{{Chae}}{{Chae}}{2003}]{Chae2003}
{Chae} K.-H.,  2003, \mn@doi [\mnras] {10.1111/j.1365-2966.2003.07092.x}, \href
  {http://adsabs.harvard.edu/abs/2003MNRAS.346..746C} {346, 746}

\bibitem[\protect\citeauthoryear{{Chae} et~al.,}{{Chae}
  et~al.}{2002}]{Chae2002}
{Chae} K.-H.,  et~al., 2002, \mn@doi [Physical Review Letters]
  {10.1103/PhysRevLett.89.151301}, \href
  {http://adsabs.harvard.edu/abs/2002PhRvL..89o1301C} {89, 151301}

\bibitem[\protect\citeauthoryear{{Chi}, {Barthel}  \& {Garrett}}{{Chi}
  et~al.}{2013}]{Chi2013}
{Chi} S.,  {Barthel} P.~D.,   {Garrett} M.~A.,  2013, \mn@doi [\aap]
  {10.1051/0004-6361/201220783}, \href
  {http://adsabs.harvard.edu/abs/2013A%26A...550A..68C} {550, A68}

\bibitem[\protect\citeauthoryear{{Collett}}{{Collett}}{2015}]{Collett2015}
{Collett} T.~E.,  2015, \mn@doi [\apj] {10.1088/0004-637X/811/1/20}, \href
  {http://adsabs.harvard.edu/abs/2015ApJ...811...20C} {811, 20}

\bibitem[\protect\citeauthoryear{{Dalal} \& {Kochanek}}{{Dalal} \&
  {Kochanek}}{2002}]{Dalal2002}
{Dalal} N.,  {Kochanek} C.~S.,  2002, \mn@doi [\apj] {10.1086/340303}, \href
  {http://adsabs.harvard.edu/abs/2002ApJ...572...25D} {572, 25}

\bibitem[\protect\citeauthoryear{{Deller} \& {Middelberg}}{{Deller} \&
  {Middelberg}}{2014}]{Deller2014}
{Deller} A.~T.,  {Middelberg} E.,  2014, \mn@doi [\aj]
  {10.1088/0004-6256/147/1/14}, \href
  {http://adsabs.harvard.edu/abs/2014AJ....147...14D} {147, 14}

\bibitem[\protect\citeauthoryear{{Deller}, {Tingay}, {Bailes}  \&
  {West}}{{Deller} et~al.}{2007}]{Deller2007}
{Deller} A.~T.,  {Tingay} S.~J.,  {Bailes} M.,   {West} C.,  2007, \mn@doi
  [\pasp] {10.1086/513572}, \href
  {http://adsabs.harvard.edu/abs/2007PASP..119..318D} {119, 318}

\bibitem[\protect\citeauthoryear{{Deller} et~al.,}{{Deller}
  et~al.}{2011}]{Deller2011}
{Deller} A.~T.,  et~al., 2011, \mn@doi [\pasp] {10.1086/658907}, \href
  {http://adsabs.harvard.edu/abs/2011PASP..123..275D} {123, 275}

\bibitem[\protect\citeauthoryear{{Despali}, {Vegetti}, {White}, {Giocoli}  \&
  {van den Bosch}}{{Despali} et~al.}{2018}]{Despali2018}
{Despali} G.,  {Vegetti} S.,  {White} S.~D.~M.,  {Giocoli} C.,   {van den
  Bosch} F.~C.,  2018, \mn@doi [\mnras] {10.1093/mnras/sty159}, \href
  {http://adsabs.harvard.edu/abs/2018MNRAS.475.5424D} {475, 5424}

\bibitem[\protect\citeauthoryear{{Fassnacht}, {Xanthopoulos}, {Koopmans}  \&
  {Rusin}}{{Fassnacht} et~al.}{2002}]{Fassnacht2002}
{Fassnacht} C.~D.,  {Xanthopoulos} E.,  {Koopmans} L.~V.~E.,   {Rusin} D.,
  2002, \mn@doi [\apj] {10.1086/344368}, \href
  {http://adsabs.harvard.edu/abs/2002ApJ...581..823F} {581, 823}

\bibitem[\protect\citeauthoryear{{Flewelling} et~al.,}{{Flewelling}
  et~al.}{2016}]{Flewelling2016Panstarss}
{Flewelling} H.~A.,  et~al., 2016, preprint, \href
  {http://adsabs.harvard.edu/abs/2016arXiv161205243F} {} (\mn@eprint {arXiv}
  {1612.05243})

\bibitem[\protect\citeauthoryear{{Garrett} et~al.,}{{Garrett}
  et~al.}{2001}]{Garrett2001}
{Garrett} M.~A.,  et~al., 2001, \mn@doi [\aap] {10.1051/0004-6361:20000537},
  \href {http://adsabs.harvard.edu/abs/2001A%26A...366L...5G} {366, L5}

\bibitem[\protect\citeauthoryear{{Garrett}, {Wrobel}  \& {Morganti}}{{Garrett}
  et~al.}{2005}]{Garrett2005}
{Garrett} M.~A.,  {Wrobel} J.~M.,   {Morganti} R.,  2005, \mn@doi [\apj]
  {10.1086/426424}, \href {http://adsabs.harvard.edu/abs/2005ApJ...619..105G}
  {619, 105}

\bibitem[\protect\citeauthoryear{{Gilman}, {Birrer}, {Treu}  \&
  {Keeton}}{{Gilman} et~al.}{2018}]{Gilman2018}
{Gilman} D.,  {Birrer} S.,  {Treu} T.,   {Keeton} C.~R.,  2018, preprint, \href
  {http://adsabs.harvard.edu/abs/2017arXiv171204945G} {} (\mn@eprint {arXiv}
  {1712.04945})

\bibitem[\protect\citeauthoryear{{G{\"u}rkan}, {Jackson}, {Koopmans},
  {Fassnacht}  \& {Berciano Alba}}{{G{\"u}rkan} et~al.}{2014}]{Gurkan2014}
{G{\"u}rkan} G.,  {Jackson} N.,  {Koopmans} L.~V.~E.,  {Fassnacht} C.~D.,
  {Berciano Alba} A.,  2014, \mn@doi [\mnras] {10.1093/mnras/stu557}, \href
  {http://adsabs.harvard.edu/abs/2014MNRAS.441..127G} {441, 127}

\bibitem[\protect\citeauthoryear{{Henstock}, {Browne}, {Wilkinson}  \&
  {McMahon}}{{Henstock} et~al.}{1997}]{Henstock1997}
{Henstock} D.~R.,  {Browne} I.~W.~A.,  {Wilkinson} P.~N.,   {McMahon} R.~G.,
  1997, \mn@doi [\mnras] {10.1093/mnras/290.2.380}, \href
  {http://adsabs.harvard.edu/abs/1997MNRAS.290..380H} {290, 380}

\bibitem[\protect\citeauthoryear{{Herrera Ruiz} et~al.,}{{Herrera Ruiz}
  et~al.}{2017}]{Herrera-Ruiz2017}
{Herrera Ruiz} N.,  et~al., 2017, \mn@doi [\aap] {10.1051/0004-6361/201731163},
  \href {http://adsabs.harvard.edu/abs/2017A%26A...607A.132H} {607, A132}

\bibitem[\protect\citeauthoryear{{Hewitt}, {Turner}, {Schneider}, {Burke}  \&
  {Langston}}{{Hewitt} et~al.}{1988}]{Hewitt1988}
{Hewitt} J.~N.,  {Turner} E.~L.,  {Schneider} D.~P.,  {Burke} B.~F.,
  {Langston} G.~I.,  1988, \mn@doi [\nat] {10.1038/333537a0}, \href
  {http://adsabs.harvard.edu/abs/1988Natur.333..537H} {333, 537}

\bibitem[\protect\citeauthoryear{{Hsueh}, {Fassnacht}, {Vegetti}, {McKean},
  {Spingola}, {Auger}, {Koopmans}  \& {Lagattuta}}{{Hsueh}
  et~al.}{2016}]{Hsueh2016}
{Hsueh} J.-W.,  {Fassnacht} C.~D.,  {Vegetti} S.,  {McKean} J.~P.,  {Spingola}
  C.,  {Auger} M.~W.,  {Koopmans} L.~V.~E.,   {Lagattuta} D.~J.,  2016, \mn@doi
  [\mnras] {10.1093/mnrasl/slw146}, \href
  {http://adsabs.harvard.edu/abs/2016MNRAS.463L..51H} {463, L51}

\bibitem[\protect\citeauthoryear{{Hsueh} et~al.,}{{Hsueh}
  et~al.}{2017}]{Hsueh2017}
{Hsueh} J.-W.,  et~al., 2017, \mn@doi [\mnras] {10.1093/mnras/stx1082}, \href
  {http://adsabs.harvard.edu/abs/2017MNRAS.469.3713H} {469, 3713}

\bibitem[\protect\citeauthoryear{{Impellizzeri}, {McKean}, {Castangia}, {Roy},
  {Henkel}, {Brunthaler}  \& {Wucknitz}}{{Impellizzeri}
  et~al.}{2008}]{Impellizzeri2008}
{Impellizzeri} C.~M.~V.,  {McKean} J.~P.,  {Castangia} P.,  {Roy} A.~L.,
  {Henkel} C.,  {Brunthaler} A.,   {Wucknitz} O.,  2008, \mn@doi [\nat]
  {10.1038/nature07544}, \href
  {http://adsabs.harvard.edu/abs/2008Natur.456..927I} {456, 927}

\bibitem[\protect\citeauthoryear{{Inada} et~al.,}{{Inada}
  et~al.}{2010}]{Inada2010}
{Inada} N.,  et~al., 2010, \mn@doi [\aj] {10.1088/0004-6256/140/2/403}, \href
  {http://adsabs.harvard.edu/abs/2010AJ....140..403I} {140, 403}

\bibitem[\protect\citeauthoryear{{Jackson} \& {Browne}}{{Jackson} \&
  {Browne}}{2007}]{Jackson2007}
{Jackson} N.,  {Browne} I.~W.~A.,  2007, \mn@doi [\mnras]
  {10.1111/j.1365-2966.2006.11126.x}, \href
  {http://adsabs.harvard.edu/abs/2007MNRAS.374..168J} {374, 168}

\bibitem[\protect\citeauthoryear{{Keeton}}{{Keeton}}{2001b}]{Keeton2001b}
{Keeton} C.~R.,  2001b, ArXiv Astrophysics e-prints, \href
  {http://adsabs.harvard.edu/abs/2001astro.ph..2341K} {}

\bibitem[\protect\citeauthoryear{{Keeton}}{{Keeton}}{2001a}]{Keeton2001a}
{Keeton} C.~R.,  2001a, ArXiv Astrophysics e-prints, \href
  {http://adsabs.harvard.edu/abs/2001astro.ph..2340K} {}

\bibitem[\protect\citeauthoryear{{King}, {Browne}, {Marlow}, {Patnaik}  \&
  {Wilkinson}}{{King} et~al.}{1999}]{King1999}
{King} L.~J.,  {Browne} I.~W.~A.,  {Marlow} D.~R.,  {Patnaik} A.~R.,
  {Wilkinson} P.~N.,  1999, \mn@doi [\mnras]
  {10.1046/j.1365-8711.1999.02328.x}, \href
  {http://adsabs.harvard.edu/abs/1999MNRAS.307..225K} {307, 225}

\bibitem[\protect\citeauthoryear{{Koopmans} et~al.,}{{Koopmans}
  et~al.}{1999}]{Koopmans1999}
{Koopmans} L.~V.~E.,  et~al., 1999, \mn@doi [\mnras]
  {10.1046/j.1365-8711.1999.02342.x}, \href
  {http://adsabs.harvard.edu/abs/1999MNRAS.303..727K} {303, 727}

\bibitem[\protect\citeauthoryear{{Koopmans}, {de Bruyn}, {Xanthopoulos}  \&
  {Fassnacht}}{{Koopmans} et~al.}{2000}]{Koopmans2000}
{Koopmans} L.~V.~E.,  {de Bruyn} A.~G.,  {Xanthopoulos} E.,   {Fassnacht}
  C.~D.,  2000, \aap, \href
  {http://adsabs.harvard.edu/abs/2000A%26A...356..391K} {356, 391}

\bibitem[\protect\citeauthoryear{{Koopmans} et~al.,}{{Koopmans}
  et~al.}{2003}]{Koopmans2003}
{Koopmans} L.~V.~E.,  et~al., 2003, \mn@doi [\apj] {10.1086/377434}, \href
  {http://adsabs.harvard.edu/abs/2003ApJ...595..712K} {595, 712}

\bibitem[\protect\citeauthoryear{{Koopmans}, {Browne}  \& {Jackson}}{{Koopmans}
  et~al.}{2004}]{Koopmans2004SKA}
{Koopmans} L.~V.~E.,  {Browne} I.~W.~A.,   {Jackson} N.~J.,  2004, \mn@doi
  [\nar] {10.1016/j.newar.2004.09.047}, \href
  {http://adsabs.harvard.edu/abs/2004NewAR..48.1085K} {48, 1085}

\bibitem[\protect\citeauthoryear{{MacLeod}, {Jones}, {Agol}  \&
  {Kochanek}}{{MacLeod} et~al.}{2013}]{Macleod2013}
{MacLeod} C.~L.,  {Jones} R.,  {Agol} E.,   {Kochanek} C.~S.,  2013, \mn@doi
  [\apj] {10.1088/0004-637X/773/1/35}, \href
  {http://adsabs.harvard.edu/abs/2013ApJ...773...35M} {773, 35}

\bibitem[\protect\citeauthoryear{{Mao} \& {Schneider}}{{Mao} \&
  {Schneider}}{1998}]{Mao1998}
{Mao} S.,  {Schneider} P.,  1998, \mn@doi [\mnras]
  {10.1046/j.1365-8711.1998.01319.x}, \href
  {http://adsabs.harvard.edu/abs/1998MNRAS.295..587M} {295, 587}

\bibitem[\protect\citeauthoryear{{Mao} et~al.,}{{Mao} et~al.}{2017}]{Mao2017}
{Mao} S.~A.,  et~al., 2017, \mn@doi [Nature Astronomy]
  {10.1038/s41550-017-0218-x}, \href
  {http://adsabs.harvard.edu/abs/2017NatAs...1..621M} {1, 621}

\bibitem[\protect\citeauthoryear{{Marlow}, {Rusin}, {Jackson}, {Wilkinson},
  {Browne}  \& {Koopmans}}{{Marlow} et~al.}{2000}]{Marlow2000}
{Marlow} D.~R.,  {Rusin} D.,  {Jackson} N.,  {Wilkinson} P.~N.,  {Browne}
  I.~W.~A.,   {Koopmans} L.,  2000, \mn@doi [\aj] {10.1086/301375}, \href
  {http://adsabs.harvard.edu/abs/2000AJ....119.2629M} {119, 2629}

\bibitem[\protect\citeauthoryear{{McKean}}{{McKean}}{2011}]{McKean2011}
{McKean} J.~P.,  2011, \mn@doi [\mnras] {10.1111/j.1365-2966.2010.17954.x},
  \href {http://adsabs.harvard.edu/abs/2011MNRAS.412..900M} {412, 900}

\bibitem[\protect\citeauthoryear{{McKean} et~al.,}{{McKean}
  et~al.}{2005}]{McKean2005}
{McKean} J.~P.,  et~al., 2005, \mn@doi [\mnras]
  {10.1111/j.1365-2966.2004.08516.x}, \href
  {http://adsabs.harvard.edu/abs/2005MNRAS.356.1009M} {356, 1009}

\bibitem[\protect\citeauthoryear{{McKean}, {Browne}, {Jackson}, {Fassnacht}  \&
  {Helbig}}{{McKean} et~al.}{2007a}]{McKean2007b}
{McKean} J.~P.,  {Browne} I.~W.~A.,  {Jackson} N.~J.,  {Fassnacht} C.~D.,
  {Helbig} P.,  2007a, \mn@doi [\mnras] {10.1111/j.1365-2966.2007.11618.x},
  \href {http://adsabs.harvard.edu/abs/2007MNRAS.377..430M} {377, 430}

\bibitem[\protect\citeauthoryear{{McKean} et~al.,}{{McKean}
  et~al.}{2007b}]{McKean2007a}
{McKean} J.~P.,  et~al., 2007b, \mn@doi [\mnras]
  {10.1111/j.1365-2966.2007.11744.x}, \href
  {http://adsabs.harvard.edu/abs/2007MNRAS.378..109M} {378, 109}

\bibitem[\protect\citeauthoryear{{McKean} et~al.,}{{McKean}
  et~al.}{2015}]{McKean2015SKA}
{McKean} J.,  et~al., 2015, Advancing Astrophysics with the Square Kilometre
  Array (AASKA14), \href {http://adsabs.harvard.edu/abs/2015aska.confE..84M}
  {p.~84}

\bibitem[\protect\citeauthoryear{{Metcalf}}{{Metcalf}}{2002}]{Metcalf2002}
{Metcalf} R.~B.,  2002, \mn@doi [\apj] {10.1086/343766}, \href
  {http://adsabs.harvard.edu/abs/2002ApJ...580..696M} {580, 696}

\bibitem[\protect\citeauthoryear{{Mittal}, {Porcas}  \& {Wucknitz}}{{Mittal}
  et~al.}{2007}]{Mittal2007}
{Mittal} R.,  {Porcas} R.,   {Wucknitz} O.,  2007, \mn@doi [\aap]
  {10.1051/0004-6361:20066127}, \href
  {http://adsabs.harvard.edu/abs/2007A%26A...465..405M} {465, 405}

\bibitem[\protect\citeauthoryear{{More}, {McKean}, {Muxlow}, {Porcas},
  {Fassnacht}  \& {Koopmans}}{{More} et~al.}{2008}]{More2008}
{More} A.,  {McKean} J.~P.,  {Muxlow} T.~W.~B.,  {Porcas} R.~W.,  {Fassnacht}
  C.~D.,   {Koopmans} L.~V.~E.,  2008, \mn@doi [\mnras]
  {10.1111/j.1365-2966.2007.12831.x}, \href
  {http://adsabs.harvard.edu/abs/2008MNRAS.384.1701M} {384, 1701}

\bibitem[\protect\citeauthoryear{{More}, {McKean}, {More}, {Porcas}, {Koopmans}
   \& {Garrett}}{{More} et~al.}{2009}]{More2009}
{More} A.,  {McKean} J.~P.,  {More} S.,  {Porcas} R.~W.,  {Koopmans} L.~V.~E.,
   {Garrett} M.~A.,  2009, \mn@doi [\mnras] {10.1111/j.1365-2966.2008.14342.x},
  \href {http://adsabs.harvard.edu/abs/2009MNRAS.394..174M} {394, 174}

\bibitem[\protect\citeauthoryear{{Morgan}, {Mantovani}, {Deller}, {Brisken},
  {Alef}, {Middelberg}, {Nanni}  \& {Tingay}}{{Morgan}
  et~al.}{2011}]{Morgan2011}
{Morgan} J.~S.,  {Mantovani} F.,  {Deller} A.~T.,  {Brisken} W.,  {Alef} W.,
  {Middelberg} E.,  {Nanni} M.,   {Tingay} S.~J.,  2011, \mn@doi [\aap]
  {10.1051/0004-6361/201015775}, \href
  {http://adsabs.harvard.edu/abs/2011A%26A...526A.140M} {526, A140}

\bibitem[\protect\citeauthoryear{{Myers} et~al.,}{{Myers}
  et~al.}{2003}]{Myers2003}
{Myers} S.~T.,  et~al., 2003, \mn@doi [\mnras]
  {10.1046/j.1365-8711.2003.06256.x}, \href
  {http://adsabs.harvard.edu/abs/2003MNRAS.341....1M} {341, 1}

\bibitem[\protect\citeauthoryear{{Negrello} et~al.,}{{Negrello}
  et~al.}{2017}]{Negrello2017}
{Negrello} M.,  et~al., 2017, \mn@doi [\mnras] {10.1093/mnras/stw2911}, \href
  {http://adsabs.harvard.edu/abs/2017MNRAS.465.3558N} {465, 3558}

\bibitem[\protect\citeauthoryear{{Oguri}}{{Oguri}}{2006}]{Oguri2006}
{Oguri} M.,  2006, \mn@doi [\mnras] {10.1111/j.1365-2966.2006.10043.x}, \href
  {http://adsabs.harvard.edu/abs/2006MNRAS.367.1241O} {367, 1241}

\bibitem[\protect\citeauthoryear{{Paragi} et~al.,}{{Paragi}
  et~al.}{2015}]{Paragi2015}
{Paragi} Z.,  et~al., 2015, Advancing Astrophysics with the Square Kilometre
  Array (AASKA14), \href {http://adsabs.harvard.edu/abs/2015aska.confE.143P}
  {p.~143}

\bibitem[\protect\citeauthoryear{{Phillips} et~al.,}{{Phillips}
  et~al.}{2001}]{Phillips2001}
{Phillips} P.~M.,  et~al., 2001, \mn@doi [\mnras]
  {10.1046/j.1365-8711.2001.04601.x}, \href
  {http://adsabs.harvard.edu/abs/2001MNRAS.328.1001P} {328, 1001}

\bibitem[\protect\citeauthoryear{{Planck Collaboration} et~al.,}{{Planck
  Collaboration} et~al.}{2016}]{Planck2016}
{Planck Collaboration} et~al., 2016, \mn@doi [\aap]
  {10.1051/0004-6361/201525830}, \href
  {http://adsabs.harvard.edu/abs/2016A%26A...594A..13P} {594, A13}

\bibitem[\protect\citeauthoryear{{Quinn} et~al.,}{{Quinn}
  et~al.}{2016}]{Quinn2016}
{Quinn} J.,  et~al., 2016, \mn@doi [\mnras] {10.1093/mnras/stw773}, \href
  {http://adsabs.harvard.edu/abs/2016MNRAS.459.2394Q} {459, 2394}

\bibitem[\protect\citeauthoryear{{Radcliffe}, {Garrett}, {Beswick}, {Muxlow},
  {Barthel}, {Deller}  \& {Middelberg}}{{Radcliffe}
  et~al.}{2016}]{Radcliffe2016}
{Radcliffe} J.~F.,  {Garrett} M.~A.,  {Beswick} R.~J.,  {Muxlow} T.~W.~B.,
  {Barthel} P.~D.,  {Deller} A.~T.,   {Middelberg} E.,  2016, \mn@doi [\aap]
  {10.1051/0004-6361/201527980}, \href
  {http://adsabs.harvard.edu/abs/2016A%26A...587A..85R} {587, A85}

\bibitem[\protect\citeauthoryear{{Riechers} et~al.,}{{Riechers}
  et~al.}{2011}]{Riechers2011}
{Riechers} D.~A.,  et~al., 2011, \mn@doi [\apjl] {10.1088/2041-8205/739/1/L32},
  \href {http://adsabs.harvard.edu/abs/2011ApJ...739L..32R} {739, L32}

\bibitem[\protect\citeauthoryear{{Rumbaugh}, {Fassnacht}, {McKean}, {Koopmans},
  {Auger}  \& {Suyu}}{{Rumbaugh} et~al.}{2015}]{Rumbaugh2015}
{Rumbaugh} N.,  {Fassnacht} C.~D.,  {McKean} J.~P.,  {Koopmans} L.~V.~E.,
  {Auger} M.~W.,   {Suyu} S.~H.,  2015, \mn@doi [\mnras]
  {10.1093/mnras/stv672}, \href
  {http://adsabs.harvard.edu/abs/2015MNRAS.450.1042R} {450, 1042}

\bibitem[\protect\citeauthoryear{{Rusin} et~al.,}{{Rusin}
  et~al.}{2001}]{Rusin2001}
{Rusin} D.,  et~al., 2001, \mn@doi [\aj] {10.1086/321156}, \href
  {http://adsabs.harvard.edu/abs/2001AJ....122..591R} {122, 591}

\bibitem[\protect\citeauthoryear{{Sharon}, {Riechers}, {Hodge}, {Carilli},
  {Walter}, {Wei{\ss}}, {Knudsen}  \& {Wagg}}{{Sharon}
  et~al.}{2016}]{Sharon2016}
{Sharon} C.~E.,  {Riechers} D.~A.,  {Hodge} J.,  {Carilli} C.~L.,  {Walter} F.,
   {Wei{\ss}} A.,  {Knudsen} K.~K.,   {Wagg} J.,  2016, \mn@doi [\apj]
  {10.3847/0004-637X/827/1/18}, \href
  {http://adsabs.harvard.edu/abs/2016ApJ...827...18S} {827, 18}

\bibitem[\protect\citeauthoryear{{Spingola}, {McKean}, {Auger}, {Fassnacht},
  {Koopmans}, {Lagattuta}  \& {Vegetti}}{{Spingola}
  et~al.}{2018}]{Spingola2018}
{Spingola} C.,  {McKean} J.~P.,  {Auger} M.~W.,  {Fassnacht} C.~D.,  {Koopmans}
  L.~V.~E.,  {Lagattuta} D.~J.,   {Vegetti} S.,  2018, \mn@doi [\mnras]
  {10.1093/mnras/sty1326}, \href
  {http://adsabs.harvard.edu/abs/2018MNRAS.478.4816S} {478, 4816}

\bibitem[\protect\citeauthoryear{{Stacey} et~al.,}{{Stacey}
  et~al.}{2018}]{Stacey2018}
{Stacey} H.~R.,  et~al., 2018, \mn@doi [\mnras] {10.1093/mnras/sty458}, \href
  {http://adsabs.harvard.edu/abs/2018MNRAS.476.5075S} {476, 5075}

\bibitem[\protect\citeauthoryear{{Suyu}, {Marshall}, {Auger}, {Hilbert},
  {Blandford}, {Koopmans}, {Fassnacht}  \& {Treu}}{{Suyu}
  et~al.}{2010}]{Suyu2010}
{Suyu} S.~H.,  {Marshall} P.~J.,  {Auger} M.~W.,  {Hilbert} S.,  {Blandford}
  R.~D.,  {Koopmans} L.~V.~E.,  {Fassnacht} C.~D.,   {Treu} T.,  2010, \mn@doi
  [\apj] {10.1088/0004-637X/711/1/201}, \href
  {http://adsabs.harvard.edu/abs/2010ApJ...711..201S} {711, 201}

\bibitem[\protect\citeauthoryear{{Suyu} et~al.,}{{Suyu}
  et~al.}{2012}]{Suyu2012}
{Suyu} S.~H.,  et~al., 2012, \mn@doi [\apj] {10.1088/0004-637X/750/1/10}, \href
  {http://adsabs.harvard.edu/abs/2012ApJ...750...10S} {750, 10}

\bibitem[\protect\citeauthoryear{{Treu}}{{Treu}}{2010}]{Treu2010}
{Treu} T.,  2010, \mn@doi [\araa] {10.1146/annurev-astro-081309-130924}, \href
  {http://adsabs.harvard.edu/abs/2010ARA%26A..48...87T} {48, 87}

\bibitem[\protect\citeauthoryear{{Turner}, {Ostriker}  \& {Gott}}{{Turner}
  et~al.}{1984}]{Turner1984}
{Turner} E.~L.,  {Ostriker} J.~P.,   {Gott} III J.~R.,  1984, \mn@doi [\apj]
  {10.1086/162379}, \href {http://adsabs.harvard.edu/abs/1984ApJ...284....1T}
  {284, 1}

\bibitem[\protect\citeauthoryear{{Vegetti}, {Lagattuta}, {McKean}, {Auger},
  {Fassnacht}  \& {Koopmans}}{{Vegetti} et~al.}{2012}]{Vegetti2012}
{Vegetti} S.,  {Lagattuta} D.~J.,  {McKean} J.~P.,  {Auger} M.~W.,  {Fassnacht}
  C.~D.,   {Koopmans} L.~V.~E.,  2012, \mn@doi [\nat] {10.1038/nature10669},
  \href {http://adsabs.harvard.edu/abs/2012Natur.481..341V} {481, 341}

\bibitem[\protect\citeauthoryear{{Vieira} et~al.,}{{Vieira}
  et~al.}{2013}]{Vieira2013}
{Vieira} J.~D.,  et~al., 2013, \mn@doi [\nat] {10.1038/nature12001}, \href
  {http://adsabs.harvard.edu/abs/2013Natur.495..344V} {495, 344}

\bibitem[\protect\citeauthoryear{{Walsh}, {Carswell}  \& {Weymann}}{{Walsh}
  et~al.}{1979}]{Walsh1979}
{Walsh} D.,  {Carswell} R.~F.,   {Weymann} R.~J.,  1979, \mn@doi [\nat]
  {10.1038/279381a0}, \href {http://adsabs.harvard.edu/abs/1979Natur.279..381W}
  {279, 381}

\bibitem[\protect\citeauthoryear{{Wambsganss}}{{Wambsganss}}{1998}]{Wambsganss1998}
{Wambsganss} J.,  1998, \mn@doi [Living Reviews in Relativity]
  {10.12942/lrr-1998-12}, \href
  {http://adsabs.harvard.edu/abs/1998LRR.....1...12W} {1, 12}

\bibitem[\protect\citeauthoryear{{Wilkinson} et~al.,}{{Wilkinson}
  et~al.}{2001}]{Wilkinson2001}
{Wilkinson} P.~N.,  et~al., 2001, \mn@doi [Physical Review Letters]
  {10.1103/PhysRevLett.86.584}, \href
  {http://adsabs.harvard.edu/abs/2001PhRvL..86..584W} {86, 584}

\bibitem[\protect\citeauthoryear{{Winn} et~al.,}{{Winn}
  et~al.}{2000}]{Winn2000}
{Winn} J.~N.,  et~al., 2000, \mn@doi [\aj] {10.1086/316874}, \href
  {http://adsabs.harvard.edu/abs/2000AJ....120.2868W} {120, 2868}

\bibitem[\protect\citeauthoryear{{Winn}, {Rusin}  \& {Kochanek}}{{Winn}
  et~al.}{2004}]{Winn2004}
{Winn} J.~N.,  {Rusin} D.,   {Kochanek} C.~S.,  2004, \mn@doi [\nat]
  {10.1038/nature02279}, \href
  {http://adsabs.harvard.edu/abs/2004Natur.427..613W} {427, 613}

\bibitem[\protect\citeauthoryear{{Wrobel}, {Garrett}, {Condon}  \&
  {Morganti}}{{Wrobel} et~al.}{2004}]{Wrobel2004}
{Wrobel} J.~M.,  {Garrett} M.~A.,  {Condon} J.~J.,   {Morganti} R.,  2004,
  \mn@doi [\aj] {10.1086/421743}, \href
  {http://adsabs.harvard.edu/abs/2004AJ....128..103W} {128, 103}

\bibitem[\protect\citeauthoryear{{Wucknitz}, {Biggs}  \& {Browne}}{{Wucknitz}
  et~al.}{2004}]{Wucknitz2004}
{Wucknitz} O.,  {Biggs} A.~D.,   {Browne} I.~W.~A.,  2004, \mn@doi [\mnras]
  {10.1111/j.1365-2966.2004.07514.x}, \href
  {http://adsabs.harvard.edu/abs/2004MNRAS.349...14W} {349, 14}

\bibitem[\protect\citeauthoryear{{Zhang}, {Jackson}, {Porcas}  \&
  {Browne}}{{Zhang} et~al.}{2007}]{Zhang2007}
{Zhang} M.,  {Jackson} N.,  {Porcas} R.~W.,   {Browne} I.~W.~A.,  2007, \mn@doi
  [\mnras] {10.1111/j.1365-2966.2007.11718.x}, \href
  {http://adsabs.harvard.edu/abs/2007MNRAS.377.1623Z} {377, 1623}

\makeatother
\end{thebibliography}



\appendix

\section{Supplementary figures and table}

\begin{figure*}
\centering
\includegraphics[scale = 1.25]{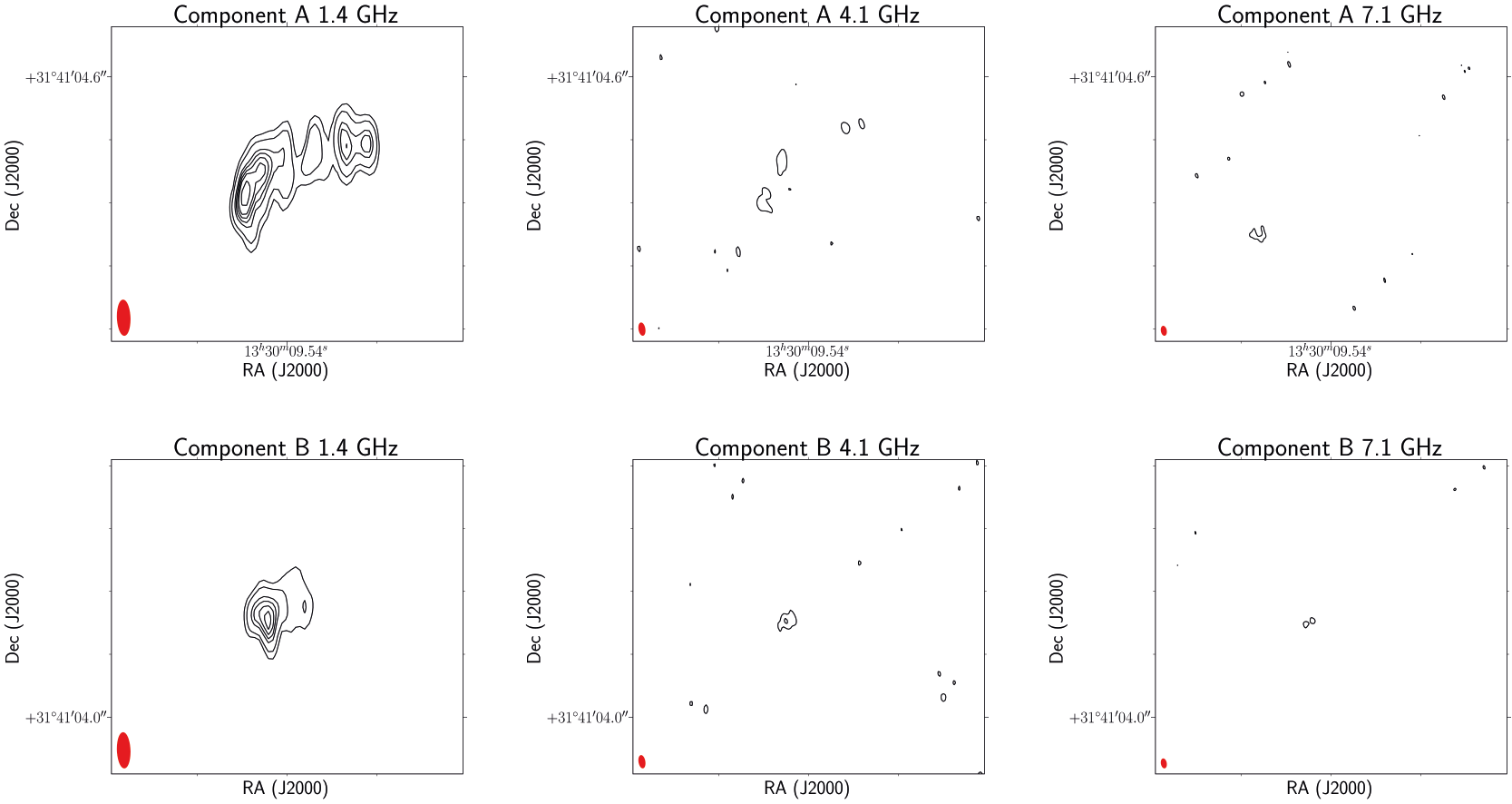}
\caption{Multi-frequency images of the lens candidate MJV00019 at 1.4 (left), 4.1 (centre) and 7.1 GHz (right). Component A is shown on the top row, component B is shown on the bottom row. The beam is plotted in red on the bottom left corner of each image. Contour levels increase by a factor of 3, where the first contour is 3 times the off-source noise level and the cutouts are 0.08 arcsec $\times$ 0.08 arcsec. The cutouts are centred at the coordinates listed in Table \ref{Tab:fluxdensities}. }\label{Fig:mjv00019-composite}
\end{figure*}

\begin{figure*}
\centering
\includegraphics[scale = 1.25]{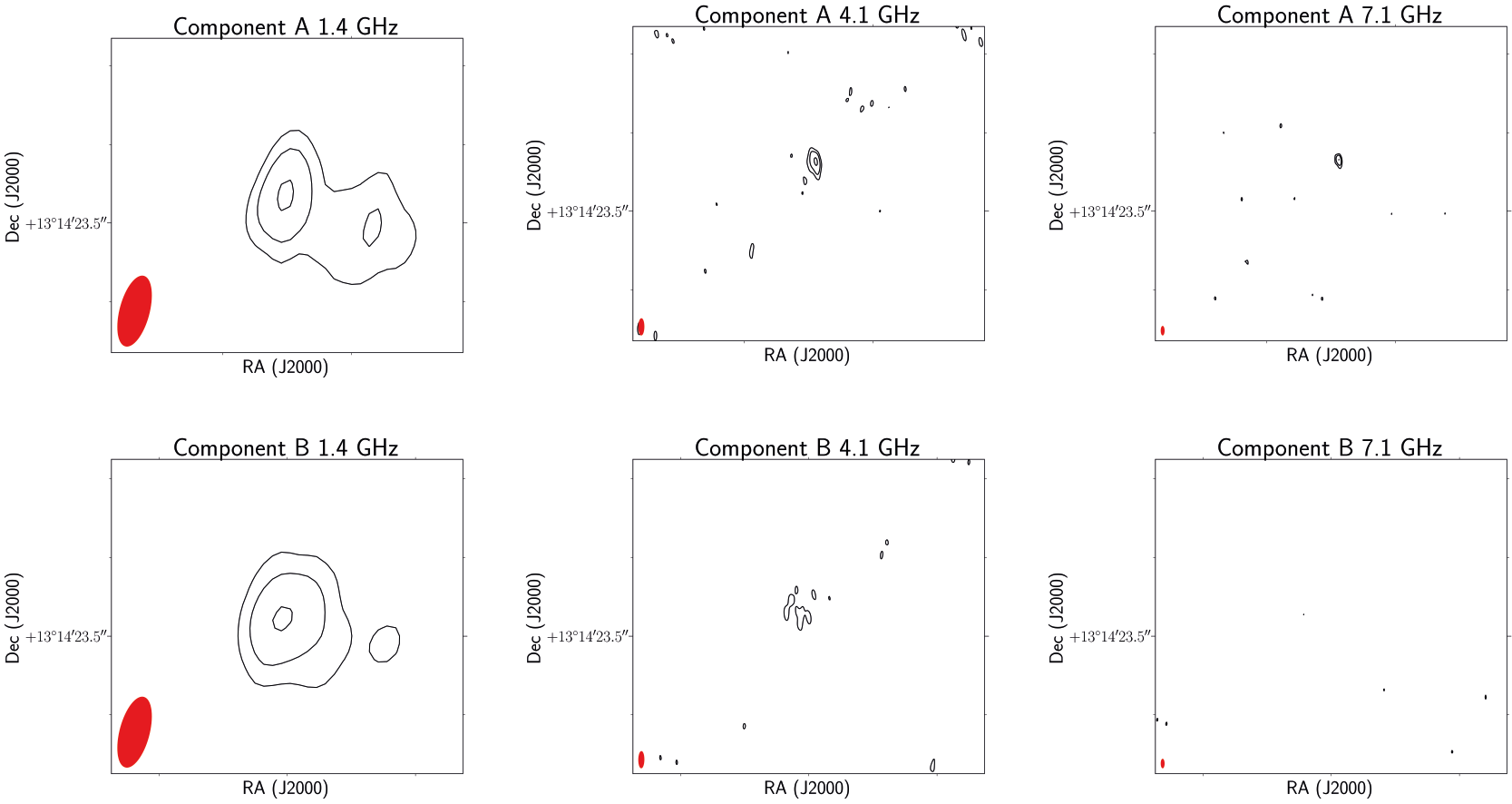}
\caption{Multi-frequency images of the lens candidate MJV00533 at 1.4 (left), 4.1 (centre) and 7.1 GHz (right). Component A is shown on the top row, component B is shown on the bottom row. The beam is plotted in red on the bottom left corner of each image. Contour levels increase by a factor of 3, where the first contour is 3 times the off-source noise level and the cutouts are 0.08 arcsec $\times$ 0.08 arcsec. The cutouts are centred at the coordinates listed in Table \ref{Tab:fluxdensities}.}\label{Fig:mjv00533-composite}
\end{figure*}

\begin{figure*}
\centering
\includegraphics[scale = 1.25]{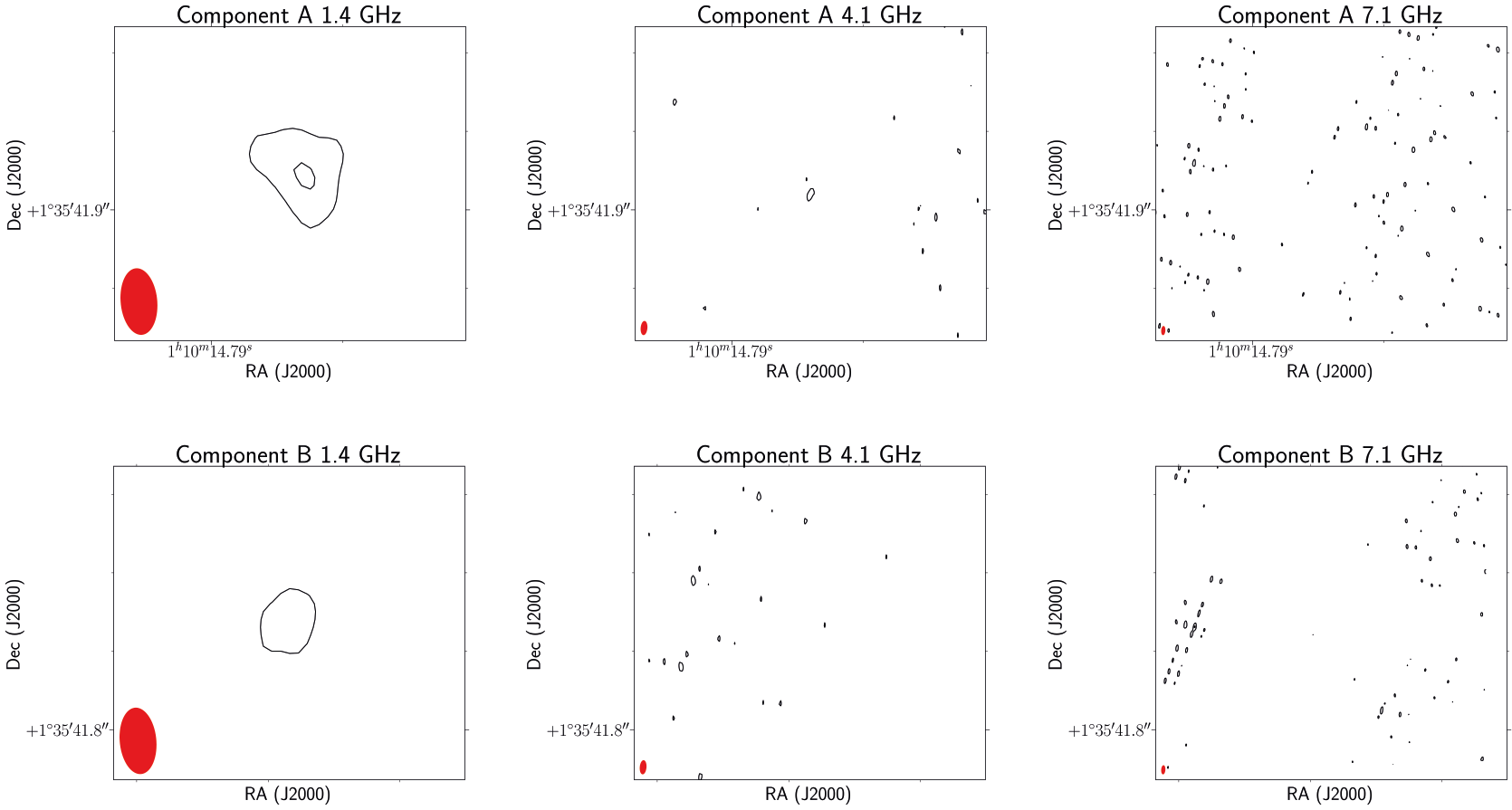}
\caption{Multi-frequency images of the lens candidate MJV02990 at 1.4 (left), 4.1 (centre) and 7.1 GHz (right). Component A is shown on the top row, component B is shown on the bottom row. The beam is plotted in red on the bottom left corner of each image. Contour levels increase by a factor of 3, where the first contour is 3 times the off-source noise level and the cutouts are 0.08 arcsec $\times$ 0.08 arcsec. The cutouts are centred at the coordinates listed in Table \ref{Tab:fluxdensities}}\label{Fig:mjv02990-composite}
\end{figure*}

\begin{figure*}
\centering
\includegraphics[scale = 1.25]{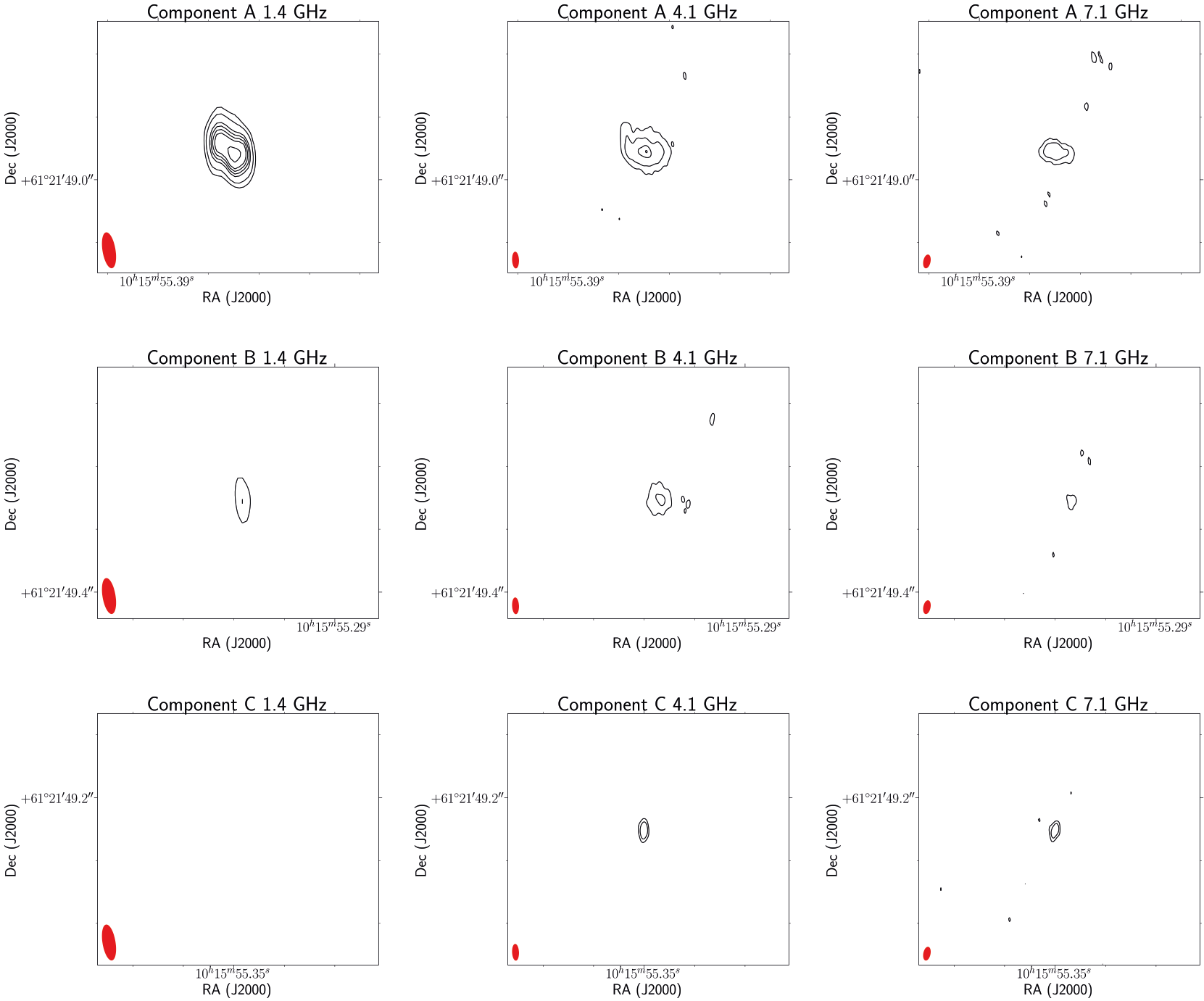}
\caption{Multi-frequency images of the lens candidate MJV04363 at 1.4 (left), 4.1 (centre) and 7.1 GHz (right). Component A is shown on the top row, component B is shown on the second row and component C is shown on the bottom row. The beam is plotted in red on the bottom left corner of each image. Contour levels increase by a factor of 3, where the first contour is 3 times the off-source noise level and the cutouts are 0.08 arcsec $\times$ 0.08 arcsec. The cutouts are centred at the coordinates listed in Table \ref{Tab:fluxdensities}}\label{Fig:mjv04363-composite}
\end{figure*}

\begin{figure*}
\centering
\includegraphics[scale = 1.25]{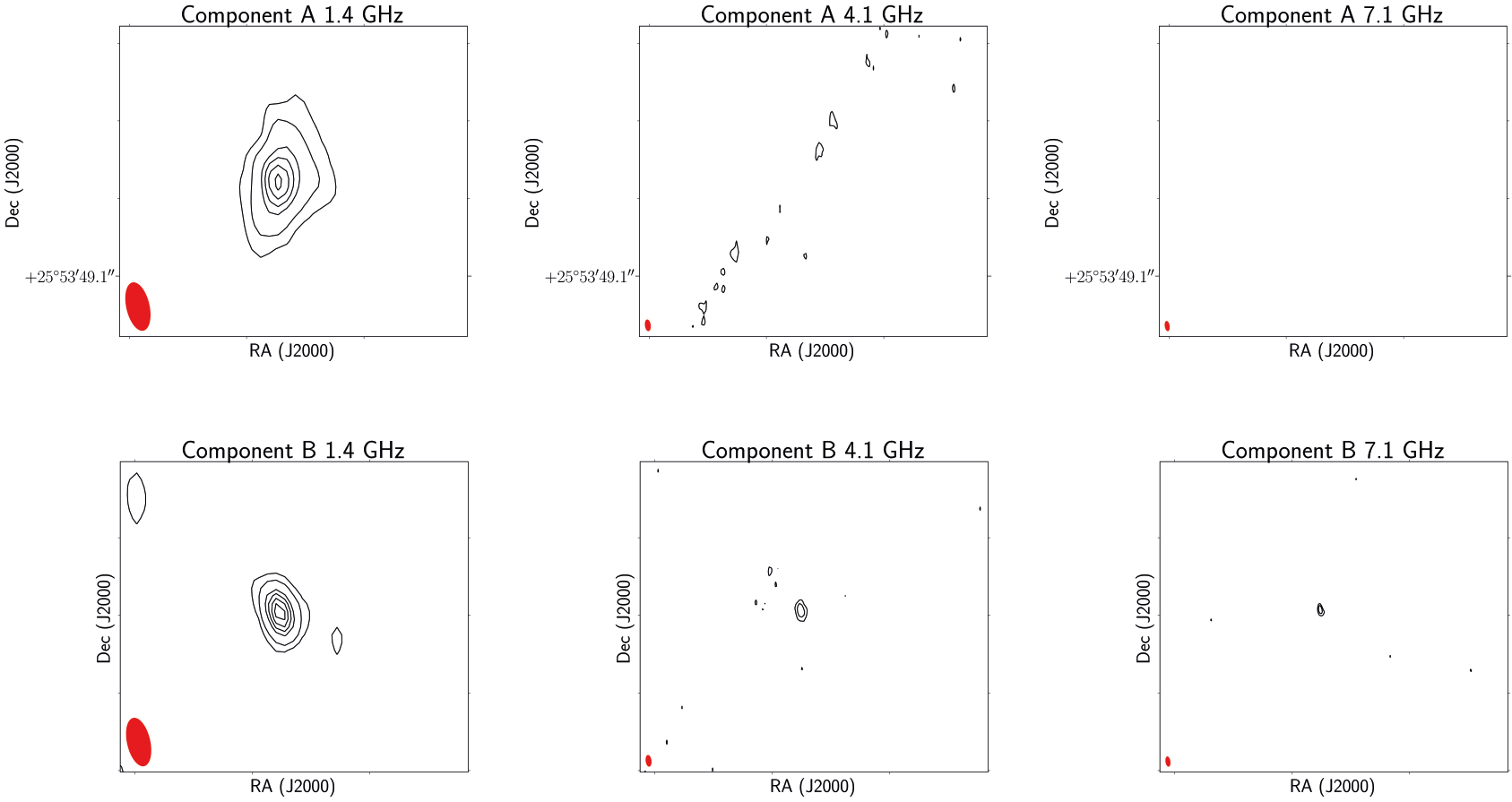}
\caption{Multi-frequency images of the lens candidate MJV06997 at 1.4 (left), 4.1 (centre) and 7.1 GHz (right). Component A is shown on the top row, component B is shown on the bottom row. The beam is plotted in red on the bottom left corner of each image. Contour levels increase by a factor of 3, where the first contour is 3 times the off-source noise level and the cutouts are 0.08 arcsec $\times$ 0.08 arcsec. The cutouts are centred at the coordinates listed in Table \ref{Tab:fluxdensities}. }\label{Fig:mjv06997-composite}
\end{figure*}

\begin{figure*}
\centering
\includegraphics[scale = 1.25]{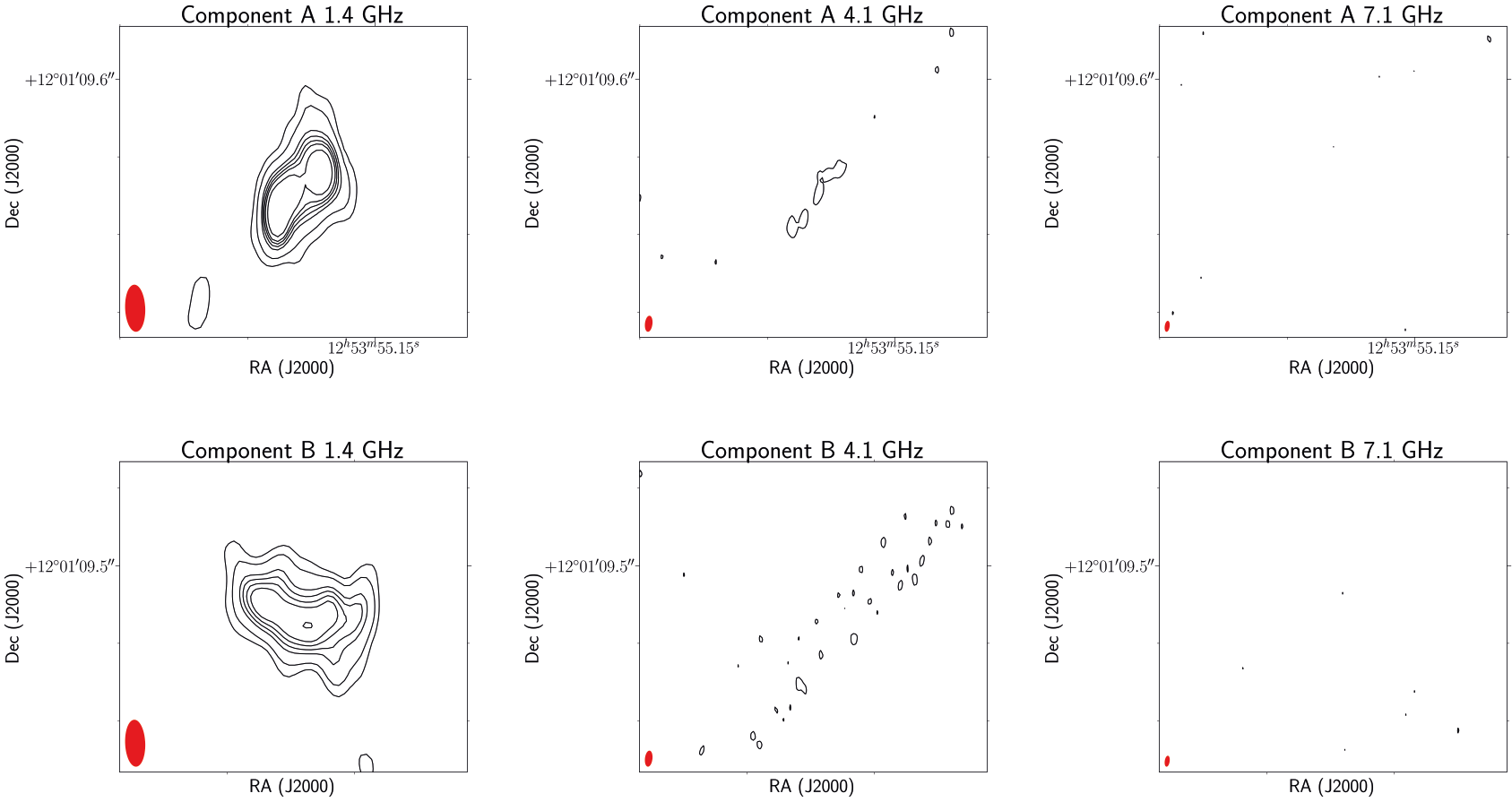}
\caption{Multi-frequency images of the lens candidate MJV07382 at 1.4 (left), 4.1 (centre) and 7.1 GHz (right). Component A is shown on the top row, component B is shown on the bottom row. The beam is plotted in red on the bottom left corner of each image. Contour levels increase by a factor of 3, where the first contour is 3 times the off-source noise level and the cutouts are 0.08 arcsec $\times$ 0.08 arcsec. The cutouts are centred at the coordinates listed in Table \ref{Tab:fluxdensities}. }\label{Fig:mjv07382-composite}
\end{figure*}

\begin{figure*}
\centering
\includegraphics[scale = 1.25]{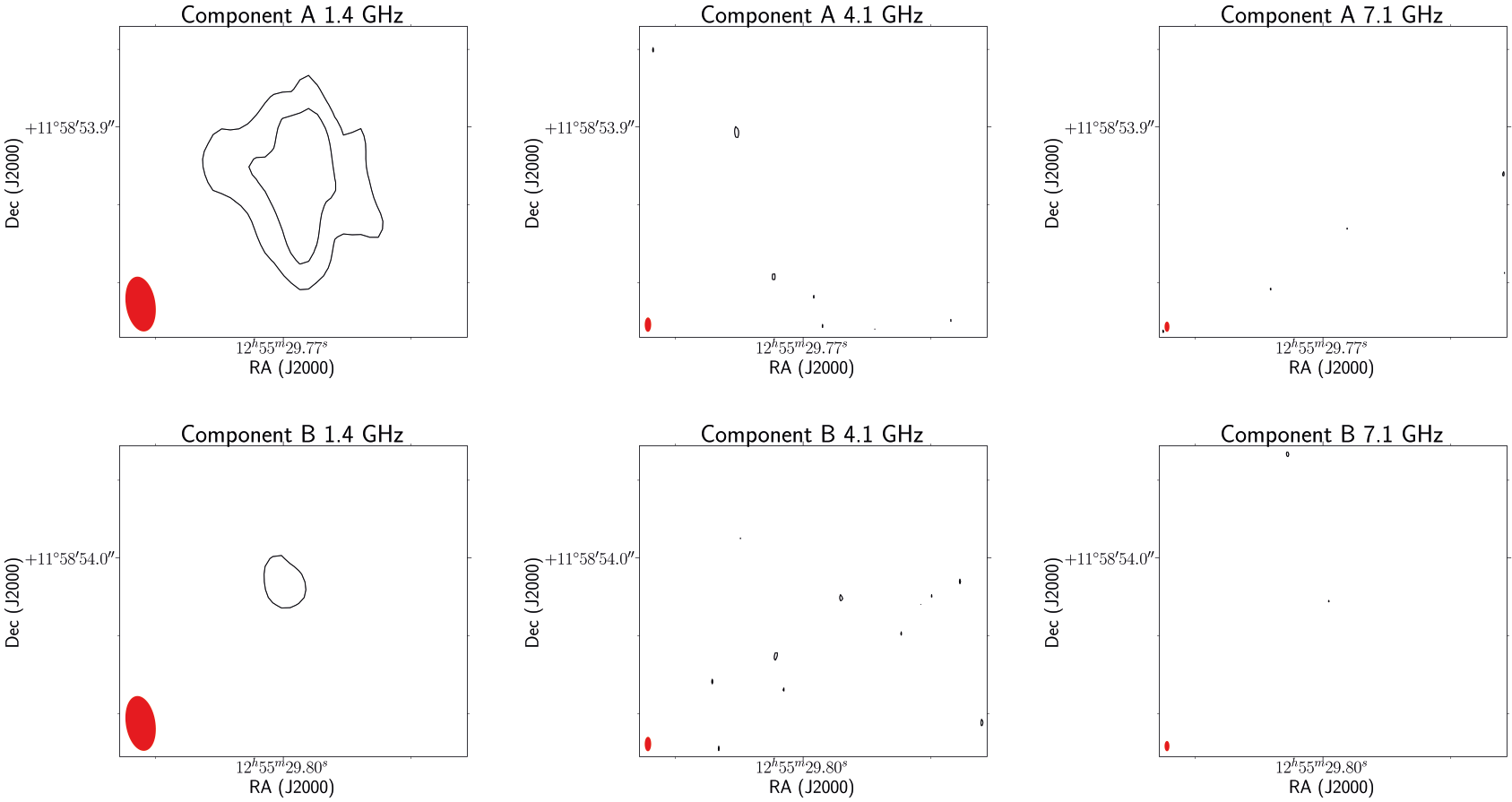}
\caption{Multi-frequency images of the lens candidate MJV07417 at 1.4 (left), 4.1 (centre) and 7.1 GHz (right). Component A is shown on the top row, component B is shown on the bottom row. The beam is plotted in red on the bottom left corner of each image. Contour levels increase by a factor of 3, where the first contour is 3 times the off-source noise level and the cutouts are 0.08 arcsec $\times$ 0.08 arcsec. The cutouts are centred at the coordinates listed in Table \ref{Tab:fluxdensities}.}\label{Fig:mjv07417-composite}
\end{figure*}

\begin{figure*}
\centering
\includegraphics[scale = 1.25]{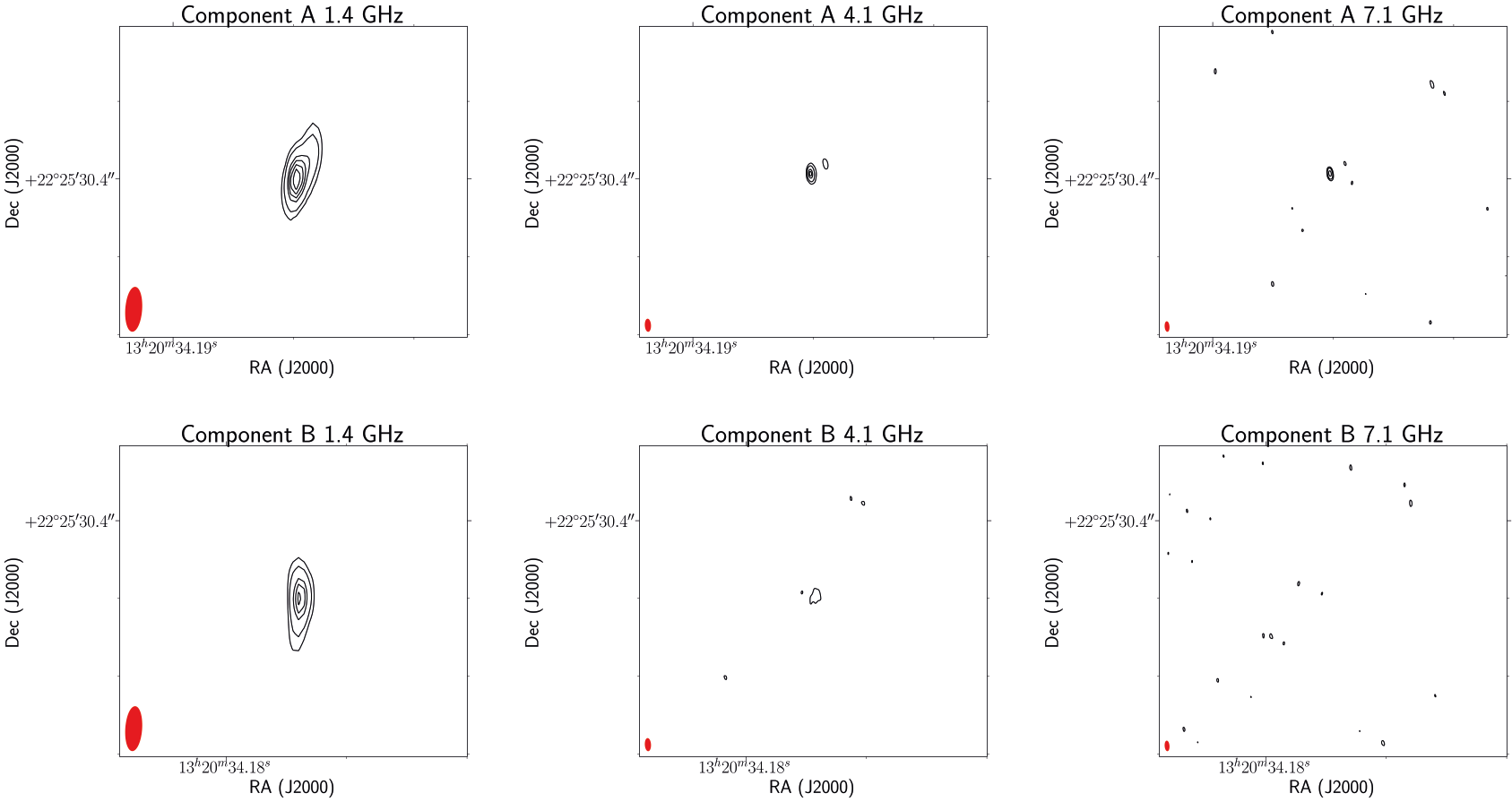}
\caption{Multi-frequency images of the lens candidate MJV07467 at 1.4 (left), 4.1 (centre) and 7.1 GHz (right). Component A is shown on the top row, component B is shown on the bottom row. The beam is plotted in red on the bottom left corner of each image. Contour levels increase by a factor of 3, where the first contour is 3 times the off-source noise level and the cutouts are 0.08 arcsec $\times$ 0.08 arcsec. The cutouts are centred at the coordinates listed in Table \ref{Tab:fluxdensities}.}\label{Fig:mjv07467-composite}
\end{figure*}

\begin{figure*}
\centering
\includegraphics[scale = 1.25]{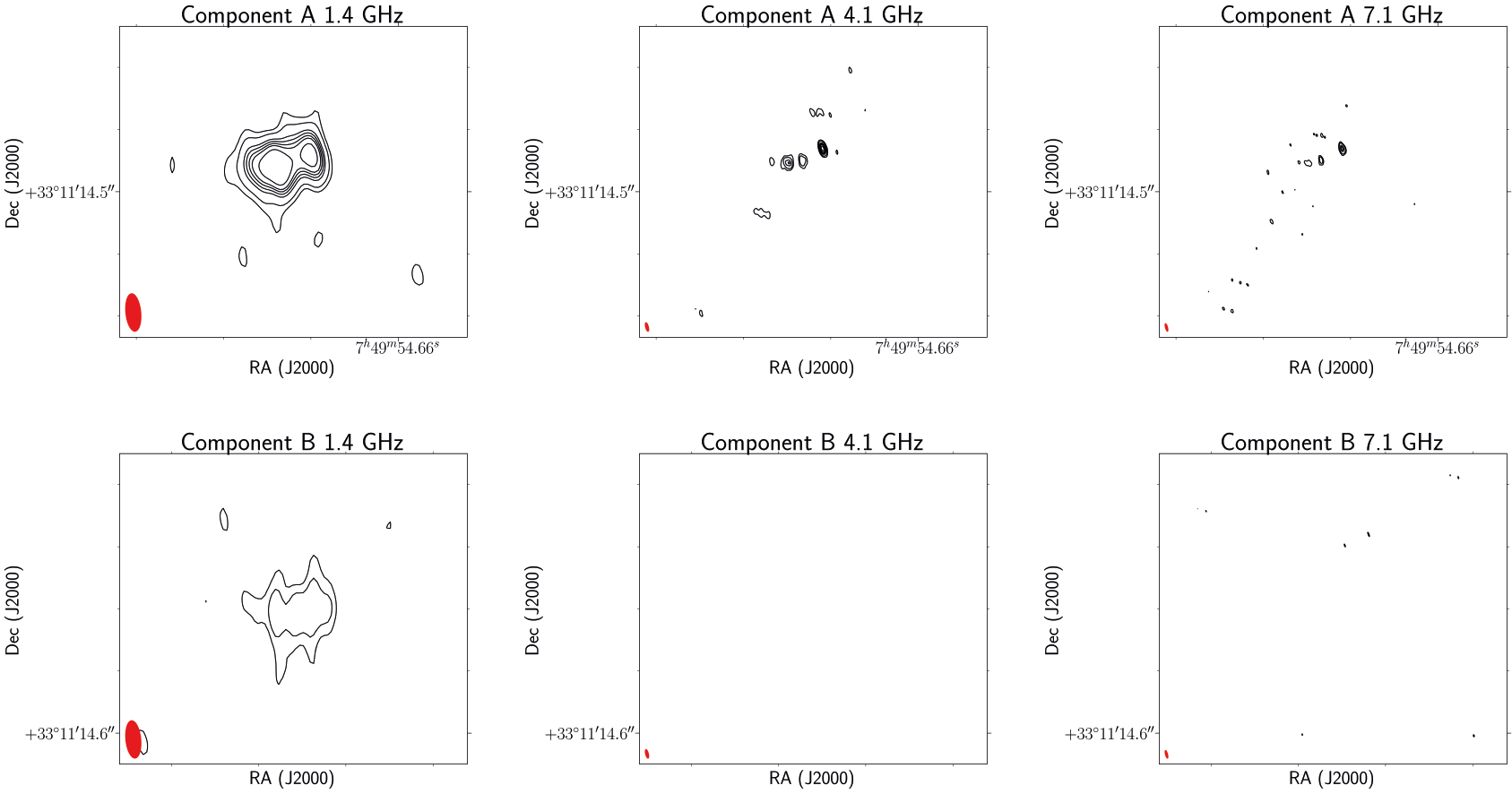}
\caption{Multi-frequency images of the lens candidate MJV11715 at 1.4 (left), 4.1 (centre) and 7.1 GHz (right). Component A is shown on the top row, component B is shown on the bottom row. The beam is plotted in red on the bottom left corner of each image. Contour levels increase by a factor of 3, where the first contour is 3 times the off-source noise level and the cutouts are 0.1 arcsec $\times$ 0.1 arcsec. The cutouts are centred at the coordinates listed in Table \ref{Tab:fluxdensities}.}\label{Fig:mjv11715-composite}
\end{figure*}

\begin{figure*}
\centering
\includegraphics[scale = 1.25]{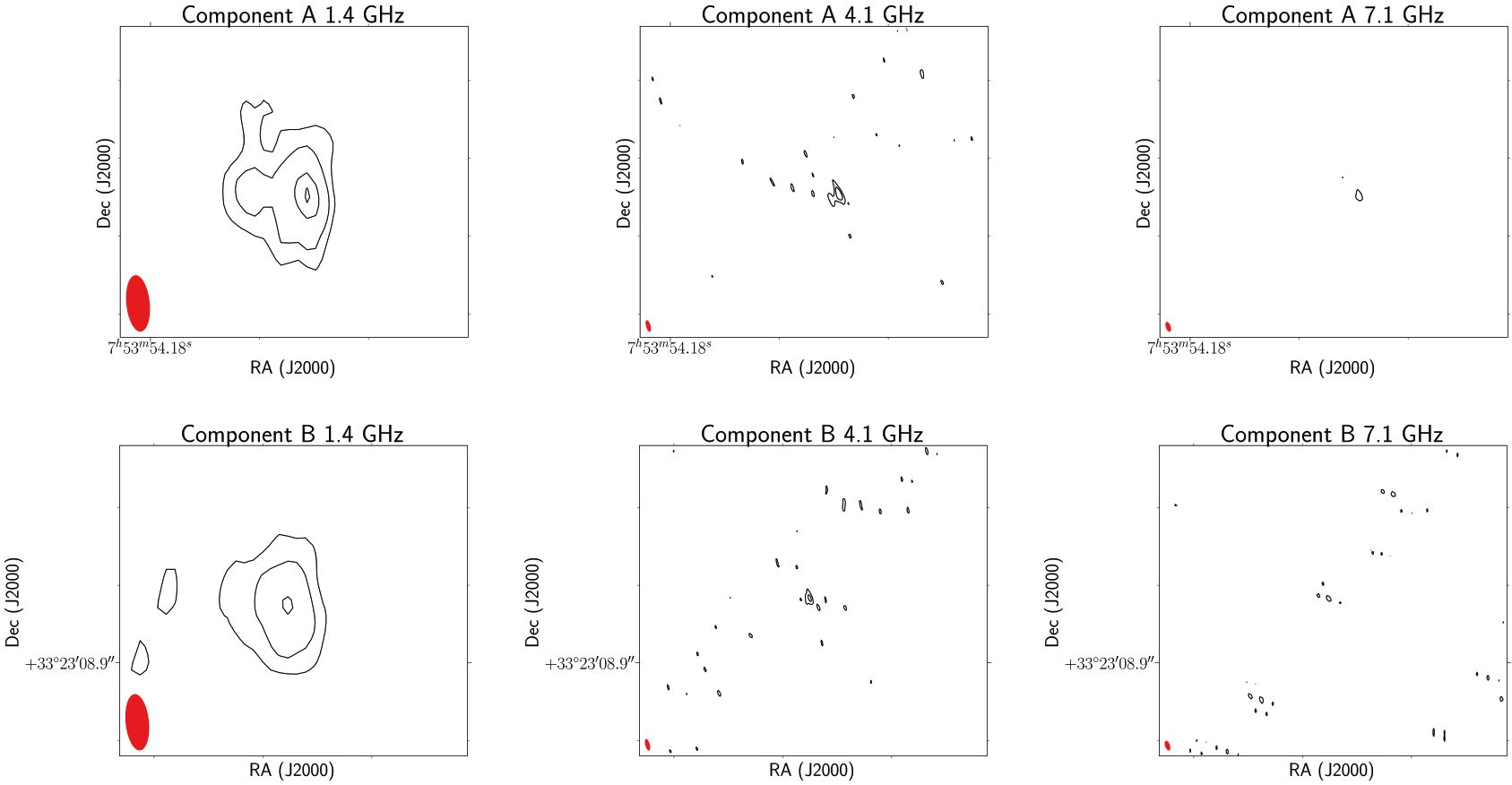}
\caption{Multi-frequency images of the lens candidate MJV11797 at 1.4 (left), 4.1 (centre) and 7.1 GHz (right). Component A is shown on the top row, component B is shown on the bottom row. The beam is plotted in red on the bottom left corner of each image. Contour levels increase by a factor of 3, where the first contour is 3 times the off-source noise level and the cutouts are 0.08 arcsec $\times$ 0.08 arcsec. The cutouts are centred at the coordinates listed in Table \ref{Tab:fluxdensities}.}\label{Fig:mjv11797-composite}
\end{figure*}

\begin{figure*}
\centering
\includegraphics[width = 0.45\textwidth]{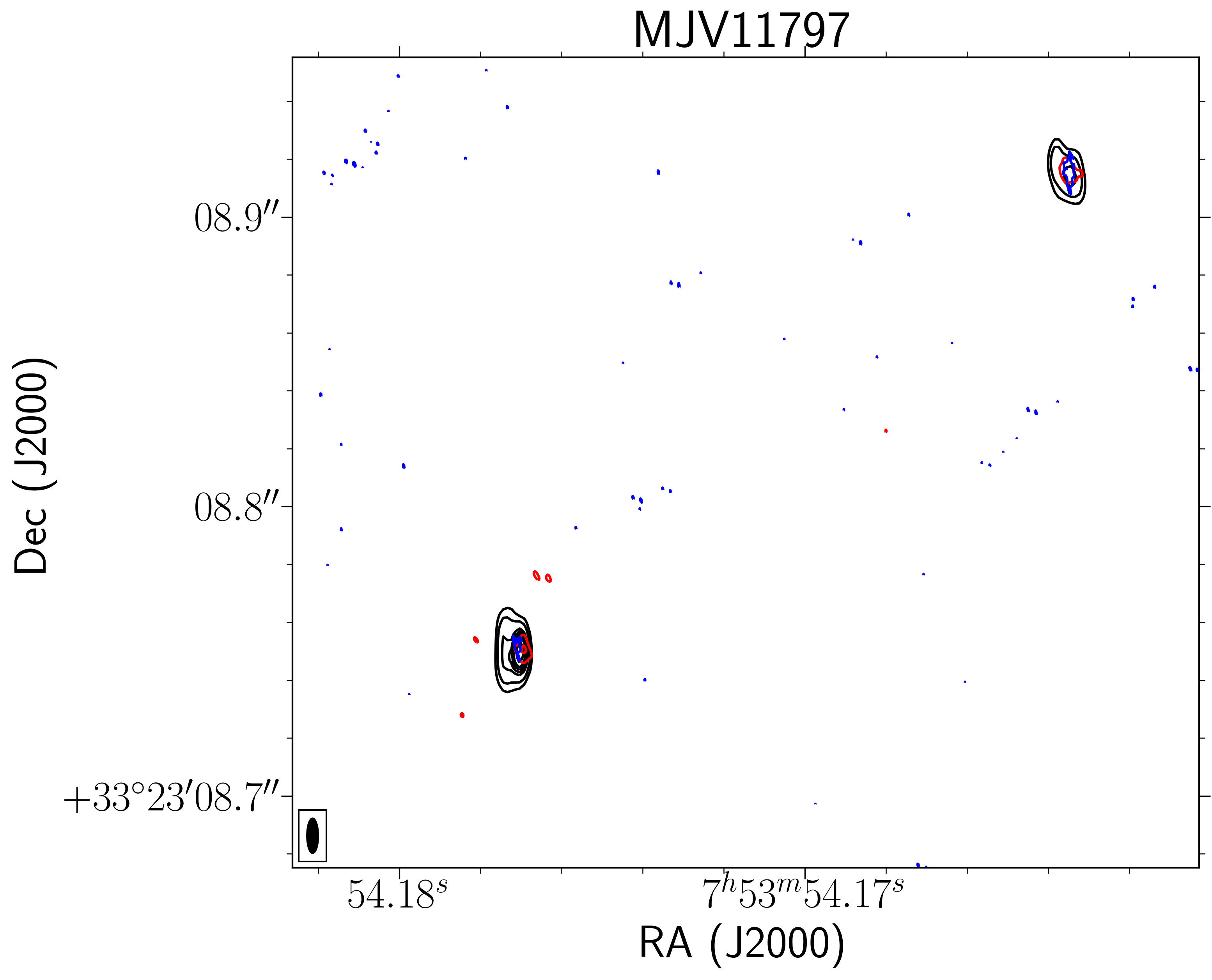}
\caption{Overlay of contours at 1.7 GHz (black),  4.1 GHz (red) and 7.1 GHz (blue) of the lens candidate MJV11797 smoothed to a resolution of 12 mas $\times$ 3 mas at P.A. of 0.150$^\circ$. The restoring beam is shown in the bottom left corner. The contour levels increase by a factor of 3, where the first contour is 3 times the off-source noise level.} \label{Fig:mjv11797-smooth}
\end{figure*}

\begin{figure*}
\centering
\includegraphics[scale = 1.25]{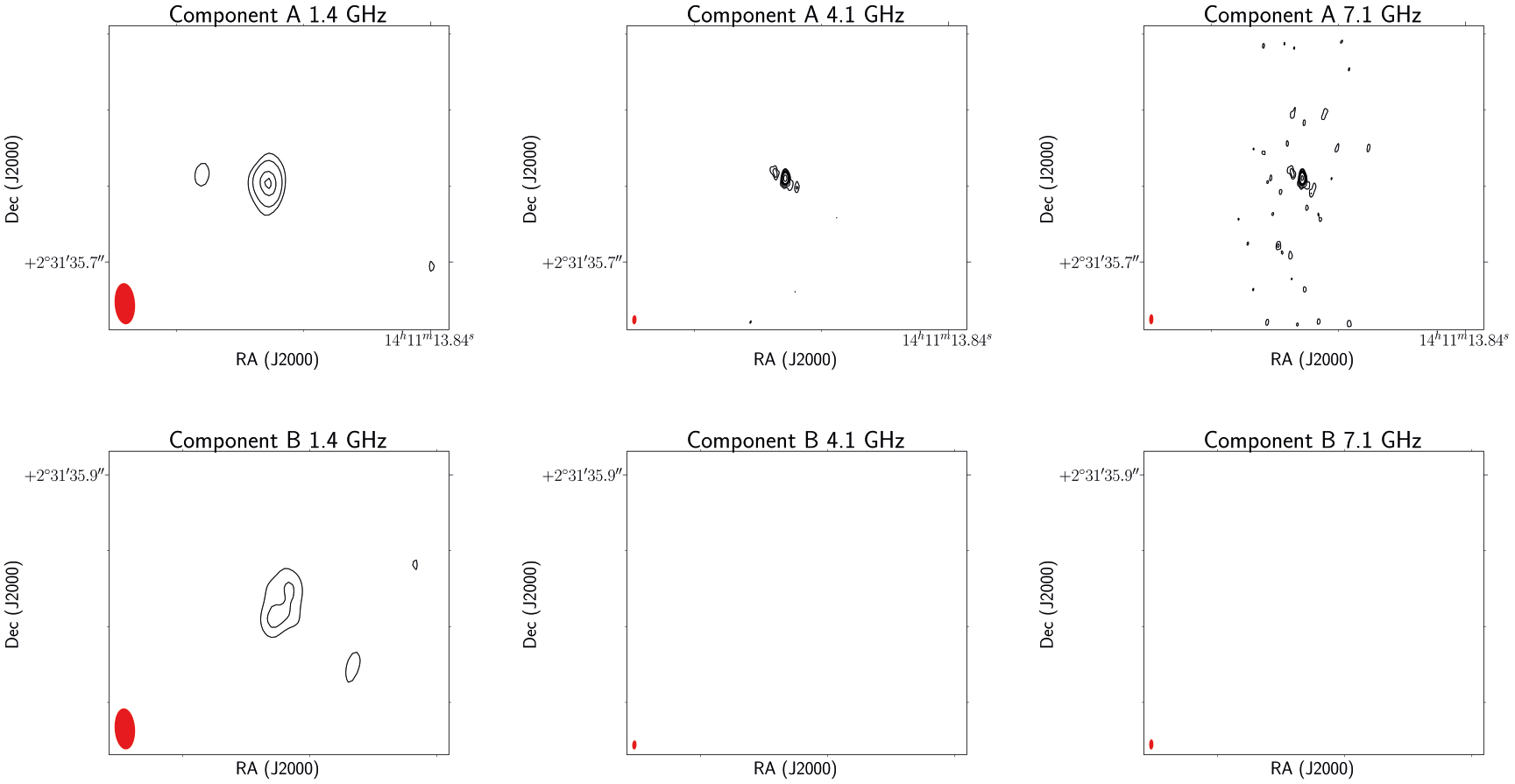}
\caption{Multi-frequency images of the lens candidate MJV14607 at 1.4 (left), 4.1 (centre) and 7.1 GHz (right). Component A is shown on the top row, component B is shown on the bottom row. The beam is plotted in red on the bottom left corner of each image. Contour levels increase by a factor of 3, where the first contour is 3 times the off-source noise level and the cutouts are 0.08 arcsec $\times$ 0.08 arcsec. The cutouts are centred at the coordinates listed in Table \ref{Tab:fluxdensities}.}\label{Fig:mjv14607-composite}
\end{figure*}

\begin{figure*}
\centering
\includegraphics[scale = 1.25]{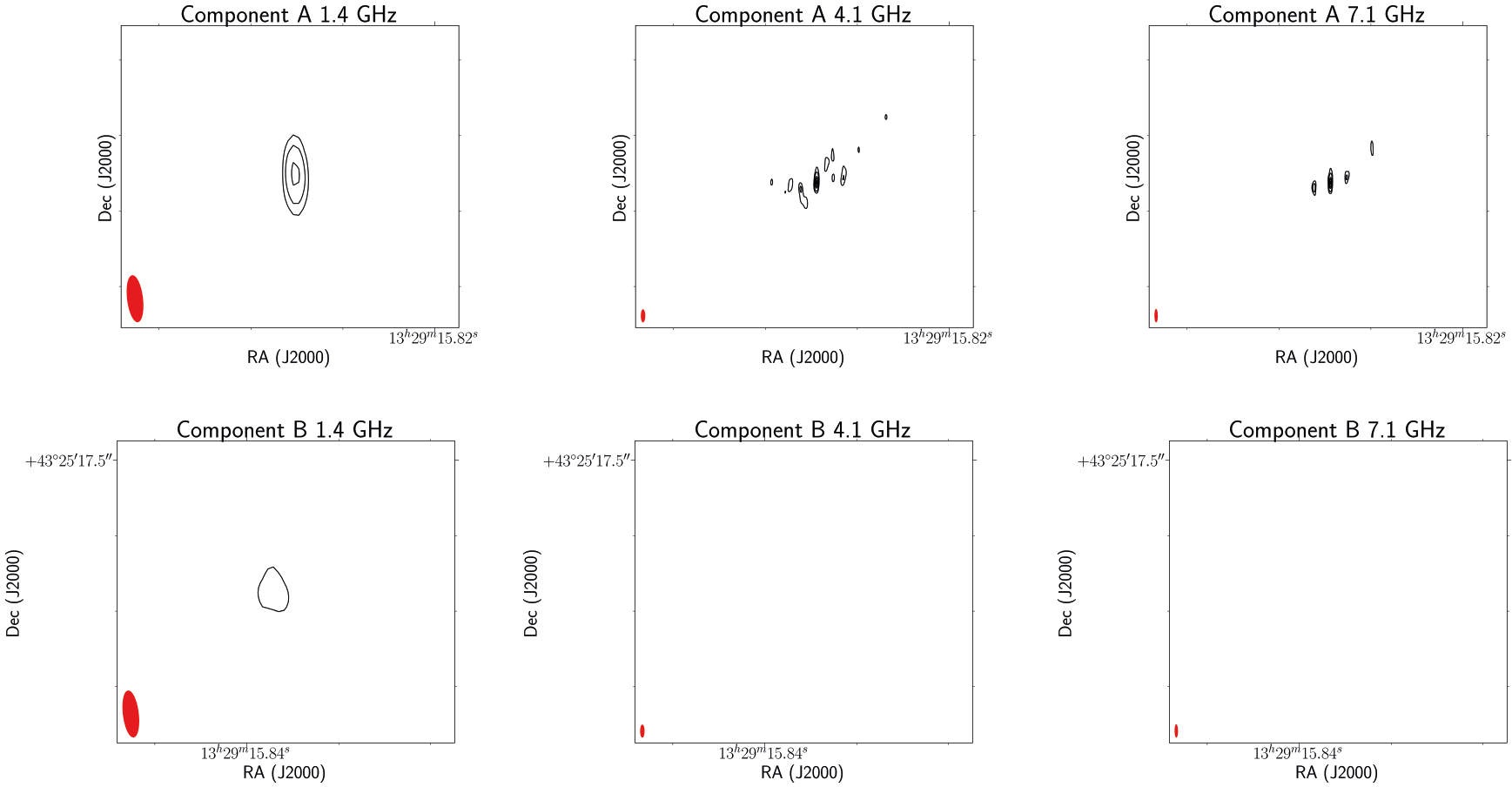}
\caption{Multi-frequency images of the lens candidate MJV16999 at 1.4 (left), 4.1 (centre) and 7.1 GHz (right). Component A is shown on the top row, component B is shown on the bottom row. The beam is plotted in red on the bottom left corner of each image. Contour levels increase by a factor of 3, where the first contour is 3 times the off-source noise level and the cutouts are 0.08 arcsec $\times$ 0.08 arcsec. The cutouts are centred at the coordinates listed in Table \ref{Tab:fluxdensities}.}\label{Fig:mjv16999-composite}
\end{figure*}

\begin{figure*}
\centering
	\includegraphics[scale = 1.1]{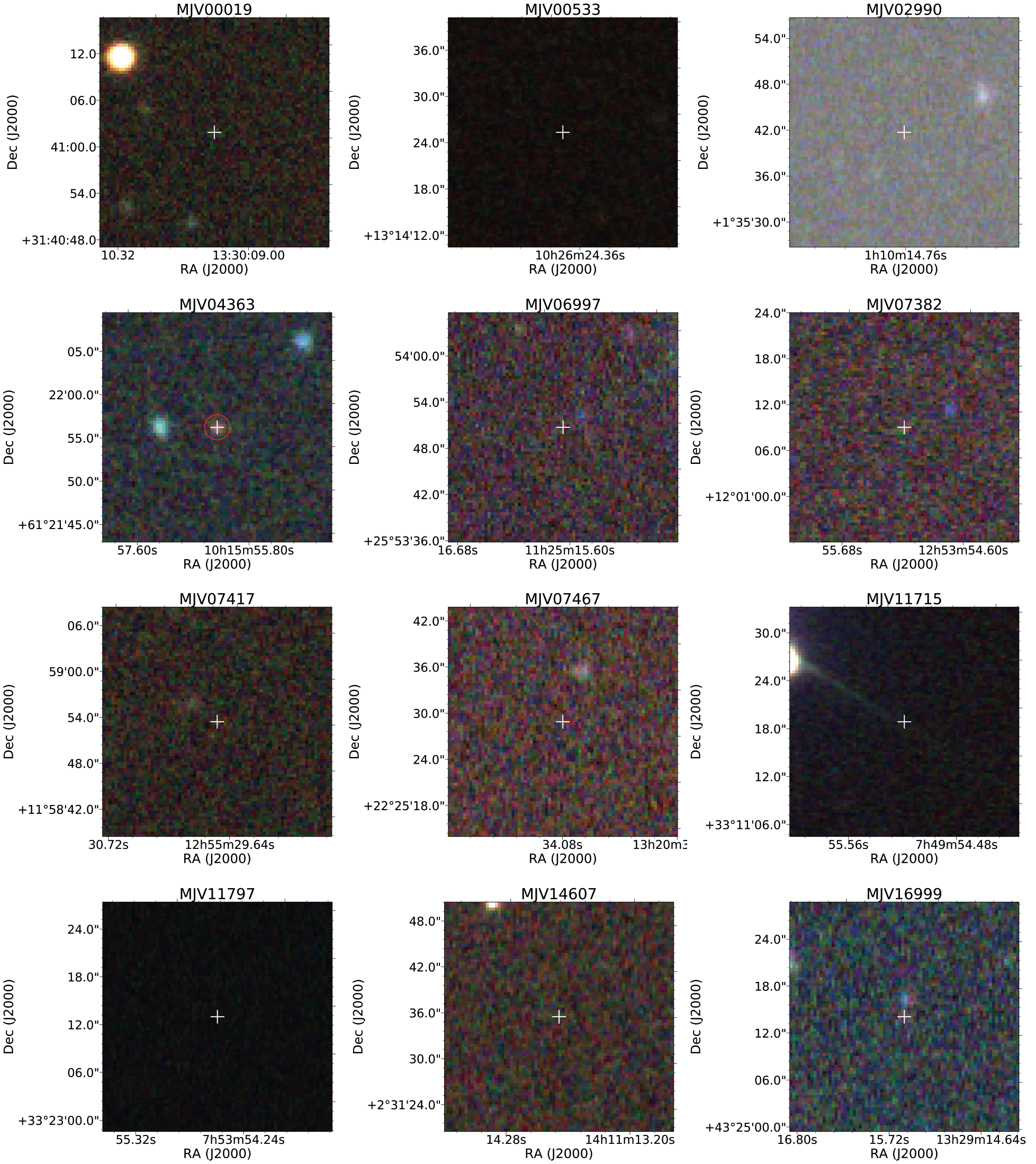}
	 \caption{RGB SDSS images of the mJIVE--20 lens candidates. Cutouts are 0.5 arcmin $\times$ 0.5 arcmin; the white cross indicates the position of component A at 1.4 GHz, the red circle indicates the possible optical counterpart of the radio detection.} \label{Fig:Panstarss}
\end{figure*}

This appendix includes the maps from the 4.1 and 7.1~GHz follow-up observations (Figs.~\ref{Fig:mjv00019-composite} to \ref{Fig:mjv16999-composite}), and the radio spectral energy distributions (Fig.~\ref{Fig:radio-spectra-1}) and the flux-ratios (Fig.~\ref{Fig:flux-ratio-1}) of the lens candidates. Optical images of the field around each target are shown in Fig.~\ref{Fig:Panstarss}.


\begin{figure*}
\centering
	 \includegraphics[scale = 0.76]{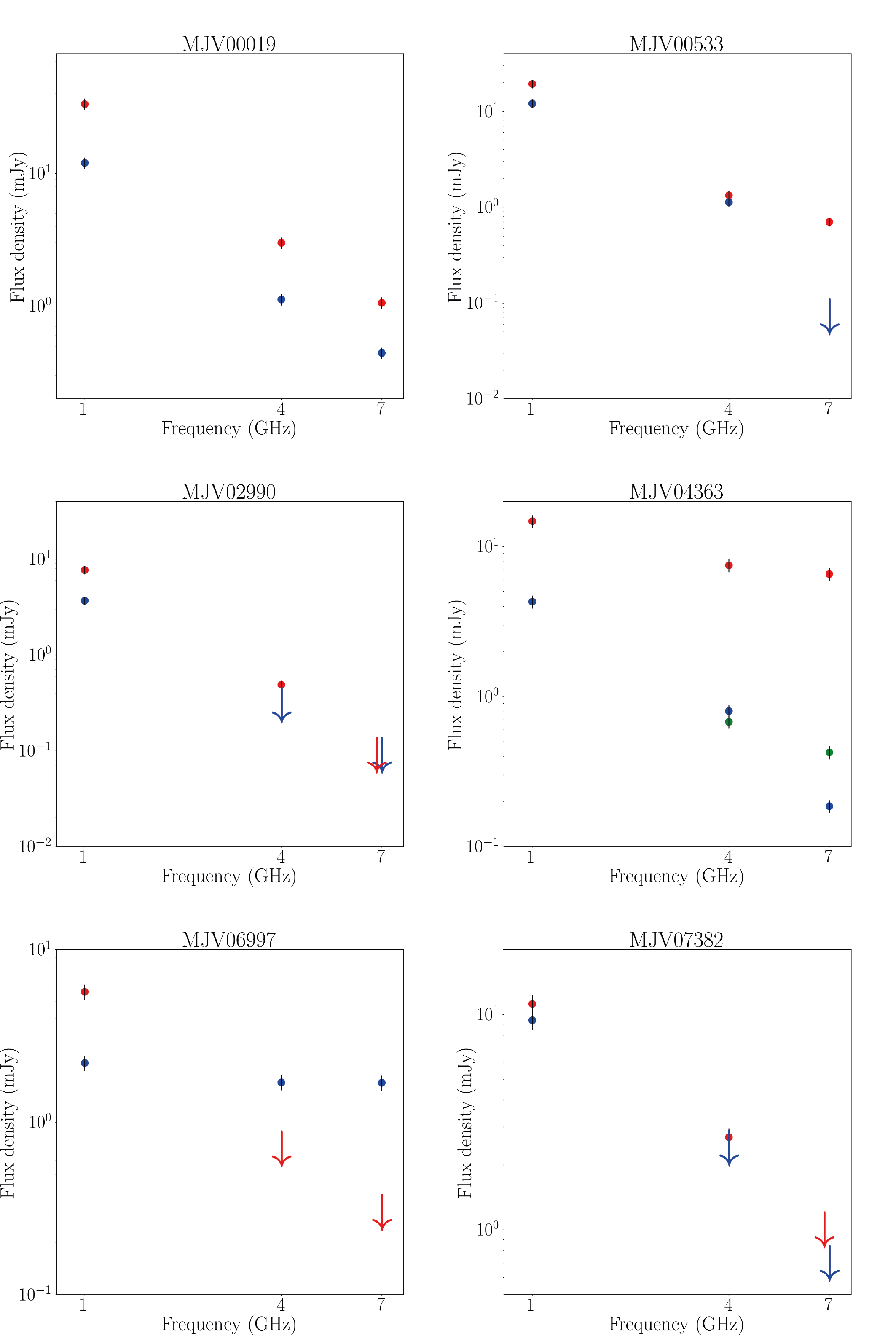}
    \caption{The spectral energy distribution of the lens candidates between 1.4 and 7.1 GHz. The filled circles indicate component A (red), B (blue) and, if detected, a third component C (green). The arrows indicate the $3\sigma$ detection limit at 4.1 and 7.1 GHz, estimated as three times the flux density within the same area of the 1.4 GHz detection of that image. The uncertainty on the flux density is assumed to be 10 per cent, which is a conservative estimate of the absolute flux density calibration of the VLBA.}
    \label{Fig:radio-spectra-1}
\end{figure*}

\begin{figure*}
\centering
	 \includegraphics[scale = 0.76]{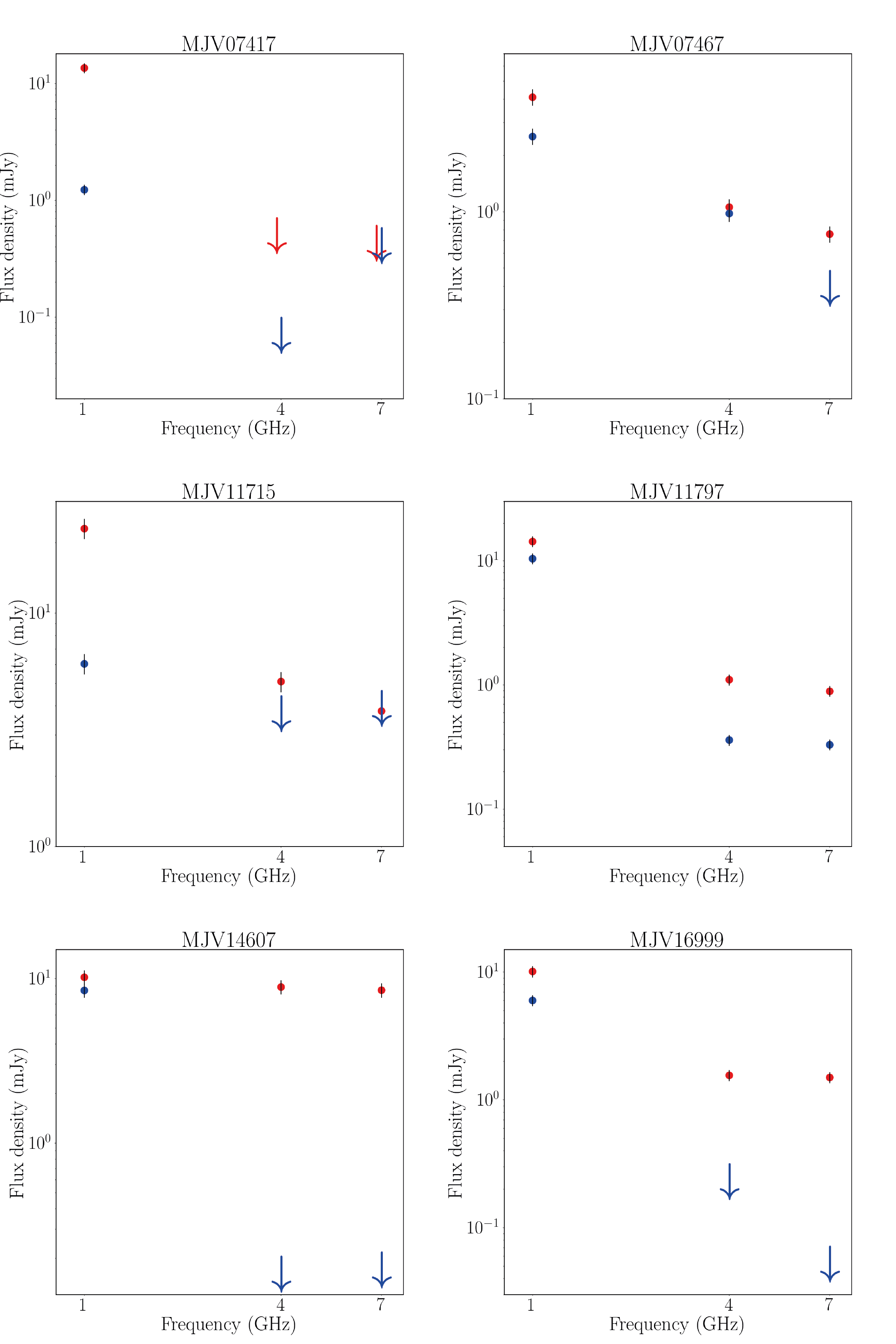}
\contcaption{}
\end{figure*}

\begin{figure*}
\centering
	 \includegraphics[scale = 0.8]{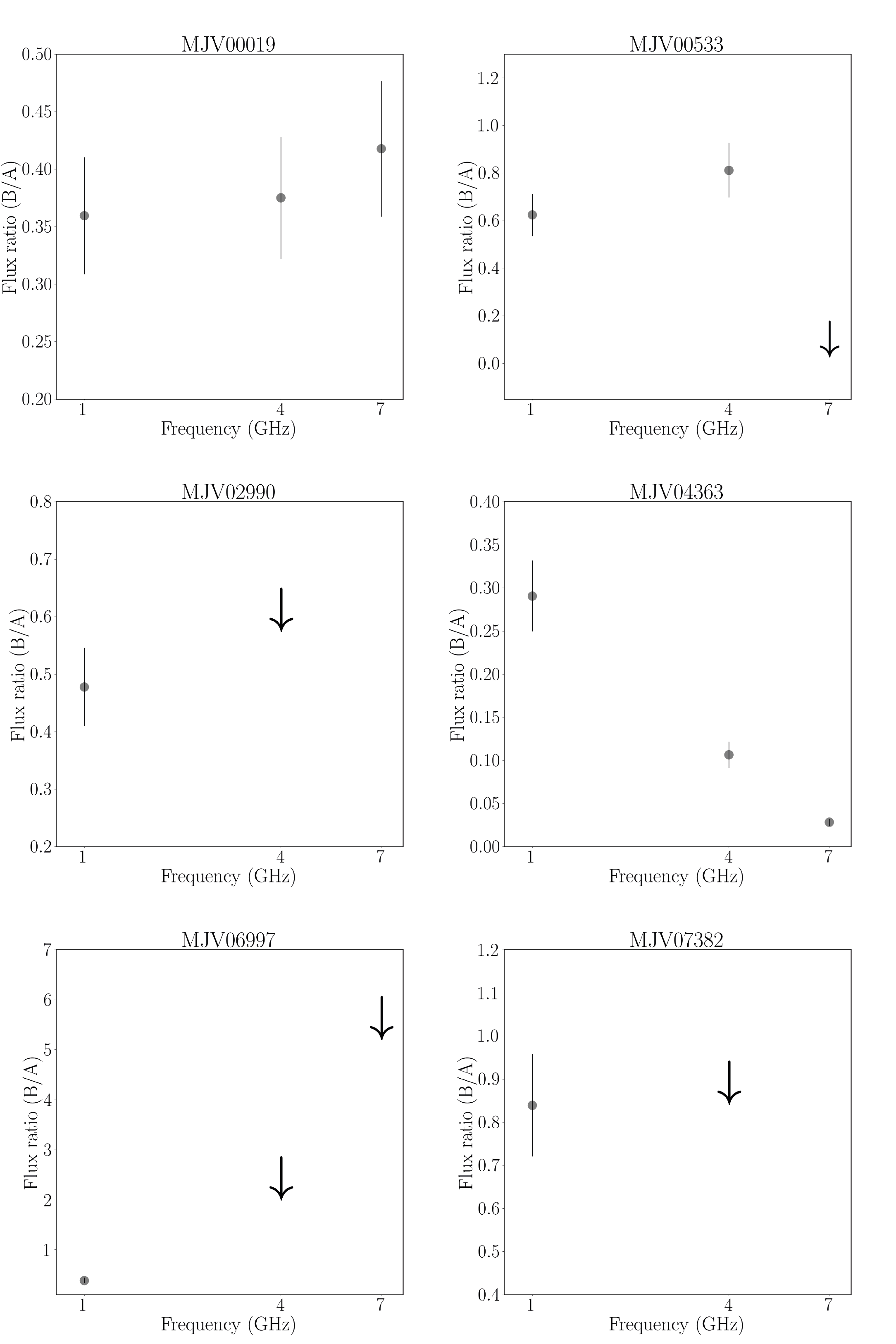}
    \caption{The flux density ratio (B/A) of the lens candidates as a function of frequency. The arrows represent the 3$\sigma$ upper limits when a detection has been made of at least one candidate lensed image.}
    \label{Fig:flux-ratio-1}
\end{figure*}

\begin{figure*}
\centering
	 \includegraphics[scale = 0.8]{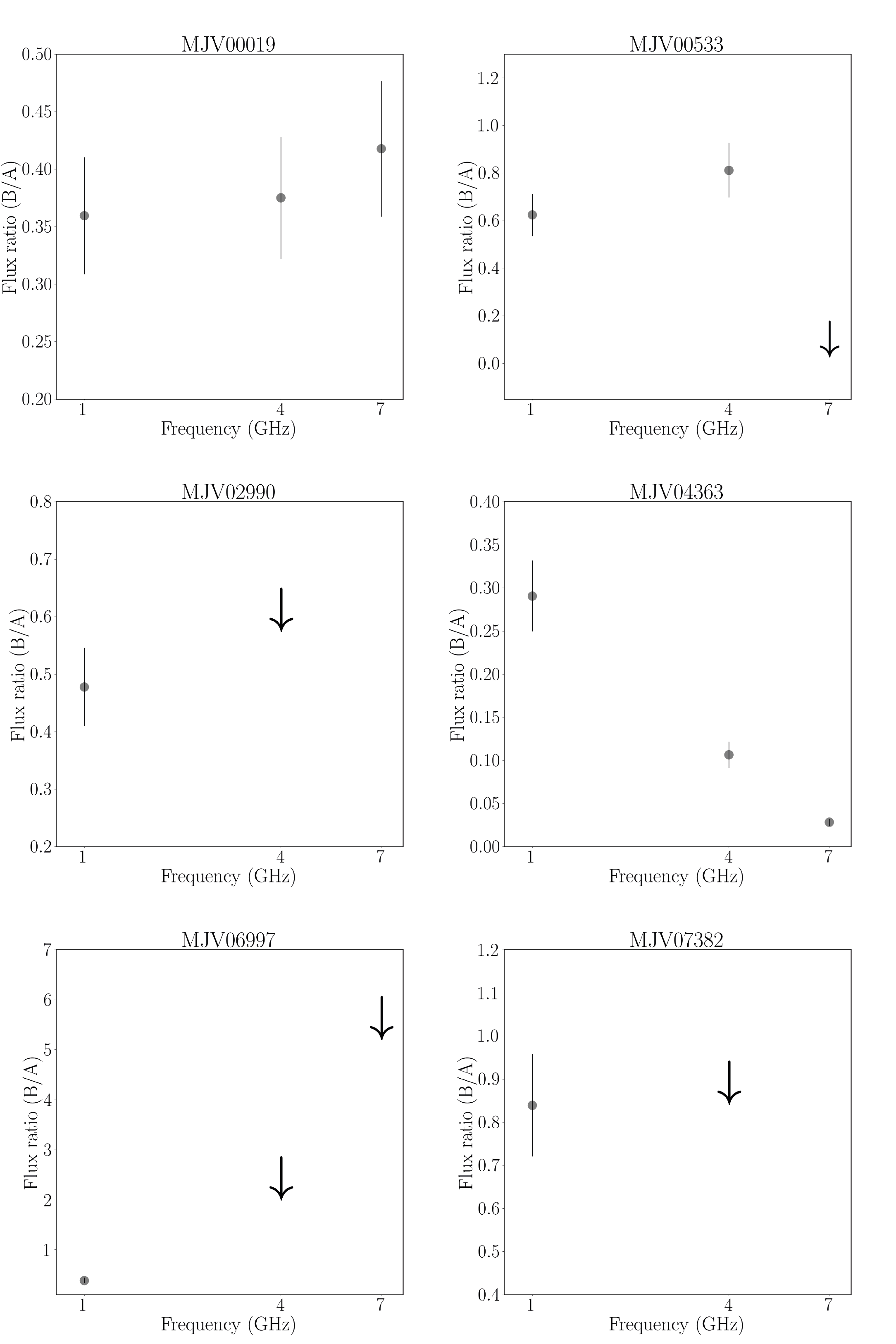}
\contcaption{}\label{Fig:flux-ratio-2}
\end{figure*}

\bsp	
\label{lastpage}
\end{document}